%% file: BW.tex
\documentclass[twocolumn,amsmath,showpacs,amsfonts,aps,prc,superscriptaddress]{revtex4} 

\usepackage{epsfig}
\usepackage{bm}

\newcommand{\2}{\frac{1}{2}}

\newcommand\bea{\begin{eqnarray}}
\newcommand\eea{\end{eqnarray}}

\newcommand{\second}{$2^{\rm nd}$}
\newcommand{\zeroth}{$0^{\rm th}$}

\newcommand{\Ro}[1]{R^2_{o,{#1}}}
\newcommand{\Rs}[1]{R^2_{s,{#1}}}
\newcommand{\Ros}[1]{R^2_{os,{#1}}}
\newcommand{\Rl}[1]{R^2_{l,{#1}}}
\newcommand{\Rmu}[1]{R^2_{\mu,{#1}}}

\begin{document}


\title{Observable implications of geometrical and dynamical aspects of freeze-out in heavy ion collisions}

\author{Fabrice Reti\`{e}re}
\affiliation{Lawrence Berkeley National Laboratory, Berkeley, CA 94720}
\author{Michael Annan Lisa}
\affiliation{Physics Department, The Ohio State University, Columbus, OH 43210}

\begin{abstract} 
Using an analytical parameterization of hadronic freeze-out in relativistic heavy ion collisions,
we present a detailed study of
the connections between features of the freeze-out configuration and physical
observables.  We focus especially on anisotropic freeze-out configurations (expected
in general for collisions at finite impact parameter), azimuthally-sensitive HBT
interferometry, and final-state interactions between non-identical particles.
Model calculations are compared with data taken in the first year of running at RHIC;
while not perfect, good agreement is found, raising the hope that a consistent
understanding of the full freeze-out scenario at RHIC is possible, an important
first step towards understanding the physics of the system prior to freeze-out.
\end{abstract}


\date{\today}
\maketitle

\input{S1_Introduction}

\input{S2_Model}

\input{S3_Observables}

\input{S4_Fits}

\input{S5_Conclusion}

\section*{Acknowledgments}
We gratefully acknowledge fruitful discussions with Drs.\ 
J. Castillo, T. Cs\"{o}rgo,
U. Heinz, P. Jacobs, P. Kolb, R. Lednicky, D. Magestro L. Ray, R. Snellings, B. Tom\'{a}\u{s}ik, S. Voloshin, R. Wells, and
U. Wiedemann.  We thank Drs. D. Magestro, K. Schweda and P. Sorensen for a 
careful reading of an early version of the text.
This work supported by the U.S. National Science Foundation under Grant No.\ PHY-0099476
and by the U.S. Department of Energy under Contract No.\ DE-AC03-76SF00098.


\end{document}

%% file: S1_Introduction.tex
\section{Introduction}

The first data from collisions between heavy nuclei at
the Relativistic Heavy Ion Collider (RHIC) have generated intense
theoretical efforts to understand the hot, dense matter generated
in the early stage of the collision~\cite{QMconferences}.
Testing these theoretical ideas relies on comparison to experimental observables.
Leptonic~\cite{leptonicObservables} or electromagnetic~\cite{gammaProbes} observables
are believed to probe directly the early, dense stage of the collision.  Most of the
early data from RHIC, however, have been on hadronic observables.  
Measurements of hadrons at high transverse momentum ($p_T$)~\cite{highpTexp} have
generated much excitement, as they may provide useful {\it probes} of the
dense medium produced at RHIC~\cite{highpTtheory}.  However, the medium {\it itself}
decays largely into the soft (low-$p_T$) hadronic sector.

Soft hadronic observables measure directly the final ``freeze-out'' stage of the
collision, when hadrons decouple from the bulk and free-stream to the detectors.
Freeze-out may correspond to a complex
configuration in the combined coordinate-momentum space, with collective components
(often called ``flow'') generating space-momentum correlations, as well as geometrical
and dynamical (flow) anisotropies.  
A detailed experimental-driven understanding of the freeze-out configuration is the
crucial first step in understanding the system's prior evolution and the physics
of hot colored matter.

In this paper, we explore in detail an analytic parameterization of the freeze-out
configuration, which includes non-trivial correlations between coordinate- and
momentum-space variables.  We discuss the connections between the physical parameters
of the model and observable quantities.  If the model, with correct choice of physical
model parameters, can adequately reproduce several independent measured quantities, then 
it might be claimed that this ``crucial first step,'' mentioned above, has been performed.

A consistent reproduction of all low-$p_T$ observations at RHIC is not achieved in 
most physical models which aim to describe the evolution of the collision.
In particular, it is difficult to reproduce momentum-space measurements while simultaneously
describing the freeze-out coordinate-space distribution probed by two-particle intensity
interferometry measurements~\cite{WH99} (also known as Hanburry-Brown-Twiss or
HBT~\cite{HanburryBrownTwiss} measurements).
Hadronic 
cascade models predict a too weak momentum azimuthal anisotropy and too 
large source sizes~\cite{CascadeFail}.  Hydrodynamic transport models
describe successfully transverse mass 
spectra and elliptic flow but fail at describing pion source radii~\cite{HK01};
some hydrodynamic models have successfully reproduced pion source radii~\cite{Hirano_HBT},
but only with different model parameters than those used to reproduce spectra and
elliptic flow~\cite{Hirano_SpectraV2}.  
Similarly, sophisticated hybrid transport models (e.g. AMPT~\cite{AMPT}) require
different model parameters~\cite{ZiWei} to reproduce data on elliptic flow~\cite{AMPTv2}
and HBT~\cite{AMPT_HBT}.
Good reproduction of observed values has been acheived in
models which adjust parameters to fit data within a given freeze-out scenario, such as in
the Buda-Lund hydro approach~\cite{BudaLund}.  The work presented here falls into this latter
category.

The parameterization used in this paper 
(``blast-wave parameterization'') is similar in form to the freeze-out configuration
obtained from hydrodynamic calculations~\cite{HKHRV01}, but we treat the physical
parameters of the configuration (e.g. temperature) as free parameters.  
Our main goal is simply to quantify the driving physical parameters of freeze-out at RHIC, and the
dependence of observables on these parameters.  

Further motivation for exploring freeze-out configurations of the type discussed here, is
that they implicitly assume a ``bulk'' system which may be described by global parameters
(temperature, flow strength, etc).  Discussions of a ``new phase of matter'' and its ``Equation of State''
are only sensible if indeed such assumptions hold.  Comparison of blast-wave calculations with
several independent measurements, then, is a crucial consistency check of these assumptions
(though, of course, a successful comparison still would not constitute a proof of their validity).

In transport models,
whether the constituents are hadrons~\cite{RQMD,Humanic}, partons~\cite{MolnarPC,BassPC},
or fluid elements~\cite{HK01,HK02,TLS01,TeaneyViscosity}, if they re-interact substantially,
pressure gradients are generated, leading to collective velocity fields (``flow''),
pushing the matter away from the hot center of the collision and into the surrounding
vacuum.
Evidence of collective flow, generated by {\it final} state re-interaction of collision products,
has been based largely on interpretations of
transverse mass spectra and transverse momentum azimuthal 
anisotropy~\cite{HK01}.  However, this scenario has been
challenged by new measurements 
of p-Au collisions~\cite{NA49Fischer} and new theoretical 
interpretations~\cite{ColorGlassCond}.  Indeed, so-called {\it initial} state effects 
such as random walk of the incoming nucleons~\cite{NA49Fischer} or Color Glass Condensate 
phenomena~\cite{ColorGlassCond} 
may offer an alternative explanation of the measured spectra and anisotropies
in transverse momentum.  This ambiguity apparently threatens the concept that
a {\it bulk} system has been created at all.
However, it is important to 
recall that collective expansion, if it exists, would manifest itself not only in momentum-space
observables, but would also generate space-momentum correlations, which can be measured
via two-particle correlations.

The possible validity of any scenario may only be claimed if a single set
of model parameters allows a successful description of {\it all} measured observables.
Here, we study,
in the context of a bulk collective flow scenario,
transverse momentum spectra, momentum-space anisotropy (``elliptic flow''),
HBT interferometry, and correlations between non-identical particles.

Similar studies have been reported
previously~\cite{TeaneyViscosity,PeitzmannBWHighPt,BurwardHoy,Peitzmann,Tomasik}.
New aspects in our study include: consideration of a more general (azimuthally-anisotropic)
freeze-out configuration, applicable to non-zero impact parameters; model studies of
azimuthally-sensitive HBT interferometry and correlations between non-identical particles;
and a multi-observable global fit to several pieces of published RHIC data.

This paper is organized as follows: 
In Section~II, we describe the blast-wave parameterization.
In Section~III, we investigate in detail the sensitivity of several
observables ($p_T$ spectra, elliptic flow, pion HBT radii, and average space-time
separation between different particle types) on the physical parameters of the
blast-wave parameterization.
In Section~IV, we perform fits to published data measured
at RHIC for Au+Au collisions at $\sqrt{s_{NN}}=130$~GeV, and, based on these fits,
describe how as-yet unpublished analyses (azimuthally-sensitive HBT interferometry,
and correlations between non-identical particles) are expected to look.
The reader primarily interested in the quality of the fit to the data and the resulting
parameters may want to skip past the details of Section~III.
In Section~V, we summarize and conclude on the relevance of the blast-wave parameterization
at RHIC.

%% file: S2_Model.tex
\section{The blast-wave Parameterization}

\subsection{Geneology and Motivation}

More than a quarter of a century ago, Westfall {\it et al}~\cite{Westfall76} introduced the 
nuclear fireball model to explain midrapidity proton spectra.  The idea was that the overlapping
nucleons of the target and projectile combined to create a hot source with velocity between
that of the target and projectile.  Protons emitted from this source were expected to be
emitted isotropically with a thermal energy distribution.

Soon thereafter, Bondorf, Garpman, and Zimanyi~\cite{BGZ78} derived a non-relativistic
expression for the energy spectra of particles emitted from a thermal {\it exploding}
source.  The radial flow in their (spherical) source results in energy spectra increasingly
different than those from a purely thermal (non-flowing) source, as the particle mass increases.
Siemens and Rasmussen~\cite{SR79} then generalized the formula with relativistic kinematics, further
simplifying by assuming a single expanding radial shell.

While a spherically-expanding source may be expected to approximate the fireball
created in lower-energy collisions, at higher energies stronger longitudinal
flow may lead to a cylindrical geometry.
A decade ago, Schnedermann {\it et al}~\cite{SSH93} introduced a simple functional
form for the phasespace density at kinetic freezeout, which approximated hydrodynamical
results assuming boost-invariant longitudinal flow~\cite{BjorkenBI}, and successfully used it to
fit $p_T$ spectra with only two parameters: a kinetic temperature, and a radial
flow strength.  The coordinate space geometry was an infinitely long solid cylinder (and
so should approximate the situation for $b=0$ collisions at midrapidity); the
transverse radial flow strength necessarily vanished along the central axis, and
is assumed maximum at the radial edge.  Most hydrodynamic calculations yield a
transverse rapidity flow field linear in the radial coordinate~\cite{TLS01}.

Huovinen {\it et al}~\cite{HKHRV01} generalized this parameterization to account for
the transversely anisotropic flow field which arises in {\it non}-central collisions,
and which generates an elliptic flow signal similar to that seen in measurements~\cite{STARv2ID}.
This added one more parameter-- the difference between the flow strength in, and out of,
the reaction plane.  The spatial geometry remained cylindrical, though it was assumed to
be a cylindrical shell, not a solid cylinder.

The measured  elliptical flow systematics as a function of $p_T$ and mass
are fairly well-fit with the Huovinen parameterization~\cite{STARv2ID}.  However,
better fits were achieved when
the STAR Collaboration generalized the model even further, adding a fourth parameter
designed to account for the anisotropic shape of the source in coordinate space~\cite{STARv2ID}.
A shell geometry was still assumed.

To calculate the spatial homogeneity lengths probed by two-particle correlation measurements~\cite{WH99},
we must revert from the unrealistic shell geometry to a solid emission region (infinite series of
elliptical shells).
Furthermore, additional parameters corresponding to the source size, emission time, and
emission duration must be included, increasing the number of parameters~\cite{FabriceBW}.
A similar generalization has been studied by Wiedemann~\cite{W98}.
Finally, in this paper, we explore
the effects of a ``hard-edge'' versus a smooth spatial density profile; similar studies
have recently been done by Tom\'{a}\u{s}ik {\it et al}~\cite{Tomasik9901} and
Peitzmann~\cite{Peitzmann} for the more restricted case of
a transversely-isotropic source.  This brings to eight the total number of parameters
which we study.

Although the blast-wave {\it functional form} was motivated by its similarity to
the freeze-out configuration of a real dynamical model (i.e. hydrodynamical solutions),
it is not necessarily true that the hydrodynamical freeze-out configuration corresponds
to the parameter set that best describes the data.  In this sense, the blast-wave model
presented here remains only a parameterization.  With eight freely-tunable parameters, 
it is clearly a toy model with little predictive power.  However, the goal is to see whether
a consistent description of the data from the soft sector at RHIC is possible within a
simple boost-invariant model with transverse collective flow.  If this turns out to be the
case, then it is worthwhile considering that the parameter values indeed characterize the
size, shape, timescales, temperature, and flow strengths of the freeze-out configuration.
A consistent parameterization in terms of such physical quantities represents a true step
forward, and provides valuable feedback to theorists constructing physical models of the
collision.

\subsection{Parameters and Quantities in the blast-wave}
\label{sec:parameters}
The eight parameters of the blast-wave parameterization described
in this paper are 
$T$, $\rho_0$, $\rho_2$, $R_y$, $R_x$, $a_s$, $\tau_0$, and $\Delta\tau$;
their physical meaning is given below.

\begin{figure}[t!]
\epsfig{file=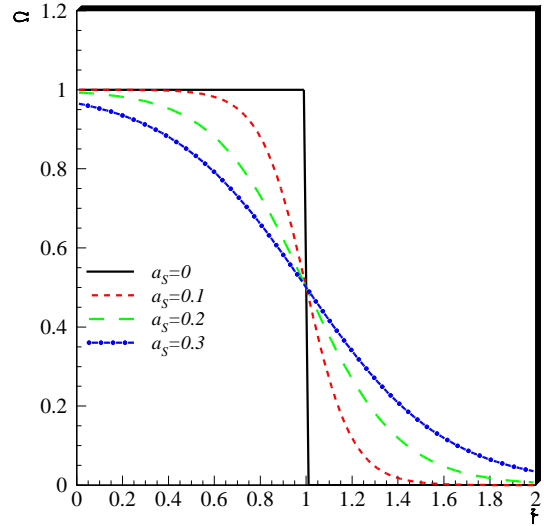,width=8cm}
\caption{(Color online)  The source weighting function $\Omega$, as a function of the normalized elliptical
radius $\tilde{r}$ for several values of the surface diffuseness parameter $a_s$.
\label{fig:surface}}
\end{figure}

The freeze-out distribution is infinite in the beam ($z$) direction, and
elliptical in the transverse ($x-y$) plane.  (The $x-z$ plane is the reaction plane.)  The transverse
shape is controlled by the radii $R_y$ and $R_x$, and the spatial weighting
of source elements is given by
\begin{equation}
\label{eq:Omega}
\Omega(r,\phi_s) = \Omega(\tilde{r}) = \frac{1}{1+e^{(\tilde{r}-1)/a_s}}
\end{equation}
where a fixed value of the ``normalized elliptical radius''
\begin{equation}
\label{eq:rtilde-def}
\tilde{r}(r,\phi_s) \equiv \sqrt{\frac{(r\cos(\phi_s))^2}{R_x^2}+\frac{(r\sin(\phi_s))^2}{R_y^2}}
\end{equation}
corresponds to a given elliptical sub-shell within the solid volume of the freeze-out distribution.

The parameter $a_s$ corresponds to a surface diffuseness of the emission source.
As shown in Figure~\ref{fig:surface}, a hard edge (``box profile'') may be assumed by
setting $a_s = 0$, while the density profile approximates a Gaussian shape for $a_s\approx0.3$.

It should be noted that the weighting function $\Omega(r,\phi_s)$ is not, in general, the source
density distribution.  In particular, as we discuss especially in Sections~\ref{sec:HBT_radii} and~\ref{sec:non-id},
non-zero collective flow induces space-momentum correlations which dominate the spatial source
density distributions.  Only for a system without flow ($\rho_0=\rho_2=0$, see below), the source
distribution is given by $\Omega$, so that, e.g., for $a_s=0$, there is a uniform density of
sources ($\frac{d^2N}{dxdy}={\rm const}$) inside the ellipse defined by $R_y$ and $R_x$, and no sources outside.

\begin{figure}[t!]
\epsfig{file=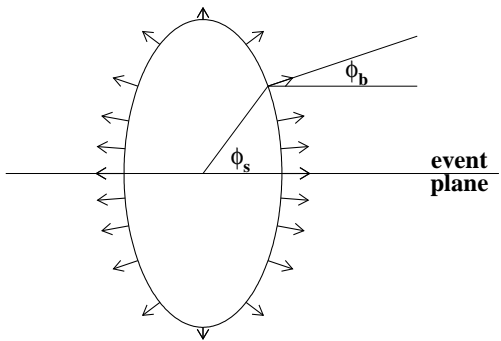,width=8cm}
\caption{Schematic illustration of an elliptical sub-shell of the source.  Here, the
source is extended out of the reaction plane ($R_y > R_x$).  Arrows represent the direction
and magnitude of the flow boost.  In this example, $\rho_2 > 0$ (see Equation~\ref{eq:rho}).
\label{fig:ellipse-cartoon}}
\end{figure}

The momentum spectrum of particles emitted from a source element at $(x,y,z)$ is given
by a fixed temperature $T$ describing the thermal kinetic motion, boosted by a transverse
rapidity $\rho(x,y)$.  This is common in models of this type.  However, unlike transversely
{\it isotropic} parameterizations, the azimuthal direction of the boost (denoted $\phi_b$) is {\it not}
necessarily identical to the spatial azimuthal angle $\phi_s = \tan^{-1}(y/x)$.  Instead, in our
model, the boost is
perpendicular to the elliptical sub-shell on which the source element is found; see Figure~\ref{fig:ellipse-cartoon}.
We believe this to be a more natural extension of an ``outward'' boost for non-isotropic
source distributions than that used by Heinz and Wong~\cite{HW02}, who used an anisotropic
shape but always assumed radial boost direction ($\phi_b=\phi_s$).
It may be shown that for our model
\begin{equation}
\label{eq:phis-vs_phib}
\tan(\phi_s) = \left(\frac{R_y}{R_x}\right)^2 \tan(\phi_b) .
\end{equation}

Hydrodynamical calculations for {\it central} collisions (i.e. azimuthally
isotropic freezeout distribution) suggest that the flow rapidity boost depends
linearly on the freeze-out radius~\cite{TLS01}.  
We assume a similar scenario, but in our more generalized parameterization,
the boost strength depends linearly on the normalized elliptical radius $\tilde{r}$ defined
in Equation~\ref{eq:rtilde-def}.
Thus, in the absence of an azimuthal dependence of the
flow (to be introduced shortly), all source elements on the outer edge of the source
boost with the same (maximum) transverse rapidity $\rho_0$ in an ``outward'' direction.

In non-central collisions, the strength of the flow boost itself may depend on azimuthal angle, as
suggested by Huovinen et al~\cite{HKHRV01}.  As those authors did, we incorporate this via a parameter
$\rho_2$, which characterizes the strength of the second-order oscillation of the transverse rapidity
as a function of $\phi_b$.  Hence, the flow rapidity is given by
\begin{equation}
\label{eq:rho}
\rho(r,\phi_s) = \tilde{r}\left(\rho_0 + \rho_2 \cos(2\phi_b)\right) .
\end{equation}

We note that source anisotropy enters into our parameterization in two independent ways,
and each contributes to, e.g., elliptic flow.  Setting $\rho_2 > 0$ means the boost is
stronger in-plane than out-of-plane, contributing to positive elliptic flow.  However,
even if $\rho_2=0$ (but $\rho_0\neq 0$), setting $R_y > R_x$ still generates positive elliptic flow, since
this means there are {\it more} sources emitting in-plane than out-of-plane (see
Figure~\ref{fig:ellipse-cartoon}).  The STAR Collaboration found that both types of anisotropy
were required to fit their elliptic flow data~\cite{STARv2ID}.  In generalizing the circular transverse geometry
parameterization of Huovinen {\it et al}~\cite{HKHRV01} (in which $\phi_b=\phi_s$), they 
added a parameter $s_2$ and weighted source elements with a given $\phi_b$ as
\begin{equation}
\label{eq:STAR-s2}
\frac{dN}{d\phi_b} = \frac{dN}{d\phi_s} \sim (1+2s_2\cos(2\phi_s))
\end{equation}
Thus, a positive value of $s_2=\langle\cos(2\phi_b)\rangle$ corresponded to more source
elements emitting in-plane, similar to setting $R_y>R_x$ in our parameterization.

To facilitate comparison of fits with the STAR model and with ours, we
relate the $s_2$ of STAR to the geometric anisotropy of our parameterization.
In the case of isotropic boost ($\rho_2=0$)~\cite{thanksSergeiVoloshin},
\begin{equation}
\label{eq:s2-in-BW}
s_2 = \langle\cos(2\phi_b)\rangle = \frac{1}{2}\frac{\left(\frac{R_y}{R_x}\right)^2-1}{\left(\frac{R_y}{R_x}\right)^2+1}.
\end{equation}
If $\rho_2\neq 0$, anisotropies in the space-momentum correlations lead to a significantly
more complicated expression.

Finally, since our model is based on a longitudinally boost-invariant assumption, it is sensible that the freezeout
occurs with a given distribution in longitudinal proper time $\tau = \sqrt{t^2-z^2}$.
We assume a Gaussian distribution
peaked at $\tau_0$ and with a width $\Delta\tau$
\begin{equation}
\label{eq:taudist}
\frac{dN}{d\tau} \sim \exp\left(-\frac{(\tau-\tau_0)^2}{2\Delta\tau^2}\right) .
\end{equation}

We note that although the source emits particles over a finite duration in proper time $\tau$,
we assume that none of the source parameters changes with $\tau$.  This is obviously
an oversimplification valid only for small $\Delta\tau$;
with time, one may expect the flow field to evolve (increase {\it or} decrease),
and it is natural to expect the transverse sizes $R_x$ and $R_y$ to change (grow {\it or} fall) with time.
However, calculation of the time dependence of these parameters requires a true dynamical model
and is outside the scope and spirit of the present work.

\subsection{The Emission Function}
Our emission function is essentially a generalization of azimuthally-isotropic emission functions
used by previous authors~\cite{AS95,WSH96,CL96,WHTW98}, and here we follow closely~\cite{WSH96}:
\begin{eqnarray}
\label{eq:firstS}
S(x,K) & = & m_T\cosh(\eta-Y) \Omega(r,\phi_s) e^{\frac{-(\tau-\tau_0)^2}{2\Delta\tau^2}} \times \nonumber \\
       &   & \frac{1}{e^{K\cdot u/T} \pm 1} \nonumber \\
       & = & m_T\cosh(\eta-Y) \Omega(r,\phi_s) e^{\frac{-(\tau-\tau_0)^2}{2\Delta\tau^2}} \times \\
       &   & \sum_{n=1}^\infty (\mp 1)^{n+1}  e^{-n K\cdot u/T} . \nonumber
\end{eqnarray}
Where the upper (lower) sign is for fermions (bosons).  Often, only the first term in the sum
in Equation~\ref{eq:firstS} is used, resulting in a Boltzmann distribution for all particles.
Below, we show that there is a small change to observables when truncating after the second term,
and negligible effect when including further terms.
The Boltzmann factor $\exp(-K\cdot u/T)$ arises from our assumption of local thermal
equilibrium within a source moving with four-velocity $u_\mu(x)$.  We assume longitudinal
boost invariance by setting the longitudinal flow velocity $v_L = z/t$
($z=\tau\sinh\eta$ and $t=\tau\cosh\eta$), so that the longitudinal
flow rapidity $\eta_{\rm flow} = \frac{1}{2}\ln\left[(1+v_L)/(1-v_L)\right]$ is identical~\cite{BjorkenBI} to the 
space-time rapidity $\eta = \frac{1}{2}\ln\left[(t+z)/(t-z)\right]$.  Thus, in cylindrical coordinates
\begin{eqnarray}
u_\mu(x) = \left( \cosh\eta \cosh\rho(r,\phi_s), \sinh\rho(r,\phi_s) \cos\phi_b,      \right. \nonumber \\
            \left.      \sinh\rho(r,\phi_s) \sin\phi_b, \sinh\eta \cosh\rho(r,\phi_s) \right)
\end{eqnarray}
and
\begin{equation}
K_\mu = \left( m_T \cosh Y, p_T\cos\phi_p, p_T\sin\phi_p, m_T \sinh Y \right) ,
\end{equation}
where the transverse momentum ($p_T$), transverse mass ($m_T$), rapidity ($Y$), and azimuthal
angle ($\phi_p$) refer to the {\it momentum of the emitted particle}, not the source element.
(Note that three azimuthal angles-- $\phi_s$,
$\phi_b$, and $\phi_p$-- are relevant to this discussion.)  Thus
\begin{eqnarray}
K_\mu u^\mu = m_T\cosh\rho(r,\phi_s)\cosh(\eta-Y) - \nonumber \\
 p_T\sinh\rho(r,\phi_s)\cos(\phi_b-\phi_p) ,
\end{eqnarray}
and the emission function (Equation~\ref{eq:firstS}) may be rewritten as
\begin{eqnarray}
\label{eq:secondS}
S(x,K) & = & S(r,\phi_s,\tau,\eta)      \nonumber \\
       & = & m_T \cosh(\eta-Y) \Omega(r,\phi_s) e^{\frac{-(\tau-\tau_0)^2}{2\Delta\tau^2}} \times   \\
       &   & \sum_{n=1}^\infty (\mp 1)^{n+1} e^{n \alpha \cos(\phi_b-\phi_p)} e^{-n \beta \cosh(\eta-Y)} . \nonumber
\end{eqnarray}
where we define
\begin{eqnarray}
\label{eq:alphadef}
\alpha \equiv \frac{p_T}{T}\sinh \rho(r,\phi_s)   \\
\label{eq:betadef}
\beta  \equiv \frac{m_T}{T}\cosh \rho(r,\phi_s) .
\end{eqnarray}

Exploiting the boost invariance and infinite longitudinal extension of our source, and focusing
on observables at midrapidity and using the longitudinally co-moving system (LCMS) for HBT
measurements, we may simplify Equation~\ref{eq:secondS} by setting $Y=0$.

\subsection{Calculating Observables}
\label{sec:integrals}
All observables which we will calculate are related to integrals of the emission function
(\ref{eq:secondS}) over phasespace $d^4x=dxdydzdt=\tau d\tau d\eta r dr d\phi_s$, weighted
with some quantity $B(x,K)$.  In all cases, the integrals over $\tau$ and $\eta$ may be
done analytically, though the result depends on whether $B(x,K)$ itself depends on $\tau$ and $\eta$.

In particular, if $B(x,K)=B^\prime(r,\phi_s,K)\tau^i\sinh^j\eta\cosh^k\eta$
 then the integrals of interest are~\cite{caviat}

\begin{eqnarray}
\label{eq:FourFoldIntegral}
 & \int_0^{2\pi}d\phi_s \int_{0}^\infty r dr \int_{-\infty}^\infty d\eta \int_{-\infty}^\infty \tau d\tau
   S(x,K) B(x,K) \nonumber \\
 & = m_T H_i \cdot \left\{ B^\prime \right\}_{j,k}(K)
\end{eqnarray}
where the $\tau$ and $\eta$ integrals are denoted
\begin{eqnarray}
\label{eq:Hs}
H_i & \equiv & \int_{-\infty}^\infty d\tau \tau^{i+1} e^{\frac{-(\tau-\tau_0)^2}{2\Delta\tau^2}} \nonumber \\
H_0 & = & \sqrt{2\pi}\cdot\Delta\tau\cdot\tau_0  \nonumber \\
H_1 & = & \sqrt{2\pi}\cdot\Delta\tau\cdot\left(\Delta\tau^2+\tau_0^2\right)   \\
H_2 & = & \sqrt{2\pi}\cdot\Delta\tau\cdot\tau_0\left(3\Delta\tau^2+\tau_0^2\right)  \nonumber 
\end{eqnarray}
and
\begin{eqnarray}
G_{j,k}(x,K) & \equiv & \int_{-\infty}^\infty d\eta e^{-\beta \cosh\eta} \sinh^j\eta \cosh^{k+1}\eta \nonumber \\
G_{0,0}(x,K) & = & 2{\rm K}_1(\beta)                                          \nonumber \\
G_{0,1}(x,K) & = & 2\left[ \frac{{\rm K}_1(\beta)}{\beta} + {\rm K}_0(\beta) \right] \nonumber \\
G_{1,0}(x,K) & = & G_{1,1}(x,K) = 0                                                 \\
G_{0,2}(x,K) & = & 2\left[ \frac{{\rm K}_2(\beta)}{\beta} + {\rm K}_1(\beta)  \right] \nonumber \\
G_{2,0}(x,K) & = & 2\frac{{\rm K}_2(\beta)}{\beta}    .            \nonumber
\end{eqnarray}
$\beta$ was defined in (\ref{eq:betadef}), and $\rm{K}_n$ are the modified Bessel functions.
For the above, we define the notation
\begin{eqnarray}
\left\{ B^\prime \right\}_{j,k}(K) \equiv \sum_{n=1}^\infty \left\{ (\mp 1)^{n+1} 
     \int_0^{2\pi}d\phi_s \int_{0}^\infty r dr \left[  \right. \right. \nonumber \\
     \left. \left.  G_{j,k}(x,n K) B^\prime(x,K) 
      \times e^{n \alpha \cos(\phi_b-\phi_p)}    \Omega(r,\phi_s) \right] \right\}
     \label{eq:bracketBetaPrime}
\end{eqnarray}
for the remaining integrals, which we perform numerically.
(Note that $G_{i,j}(x,nK)$ retains dependence on $r$ and $\phi_s$ due to its dependence on $\beta$, as
defined in Equation~\ref{eq:betadef}, and so cannot move outside the integrals in Equation~\ref{eq:bracketBetaPrime}.)

%% file: S3_Observables.tex
\section{Calculation of Hadronic Observables}
\label{sec:S3}

In this Section, we discuss how hadronic observables are calculated from the parameterized
source and illustrate the sensitivity of these observables to the various parameters presented
in Section~\ref{sec:parameters}.

With several observables depending on several parameters, it is not feasible to
explore the entire numerical parameter space.  Instead, we anticipate the results
of the next Section, in which we fit our model to existing data, and vary the parameters
by ``reasonable'' amounts about values similar to those which fit the data.
The default parameter values used in several of the calculations in this Section are listed in Table~\ref{tab:defaultParams}.

\begin{table}[t]
\begin{tabular}{lcc}
\hline
\hline
\multicolumn{1}{c}{parameter} & \multicolumn{1}{c}{round source} & \multicolumn{1}{c}{non-round source} \\
\hline
$\rho_2$         	     &   0      &   0.05 \\
$R_{x}$ (fm)     	     &   12.04  &   11   \\
$R_{y}$ (fm)     	     &   12.04  &   13   \\
$T$ (GeV)              	     &   \multicolumn{2}{c}{0.1}  \\
$\rho_{0}$        	     &   \multicolumn{2}{c}{0.9}  \\
$\tau_0$  (fm/c)    	     &   \multicolumn{2}{c}{9}    \\
$\Delta \tau$ (fm/c)	     &   \multicolumn{2}{c}{2}    \\
$a_s$                        &   \multicolumn{2}{c}{0}    \\
\hline\hline
\end{tabular}
\caption{
Default parameter values for most calculations in Section~\ref{sec:S3}.
Note that $\rho_2$, $R_x$ and $R_y$ default values depend on whether we are discussing an azimuthally isotropic (``round'')
or anisotropic (``non-round'') source.  One might expect such sources from central and peripheral collisions, respectively.
\label{tab:defaultParams}}
\end{table}

\subsection{$p_T$ spectra}
\label{sec:spectra}

In the notation of Section~\ref{sec:integrals} the (azimuthally-integrated) $p_T$
spectrum is calculated as
\begin{eqnarray}
\frac{dN}{p_T dp_T} & = & \int d\phi_p \int d^4x S(x,K)  \nonumber \\
 & \propto & m_T \int d\phi_p \left\{1\right\}_{0,0}(K)  .
\label{eq:spectra}
\end{eqnarray}
In this paper, we focus only on the shapes, not the normalizations, of the spectra.

We note that spectra calculated in the blast-wave model scale neither with $m_T$ nor $p_T$,
as both quantities enter the expression through $\alpha$ and $\beta$ (Equations~(\ref{eq:alphadef})
and~(\ref{eq:betadef})).  This breaking of $m_T$-scaling is a well-known consequence of finite
transverse flow~\cite{SSH93} ($\rho\neq 0$ in our model).

According to Equation~\ref{eq:spectra}, $m_T$ spectra calculated in the blast-wave model are insensitive
to the time parameters $\tau_0$ and $\Delta\tau$.  
The spectral shapes are furthermore insensitive to the spatial scale (i.e. $R_y$) of the source,
though, as we see below, there is some small sensitivity to the spatial shape (i.e. $\frac{R_y}{R_x}$).

\begin{figure}[h!]
\epsfig{file=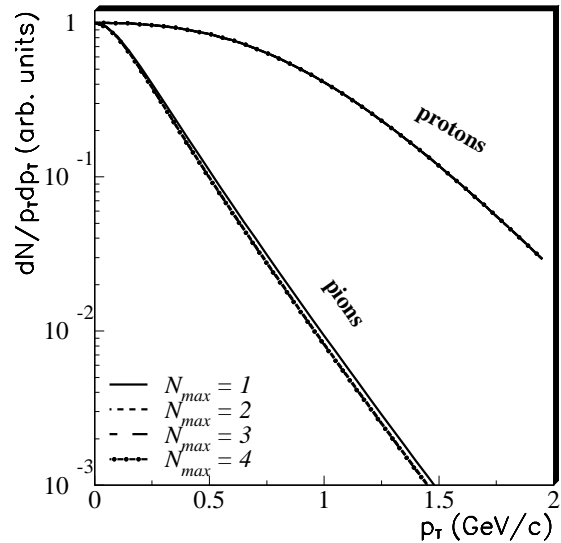,width=8cm}
\caption{Transverse momentum spectra for protons (upper curves) and pions (lower curves), as
calculated by Equation~\ref{eq:spectra}, for several values of $N$, the maximum value of $n$
taken in the summation of Equation~\ref{eq:firstS}; see text for details.
Parameters values correspond to the ``non-round'' source of Table~\ref{tab:defaultParams}.
All spectra are arbitrarily normalized to unity at $p_T=0$.
\label{fig:Spectra_QuantumExpansion}}
\end{figure}

First, we study the importance of using quantum (as opposed to classical) statistics in the source function.
Figure~\ref{fig:Spectra_QuantumExpansion} shows $p_T$ spectra for pions and protons, treated as bosons
and fermions, respectively.  Model parameters were set to the ``non-round'' values listed in Table~\ref{tab:defaultParams}.
The sum in Equation~\ref{eq:firstS} (and Equations~\ref{eq:secondS} and~\ref{eq:bracketBetaPrime}) is
over $n = 1 \ldots N$; curves are shown for $N=1,2,3,4$.  For parameter values in the range we study
here, proton spectra are essentially independent of $N$.  For $N > 1$, the pion spectra are likewise robust
against the value of $N$, though in the classical limit ($N=1$), there is relatively lower yield at low $p_T$.
(Note that all spectra are arbitrarily normalized to unity at $p_T = 0$.)  Calculations below use the truncation
$N=2$.

\subsubsection{Spectra from Central Collisions}

Focusing first on central collisions (so that the flow anisotropy parameter $\rho_2=0$ and $R_x=R_y$),
then, we need only consider the spectra sensitivity to the temperature and radial flow parameters $\rho_0$
and $T$, and to the surface diffusion $a_s$.

\begin{figure}[h!]
\epsfig{file=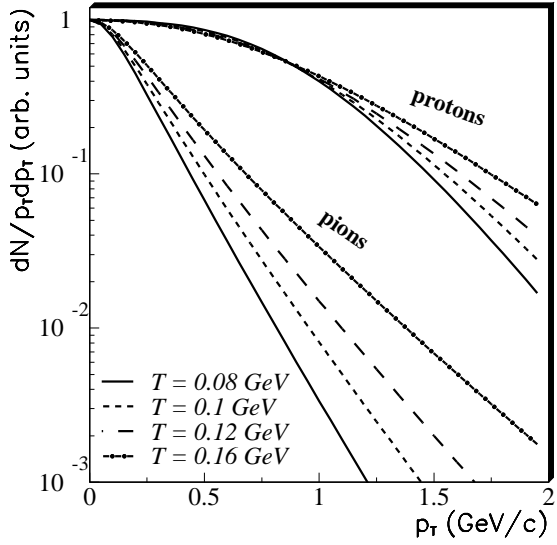,width=8cm}
\caption{Transverse momentum spectra for protons (upper curves) and pions (lower curves), as
calculated by Equation~\ref{eq:spectra}, for several values of the temperature parameter $T$.
Other parameters follow the ``round'' source defaults of Table~\ref{tab:defaultParams}.
All spectra are arbitrarily normalized to unity at $p_T=0$.
\label{fig:Spectra_vary_T}}
\end{figure}

\begin{figure}[h!]
\epsfig{file=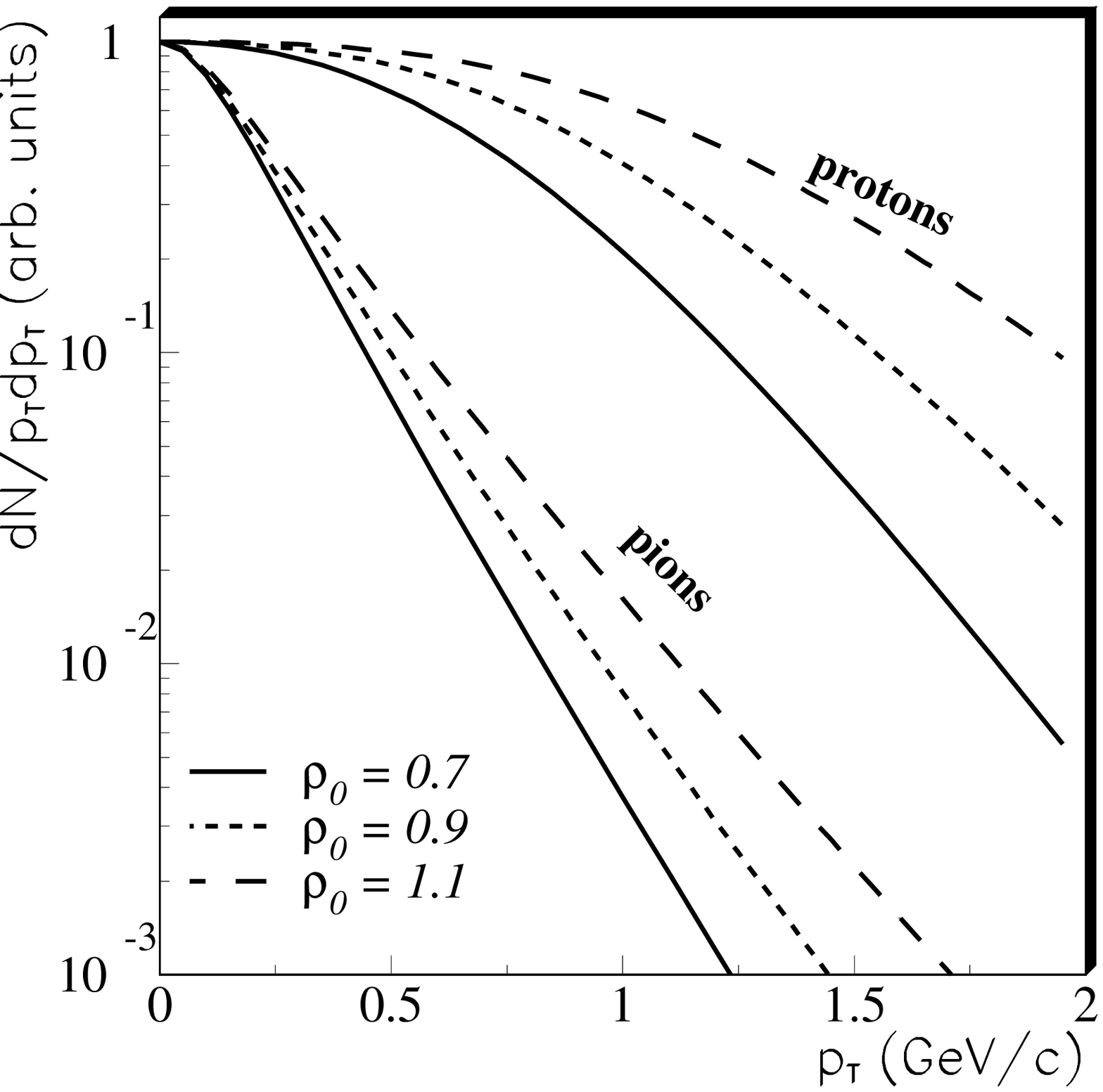,width=8cm}
\caption{
Transverse momentum spectra for protons (upper curves) and pions (lower curves), as
calculated by Equation~\ref{eq:spectra}, for several values of the radial flow parameter $\rho_0$.
Other parameters follow the ``round'' source defaults of Table~\ref{tab:defaultParams}.
All spectra are arbitrarily normalized to unity at $p_T=0$.
\label{fig:Spectra_vary_rho0}}
\end{figure}

Fixing the transverse spatial density distribution to a box profile ($a_s=0$), the evolution of spectral shapes
for pions and protons are shown in Figures~\ref{fig:Spectra_vary_T} and~\ref{fig:Spectra_vary_rho0},
as the temperature and radial flow parameter, respectively, are varied about nominal values
of $T=100$~MeV and $\rho_0=0.9$.  As has been noted previously~\cite{SSH93}, at low $p_T$, temperature
variations affect the lighter pions more strongly, while variations in the collective flow boost produce
a stronger effect on the heavier particles.

\begin{figure}[h!]
\epsfig{file=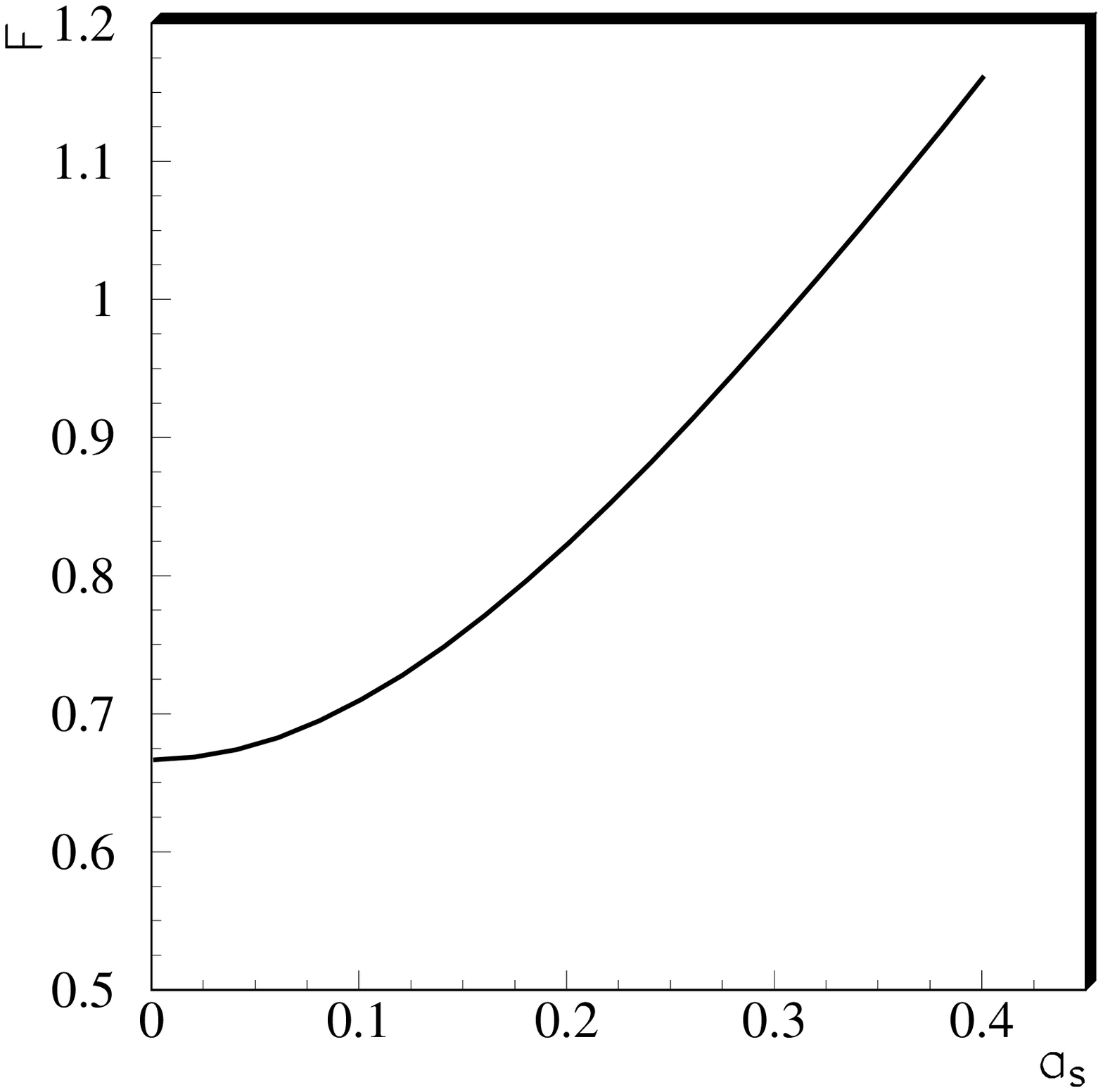,width=8cm}
\caption{The geometric constant of proportionality $F$ between the average transverse flow boost $\langle \rho \rangle$
and the blast-wave parameter $\rho_0$, as a function of the surface diffuseness $a_s$.
\label{fig:fraction_vs_as}}
\epsfig{file=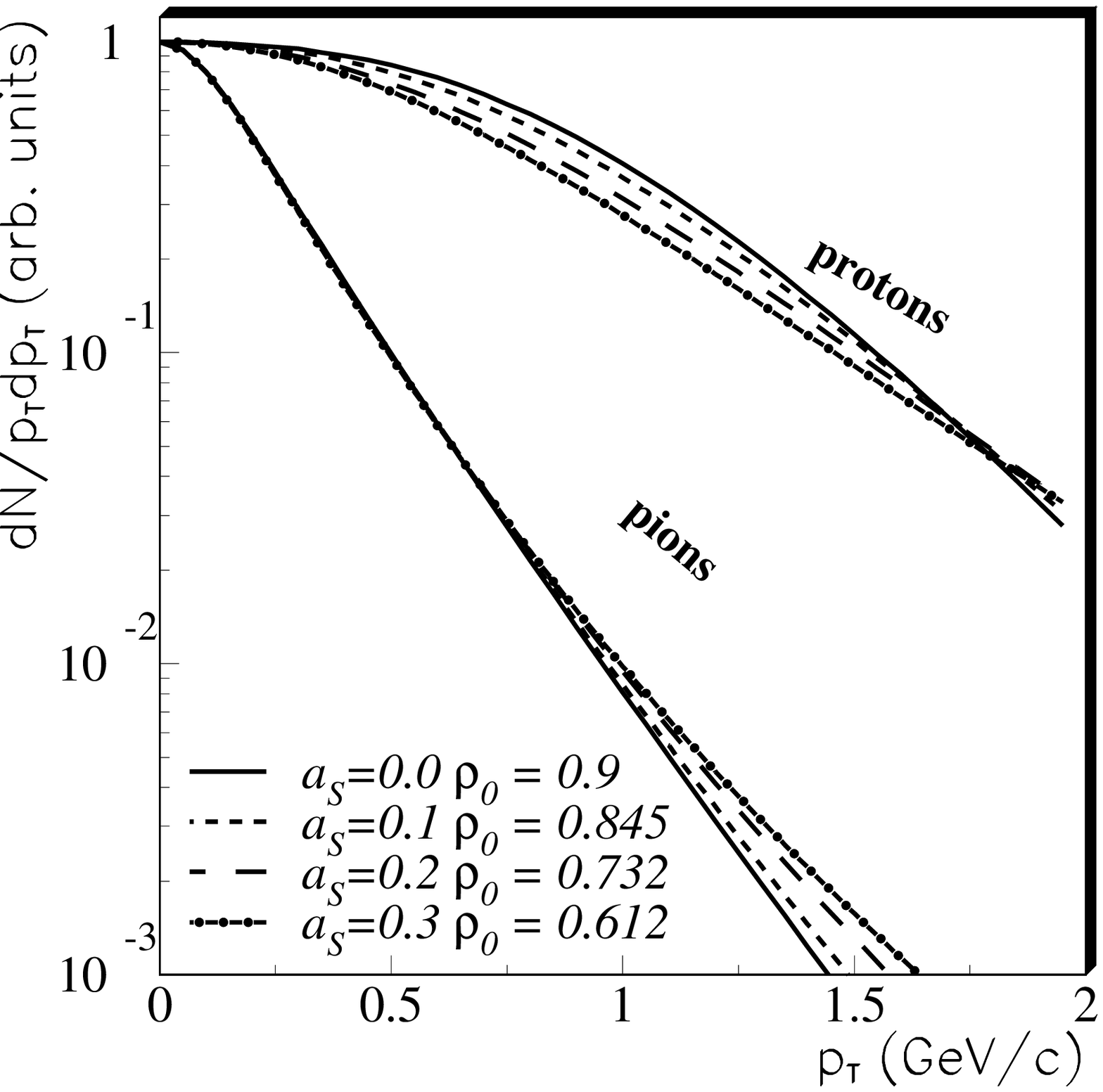,width=8cm}
\caption{
Transverse momentum spectra for protons (upper curves) and pions (lower curves), as
calculated by Equation~\ref{eq:spectra}, for several values of the surface diffuseness parameter $a_s$.
The radial flow strength $\rho_0$ is co-varied; see text for details.
Other parameters follow the ``round'' source defaults of Table~\ref{tab:defaultParams}.
All spectra are arbitrarily normalized to unity at $p_T=0$.
\label{fig:Spectra_vary_as}}
\end{figure}

Next, we consider the effect of a finite surface diffuseness parameter ($a_s \neq 0$), {\it i.e.},
using the smoother spatial density distributions of Figure~\ref{fig:surface}.  Since
we assume a transverse flow profile which increases linearly with radius (c.f. Equation~\ref{eq:rho}),
one trivial effect is that, for a fixed $\rho_0$, increasing $a_s$ will produce a larger average
flow boost $\langle \rho \rangle$.  The effect of increasing transverse flow was already explored
directly in Figure~\ref{fig:Spectra_vary_rho0}, so we avoid this trivial effect here, and
explore the effect of varying $a_s$, while keeping $\langle \rho \rangle$ constant~\cite{Peitzmann}.

The average transverse flow boost $\langle \rho \rangle = F(a_s)\cdot\rho_0$ where the geometric
proportionality constant
\begin{equation}
F(a_s) = \frac{\int_0^\infty dx \frac{x^2}{1+\exp((x-1)/a_s)}}{\int_0^\infty dx \frac{x}{1+\exp((x-1)/a_s)}}
\label{eq:fraction_vs_as}
\end{equation}
is independent of $\rho_0$ or $R=R_x=R_y$.  For the box profile, $F(a_s=0)=\frac{2}{3}$.
Figure~\ref{fig:fraction_vs_as} shows this geometric factor as a function of the surface diffuseness.

Figure~\ref{fig:Spectra_vary_as} shows the pion and proton spectra for various values of $a_s$.  The 
radial flow strength $\rho_0$ was co-varied with $a_s$ so that the average transverse flow boost was
$\langle \rho \rangle = 0.6$.  To a first approximation, the spectral shapes depend only on the
temperature $T$ and the average transverse flow boost $\langle \rho \rangle$.  
The residual dependence on the surface diffuseness parameter $a_s$ arises from the fact that while the {\it average}
boost rapidity has been held constant, the {\it spread} of boost rapidities increases with increasing 
$a_s$~\cite{Peitzmann}.  Thus,
we observe {\it qualitatively} similar variations in the spectral shapes when $a_s$ increases
(Figure~\ref{fig:Spectra_vary_as}),
as when $T$ increases (Figure~\ref{fig:Spectra_vary_T}).  The variations are not {\it quantitatively} identical since
in the present case, the velocity spread is not thermal, and the particle velocity spread evolves differently with mass,
depending on whether it arises from a boost spread or a thermal spread.

\subsubsection{Dependence of spectral shapes on source anisotropy}

Azimuthally-integrated $p_T$ spectra are often presented as a function of event centrality.
For $b\neq 0$ collisions, the emitting source may have anisotropic structure ($R_x \neq R_y$ and
$\rho_2 \neq 0$ in the present model).  Thus, it is interesting to explore possible effects of
these anisotropies.

For an azimuthally isotropic flow field ($\rho_2=0$), the spectral shapes are insensitive to spatial
anisotropies in the source (i.e. $R_x \neq R_y$).  This is because the spectral shapes are determined
by the distribution of boost velocities~\cite{Peitzmann}, which is unchanged by a shape change in our parameterization,
if $\rho_2=0$.

\begin{figure}[h!]
\epsfig{file=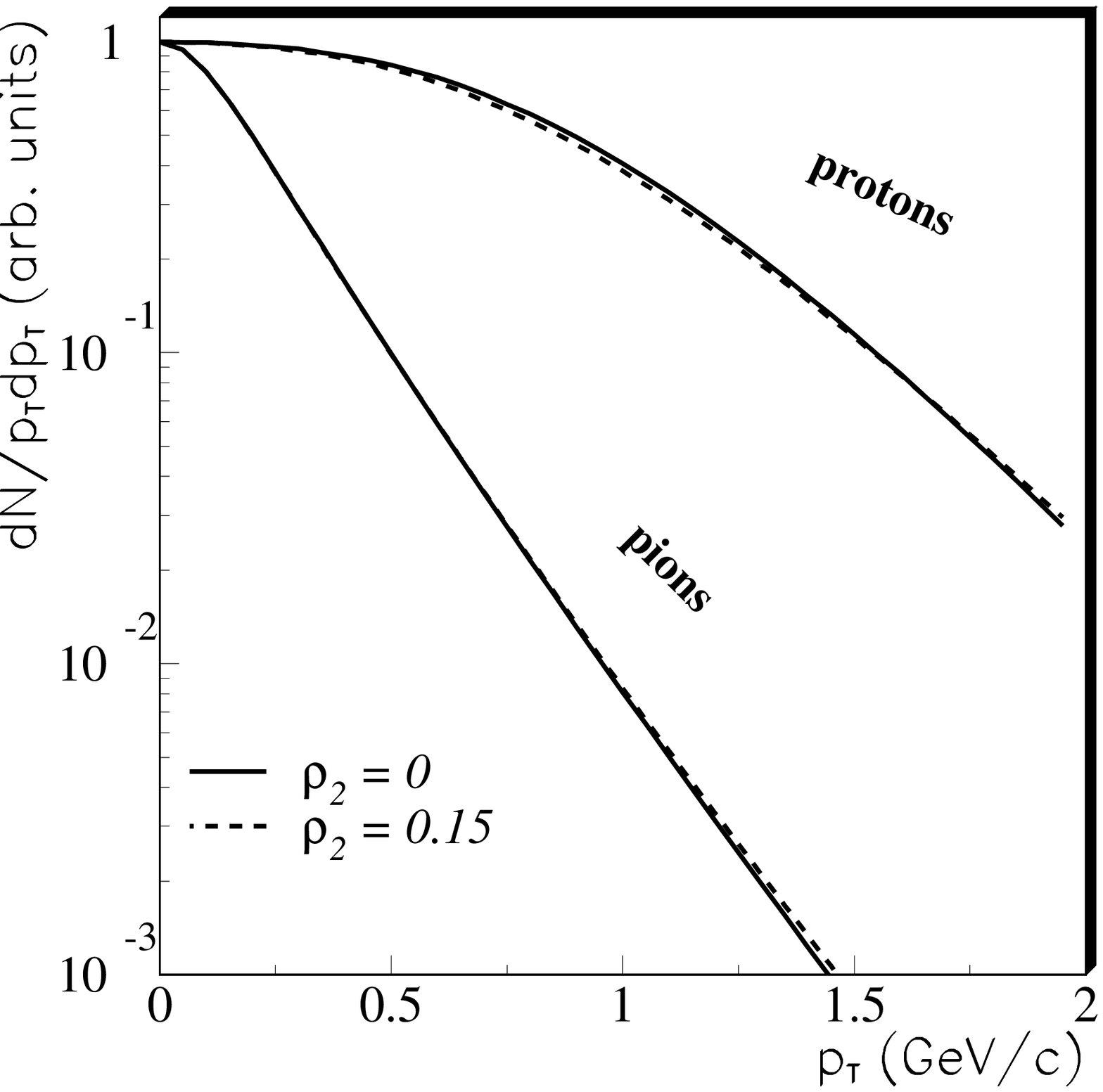,width=8cm}
\caption{
Transverse momentum spectra for protons (upper curves) and pions (lower curves), as
calculated by Equation~\ref{eq:spectra}
for an azimuthally symmetric flow field ($\rho_2=0$) and an asymmetric field ($\rho_2=0.15$).
Other parameters follow the ``round'' source defaults of Table~\ref{tab:defaultParams}.
All spectra are arbitrarily normalized to unity at $p_T=0$.
\label{fig:Spectra_vary_rhoa}}
%
\epsfig{file=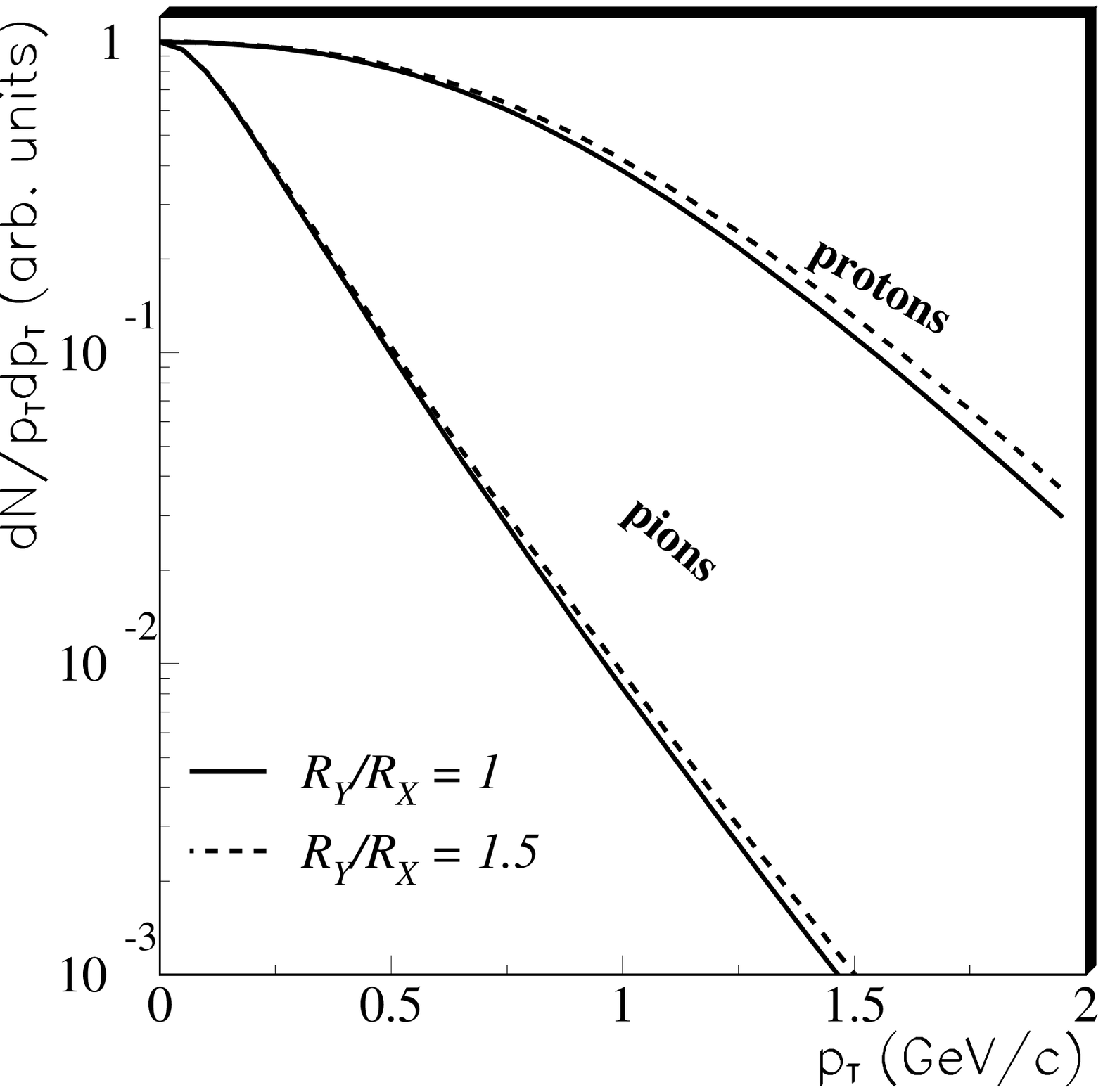,width=8cm}
\caption{
Transverse momentum spectra for protons (upper curves) and pions (lower curves), as
calculated by Equation~\ref{eq:spectra} for an azimuthally anisotropic flow field ($\rho_2=0.15$), and
a spatially isotropic ($R_y/R_x=1$) and anisotropic ($R_y/R_x=1.5$) spatial distribution.
Other parameters follow the ``round'' source defaults of Table~\ref{tab:defaultParams}.
\label{fig:Spectra_vary_RxRy_rhoa0.15}}
\end{figure}

For an azimuthally-symmetric spatial source ($R_x=R_y$), a
very small variation in the $\phi_p$-integrated
spectral shapes is observed when $\rho_2$ is changed from a value of 0.0 to 0.15, as seen in Figure~\ref{fig:Spectra_vary_rhoa}.
This, again, is due to the slightly increased spread in boost velocities; source elements emitting in-plane
boost a bit more, and out-of-plane a bit less.
This effect becomes stronger in the presence of an out-of-plane spatial anisotropy ($R_y/R_x>1$)
as shown in Figure~\ref{fig:Spectra_vary_RxRy_rhoa0.15}.
In any case, the effects of ``reasonable'' source anisotropy on the shapes of azimuthally-integrated spectra are
very small.

Thus, we conclude that azimuthally-integrated $p_T$ spectra are largely
insensitive to ``reasonable''
source anisotropies (see Section~\ref{sec:fits} for ``reasonable'' ranges)
and probe mainly the thermal motion ($T$) and average
transverse flow boost ($\langle \rho \rangle$) of the source.

\subsection{Elliptic flow versus mass and $p_T$}

In the notation of Section~\ref{sec:integrals} the elliptic flow parameter $v_2$ is
calculated as
\begin{equation}
v_2(p_T,m) = \frac{\int_0^{2\pi} d\phi_p \left\{\cos(2\phi_p)\right\}_{0,0}(K)}
      {\int_0^{2\pi} d\phi_p \left\{1\right\}_{0,0}(K)}
\label{eq:v2}
\end{equation}

A finite $v_2$ arises from azimuthal anisotropies in the source ($R_x \neq R_y$ and/or $\rho_2 \neq 0$).
As discussed below, however, the parameters $\rho_0$ and $T$ strongly affect
its value and evolution with $p_T$ and mass.  In the present parameterization, $v_2$ is not sensitive to
the overall spatial scale of the source ($R_y$) or the time parameters $\tau_0$ and $\Delta\tau$.

The $v_2$ parameter depends non-trivially on both $p_T$ and particle mass~\cite{VP00,HKHRV01,STARv2ID,VoloshinQM03}.
In this section, we explore the evolution of pion and proton $v_2$, as we vary the model
parameters from nominal ``non-round'' values for non-central collisions (cf~\ref{tab:defaultParams}).

\begin{figure}[t!]
\epsfig{file=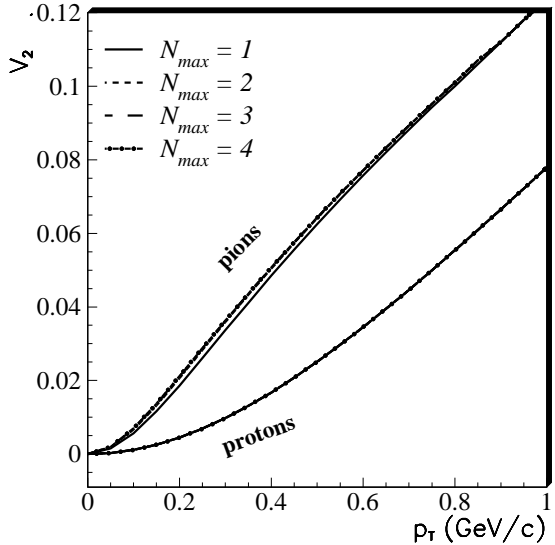,width=8cm}
\caption{
Elliptic flow parameter $v_2$ for pions (upper curves) and protons (lower curves), as a function
of transverse momentum, as calculated by Equation~\ref{eq:v2}, for several values of $N$, the maximum value of $n$
taken in the summation of Equation~\ref{eq:firstS}; see text for details.
Model parameters follow the ``non-round'' source defaults of Table~\ref{tab:defaultParams}.
\label{fig:v2_QuantumExpansion}}
\end{figure}

As with the $p_T$ spectra of Section~\ref{sec:spectra}, we first check the importance of quantum statistics.
Figure~\ref{fig:v2_QuantumExpansion} shows $v_2$ for pions and protons, for different values of $N$, where the
sum in Equation~\ref{eq:firstS} (and Equations~\ref{eq:secondS} and~\ref{eq:bracketBetaPrime}) runs
over $n=1 \ldots N$.  Again, we find only a small difference for the pions between $N=1$ (classical limit) and
$N=2$, beyond which $v_2$ is robust against further increases in $N$.  Calculations here use $N=2$.

\begin{figure}[t!]
\epsfig{file=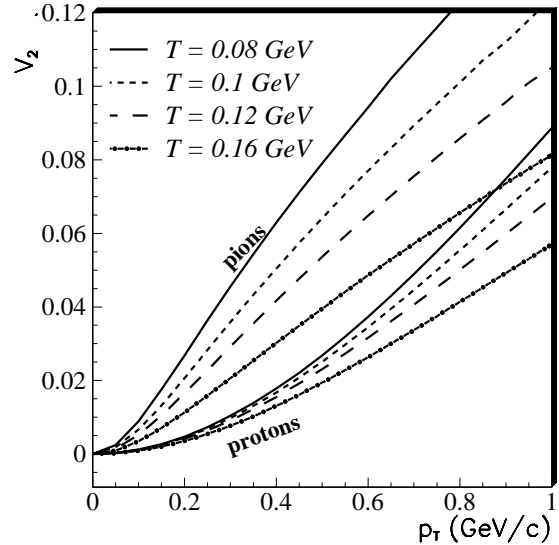,width=8cm}
\caption{
Elliptic flow parameter $v_2$ for pions (upper curves) and protons (lower curves), as a function
of transverse momentum, as calculated by Equation~\ref{eq:v2}, for various values of the temperature
parameter $T$.
Other parameters follow the ``non-round'' source defaults of Table~\ref{tab:defaultParams}.
\label{fig:v2_vary_T}}
\end{figure}

Figure~\ref{fig:v2_vary_T} shows the evolution of $v_2$ as the temperature parameter ($T$) is varied.
For both particle types shown, the increased thermal smearing in momentum space, as $T$ is increased,
leads to a reduced momentum-space anisotropy.  The effect of the thermal smearing is greater for the
lighter pions.

\begin{figure}[t!]
\epsfig{file=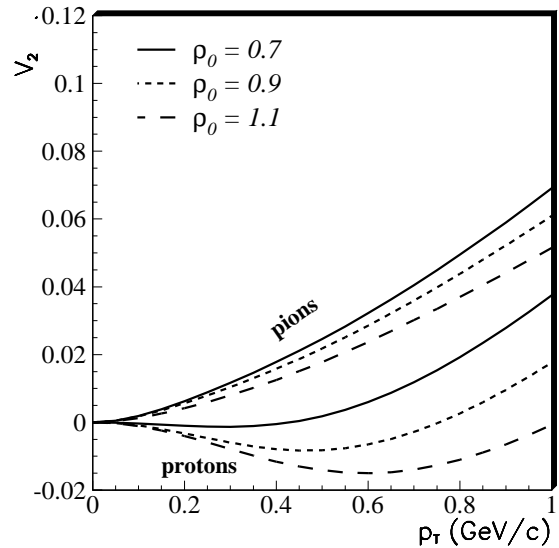,width=8cm}
\caption{
Elliptic flow parameter $v_2$ for pions (upper curves) and protons (lower curves), as a function
of transverse momentum, as calculated by Equation~\ref{eq:v2}, for various values of the transverse flow
parameter $\rho_0$.  The source spatial distribution is assumed azimuthally-anisotropic ($R_y=R_x$), but 
other parameters follow the ``non-round'' source defaults of Table~\ref{tab:defaultParams}.
\label{fig:v2_vary_rho0_RyRxrat_1.0}}
\end{figure}
\begin{figure}[h]
\epsfig{file=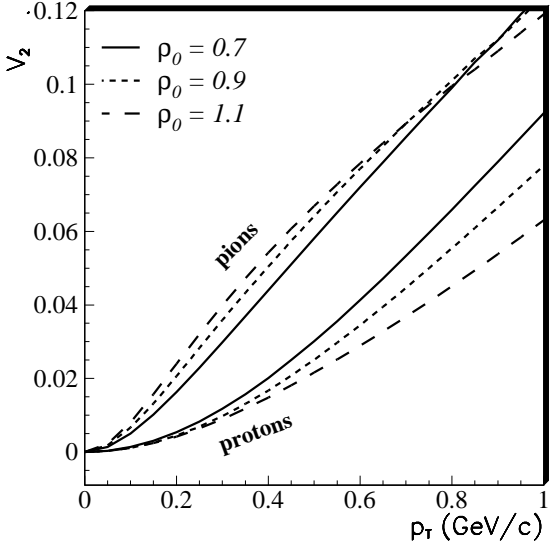,width=8cm}
\caption{
Elliptic flow parameter $v_2$ for pions (upper curves) and protons (lower curves), as a function
of transverse momentum, as calculated by Equation~\ref{eq:v2}, for various values of the transverse flow
parameter $\rho_0$.  
Other parameters follow the ``non-round'' source defaults of Table~\ref{tab:defaultParams}.
\label{fig:v2_vary_rho0_RyRxrat_1.182}}
\end{figure}

Less intuitive is the evolution of $v_2$ as the flow field ($\rho_0$ or $\rho_2$) is varied.
In Figures~\ref{fig:v2_vary_rho0_RyRxrat_1.0} and~\ref{fig:v2_vary_rho0_RyRxrat_1.182}, the
average transverse flow parameter $\rho_0$ is varied for an azimuthally isotropic ($R_y/R_x=1$) and
anisotropic ($R_y/R_x=13/11$) shape, respectively.  For the isotropic spatial distribution, we find that
$v_2$ decreases as $\rho_0$ increases, for all $p_T$ and for both particle types.  This is due
to the decreasing relative amplitude of the oscillation in the flow field.  Not surprisingly,
the effect is larger for the heavier protons.  

\begin{figure}[t!]
\epsfig{file=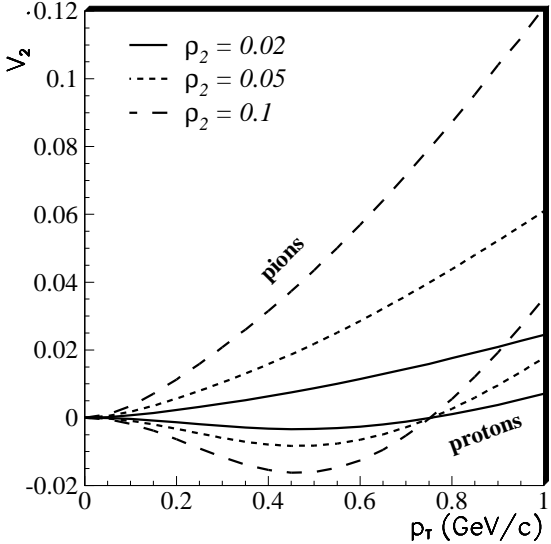,width=8cm}
\caption{
Elliptic flow parameter $v_2$ for pions (upper curves) and protons (lower curves), as a function
of transverse momentum, as calculated by Equation~\ref{eq:v2}, for various values of the modulation in
the transverse flow $\rho_2$.
Other parameters follow the {\it ``round''} (note: $R_y=R_x$) source defaults of Table~\ref{tab:defaultParams}.
\label{fig:v2_vary_rhoa_RyRxrat_1.0}}
\end{figure}
\begin{figure}[h]
\epsfig{file=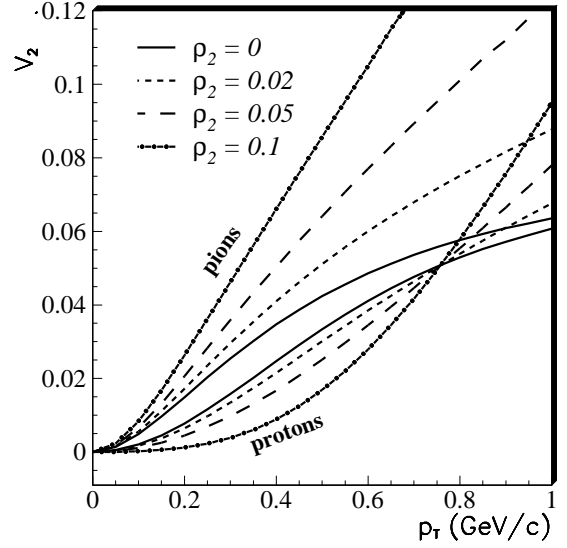,width=8cm}
\caption{
Elliptic flow parameter $v_2$ for pions (upper curves) and protons (lower curves), as a function
of transverse momentum, as calculated by Equation~\ref{eq:v2}, for various values of the modulation in
the transverse flow $\rho_2$.
Other parameters follow the ``non-round'' source defaults of Table~\ref{tab:defaultParams}.
\label{fig:v2_vary_rhoa_RyRxrat_1.182}}
\end{figure}

Indeed, it has been pointed out~\cite{Voloshin97,HKHRV01,VoloshinQM03} that
high radial flow (large $\rho_0$) can lead to negative values of $v_2$ for heavy particles; this
is clearly true for the protons in Figure~\ref{fig:v2_vary_rho0_RyRxrat_1.0}.
This negative $v_2$ reflects the {\it depletion} of the low-$p_T$ particle yield when most source
elements are highly boosted transversely; this is also the origin
of the ``shoulder arm'' in the $p_T$ spectra (c.f. Section~\ref{sec:spectra}).
This depletion is larger in-plane (since the boost is higher, i.e. $\rho_2>0$), leading to
negative $v_2$.  At higher $p_T$, and/or lower flow strengths, the competing effect is dominant:
the larger in-plane boost leads to more particles emitted in-plane, and $v_2>0$.

For a slightly anisotropic shape ($R_y/R_x=13/11$; Figure~\ref{fig:v2_vary_rho0_RyRxrat_1.182}),
the $v_2$ parameter increases 
significantly for both protons and pions, due to the larger number of source elements
boosted in-plane (c.f. Figure~\ref{fig:ellipse-cartoon}).  The finite spatial asymmetry
leads to an effect that tends to oppose the reduction of $v_2$ with increasing $\rho_0$,
discussed above, and, for pions at low $p_T$, even reverses it in this case.  Hence, at low $p_T$,
increasing $\rho_0$ increases (decreases) $v_2$ for pions (protons).

We observe similar trends when $\rho_0$ is held fixed, and $\rho_2$ is varied. 
Figure~\ref{fig:v2_vary_rhoa_RyRxrat_1.0} shows $v_2$ evolution for a spatially azimuthally-symmetric
source ($R_y/R_x=1$) and Figure~\ref{fig:v2_vary_rhoa_RyRxrat_1.182} for a slightly asymmetric
source ($R_y/R_x=13/11$).

\begin{figure}[t!]
\epsfig{file=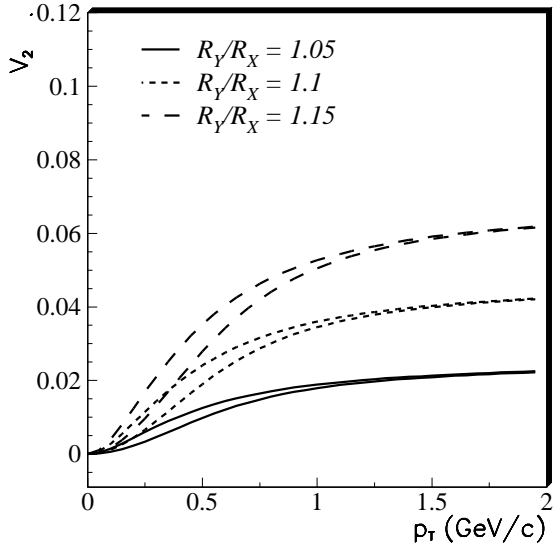,width=8cm}
\caption{
Elliptic flow parameter $v_2$ for pions (upper curves) and protons (lower curves), as a function
of transverse momentum, as calculated by Equation~\ref{eq:v2}, for various values of the
spatial anisotropy ($R_y/R_x$).
Elliptic flow parameter $v_2$ for pions (upper curves) and protons (lower curves), as a function
of transverse momentum, as calculated by Equation~\ref{eq:v2}, for various values of the modulation in
the transverse flow $\rho_2$.
Other parameters follow the {\it ``round''} (note: $\rho_2 = 0$) source defaults of Table~\ref{tab:defaultParams}.
\label{fig:v2_vary_RxRyrat_rhoa0.0}}
\end{figure}

As mentioned above, a finite value of
$v_2$ may be obtained for an azimuthally-{\it symmetric} flow field, if the spatial shape is
asymmetric.  Figure~\ref{fig:v2_vary_RxRyrat_rhoa0.0} shows the proton and pion $v_2$ for $\rho_2=0$
and $\rho_0=0.9$, for various values of $R_y/R_x$.  The larger number of sources emitting in-plane
results in more particles measured in-plane (thus $v_2>0$).  This momentum-space asymmetry saturates
at $p_T\sim 1.0$~GeV/c (for this set of parameters) and is relatively independent of particle
mass.  (We note that the $p_T$-scale in Figure~\ref{fig:v2_vary_RxRyrat_rhoa0.0} is larger than the
other figures, in order to show the $v_2$ saturation effect.)
This effect is similar to those previously discussed by Houvinen, {\it et al},~\cite{HKH02} and Shuryak~\cite{S02}.
However, we stress the non-zero $v_2$ parameter does {\it not} indicate ``elliptic flow without transverse flow,''
as suggested in~\cite{HKH02}.  If transverse flow is turned off ($\rho_0=0$), the spectra are thermal
and isotropic, and $v_2=0$ for all $p_T$; space-momentum correlations (induced by $\rho\neq 0$ in our
model) are required for a finite $v_2$.  In the Houvinen model~\cite{HKH02}
it is the implementation of the Cooper-Frye freeze-out procedure (which creates an infinitely opaque source)
which produces the space-momentum correlations, effectively generating ``flow.''  Shuryak~\cite{S02} implemented
a parameter $\kappa$ which controlled the opaqueness; $v_2$ was found to increase with $\kappa$ until the
source was essentially infinitely opaque.

\begin{figure}[t]
\epsfig{file=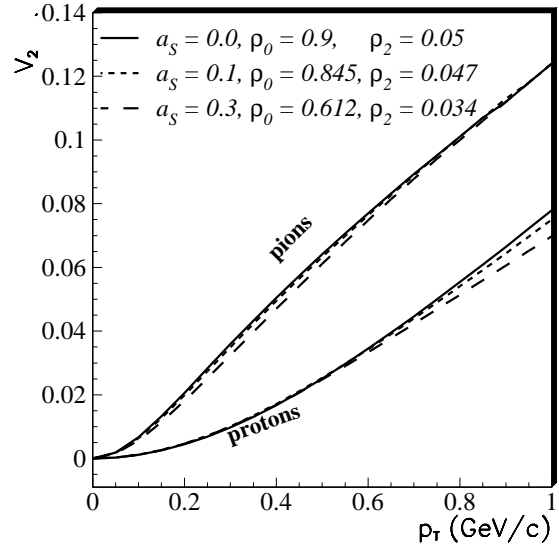,width=8cm}
\caption{
Elliptic flow parameter $v_2$ for pions (upper curves) and protons (lower curves), as a function
of transverse momentum, as calculated by Equation~\ref{eq:v2}, for various values of the
surface diffuseness parameter $a_s$.  The values of $\rho_0$ and $\rho_2$ are co-varied with
$a_s$ so that the average transverse flow and its average azimuthal modulation remain fixed; see text.
Other parameters follow the ``non-round'' source defaults of Table~\ref{tab:defaultParams}.
\label{fig:v2_vary_as}}
\end{figure}

Finally, we again explore the effect of ``softening'' the spatial source distribution $\Omega(r,\phi_s)$.
Similar to the discussion surrounding Figure~\ref{fig:Spectra_vary_as}, it is clear that
the most meaningful comparison comes from keeping
the average value of the transverse boost (and its anisotropy) constant, not keeping the parameters
$\rho_0$ and $\rho_2$ constant, as we vary $a_s$.
Similar to the discussion of the $p_T$ spectra, we find in Figure~\ref{fig:v2_vary_as}
that $v_2$ is largely insensitive to the surface diffuseness parameter $a_s$.

\subsection{HBT radii versus $p_T$ and $\phi_p$}
\label{sec:HBT_radii}

The previous subsections have discussed momentum-space observables only, even though coordinate-space considerations
came into play indirectly.  The most direct experimental probes of the space-time structure of the freeze-out
configuration are two-particle correlation functions in relative momentum~\cite{WH99,PLBNonId}.  Here, we use the
standard ``out-side-long'' coordinate system of Pratt and Bertsch~\cite{PrattBertsch}, in which the long direction
($R_l$) is parallel to the beam, the side direction ($R_s$) is perpendicular to the beam and total pair momentum, and
the out direction ($R_o$) is perpendicular to the long and side directions.

For boost-invariant sources, the HBT radii $R_{ol}^2$ and $R_{sl}^2$ vanish
by symmetry~\cite{HHLW02}.  (For the more general case, see References~\cite{W98,LHW00,HHLW02,E895HBTwrtRP,BudaLundEllipse}.)
Thus, we are left with four HBT radii, which are
related to space-time variances as~\cite{W98,WH99}
\begin{eqnarray}
\label{eq:radii-vs-xmunu}
    R_s^2 &=& \textstyle{\2}(\langle\tilde{x}^2\rangle{+}\langle\tilde{y}^2\rangle) 
            - \textstyle{\2}(\langle\tilde{x}^2\rangle{-}\langle\tilde{y}^2\rangle)\cos(2\phi_p)
  \nonumber\\*
          &&- \langle\tilde{x}\tilde{y}\rangle \sin(2\phi_p)
  \nonumber\\
    R_o^2 &=& \textstyle{\2}(\langle\tilde{x}^2\rangle{+}\langle\tilde{y}^2\rangle) 
            + \textstyle{\2}(\langle\tilde{x}^2\rangle{-}\langle\tilde{y}^2\rangle)\cos(2\phi_p)
  \nonumber\\*
          &&+ \langle\tilde{x}\tilde{y}\rangle \sin(2\phi_p)
            - 2\beta_\perp (\langle\tilde{t}\tilde{x}\rangle \cos\phi_p{+}\langle\tilde{t}\tilde{y}\rangle \sin\phi_p)
  \nonumber\\*
          && + \beta_\perp^2 \langle\tilde{t}^2\rangle, 
  \nonumber\\
    R_{os}^2 &=& \langle\tilde{x}\tilde{y}\rangle \cos(2\phi_p) 
        - \textstyle{\2} \left(\langle\tilde{x}^2\rangle{-}\langle\tilde{y}^2\rangle\right)\sin(2\phi_p)
  \nonumber\\*
             &&+ \beta_\perp (\langle\tilde{t}\tilde{x}\rangle \sin\phi_p{-}\langle\tilde{t}\tilde{y}\rangle \cos\phi_p), 
  \nonumber\\
    R_{l}^2 &=& \langle\tilde{z}^2\rangle -2 \beta_l \langle\tilde{t}\tilde{z}\rangle + \beta_l^2 \langle\tilde{t}^2\rangle, 
\end{eqnarray}
where
\begin{eqnarray}
\langle f(x) \rangle(K) \equiv \frac{\int d^4x f(x) S(x,K)}{\int d^4x S(x,K)}   \nonumber\\
\tilde{x}^\mu \equiv x^\mu - \langle x^\mu \rangle(K)
\end{eqnarray}

We restrict our attention to correlations calculated in the Longitudinally Co-Moving System (LCMS), in which
$Y = \beta_l = 0$.  In this case the last Equation~\ref{eq:radii-vs-xmunu} simplifies to
\begin{equation}
R_{l}^2 = \langle\tilde{z}^2\rangle
\end{equation}

In the notation of Section~\ref{sec:integrals}, and further defining
\begin{equation}
\label{eq:BPrime}
 \overline{ \left\{ B \right\}_{j,k} } \equiv \frac{\left\{ B \right\}_{j,k}}{\left\{ 1 \right\}_{0,0}} ,
\end{equation}
the space-time correlations of interest are
\begin{eqnarray}
\label{eq:xtildes-newnotation}
\langle \tilde{x}^2 \rangle & = & \overline{\left\{x^2\right\}_{0,0}} - \overline{\left\{x\right\}_{0,0}}^{~2}   \nonumber\\
\langle \tilde{y}^2 \rangle & = & \overline{\left\{y^2\right\}_{0,0}} - \overline{\left\{y\right\}_{0,0}}^{~2}   \nonumber\\
\langle \tilde{x}\tilde{y} \rangle & = & \overline{\left\{xy\right\}_{0,0}} -
                                       \overline{\left\{x\right\}_{0,0}} ~ \overline{\left\{y\right\}_{0,0}}   \nonumber\\
\langle \tilde{x}\tilde{t} \rangle & = & \frac{\Delta\tau^2+\tau_0^2}{\tau_0} \cdot
                                       \left(
                                          \overline{\left\{x\right\}_{0,1}} -
                                          \overline{\left\{x\right\}_{0,0}} ~ \overline{\left\{1\right\}_{0,1}}
                                       \right)                   \nonumber\\
\langle \tilde{y}\tilde{t} \rangle & = & \frac{\Delta\tau^2+\tau_0^2}{\tau_0} \cdot
                                       \left(
                                          \overline{\left\{y\right\}_{0,1}} -
                                          \overline{\left\{y\right\}_{0,0}} ~ \overline{\left\{1\right\}_{0,1}}
                                       \right)                   \nonumber\\
\langle \tilde{t}^2 \rangle & = & \left(3\Delta\tau^2 + \tau_0^2\right)\overline{\left\{1\right\}_{0,2}} - 
                     \left( \frac{\Delta\tau^2 + \tau_0^2}{\tau_0} \right)^2 \overline{\left\{1\right\}_{0,1}}^{~2}   \nonumber\\
\langle \tilde{z}^2 \rangle & = & \left(3\Delta\tau^2 + \tau_0^2\right)\overline{\left\{1\right\}_{2,0}} 
\end{eqnarray}

We note that all quantities with space-time dimensions are explicitly shown in Equations~\ref{eq:xtildes-newnotation},
and in particular, all dependence on the timescale parameters $\tau_0$ and $\Delta\tau$.

The proper time of freeze-out is often estimated (e.g.~\cite{PhenixHBT}) by fitting the
$m_T$-dependence of the measured $R_l$ radius,
to a formula motivated by Sinyukov and collaborators~\cite{MS88,AS95}, and subsequently improved
upon by others~\cite{HB95,WSH96}.  They assumed boost-invariant
longitudinal flow, but vanishing transverse flow ($\rho=0 \rightarrow \beta=\frac{m_T}{T}$)
and instantaneous freeze-out in proper time (i.e. $\Delta\tau=0$); they also simplified the
formalism by using Boltzmann statistics (i.e. using only the first term in the summation in Equation~\ref{eq:firstS}).
In this case, we find, in
agreement with References~\cite{HB95,WSH96},
\begin{equation}
R^2_l(m_T) = \tau_0^2 \frac{T}{m_T} \times \frac{\rm{K}_2(m_T/T)}{\rm{K}_1(m_T/T)} .
\end{equation}
We note that the last term, which represents a correction to the original Sinyukov formula~\cite{MS88,AS95},
approaches unity for $\frac{m_T}{T}\rightarrow\infty$ but remains sizable ($\sim 1.5$) for 
$\frac{m_T}{T}\sim2$.

\begin{figure}[h]
\epsfig{file=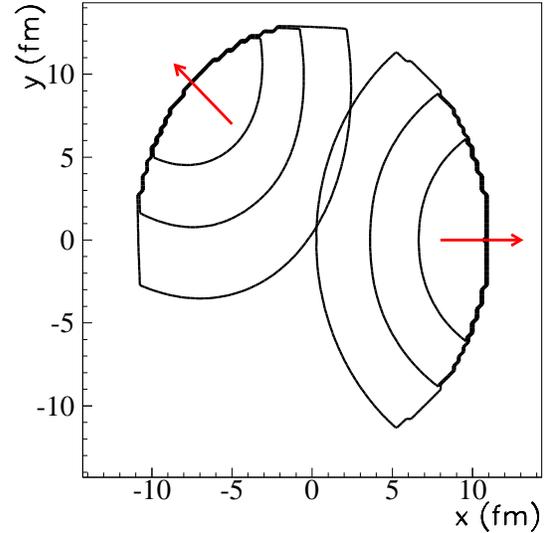,width=8cm}
\caption{
(Color online)  Emission probability contours, plotted on a linear scale,
indicate emission zones for pions with $p_T=0.3$~GeV/c at 
$\phi_p=0^\circ$ and $\phi_p=135^\circ$ (indicated by arrows),
from a blast-wave source with ``non-round'' default parameters listed in Table~\ref{tab:defaultParams}.
\label{fig:xyplots}}
\end{figure}

In general, the emission ``homogeneity region''~\cite{S95}
(characterized by the correlation coefficients of Equation~\ref{eq:xtildes-newnotation}) and the HBT radii of
Equation~\ref{eq:radii-vs-xmunu} depend on the pair momentum~\cite{WH99,W98}.  For a boost-invariant system,
this corresponds to dependences on $p_T$ and $\phi_p$.  Figure~\ref{fig:xyplots} shows
projections onto the transverse ($x-y$) plane of the emission regions for pions with $p_T=0.3$~GeV/c and
$\phi_p=0^\circ,135^\circ$, for an anisotropic source with strong transverse flow.  The flow generates strong
space-momentum correlations, so that particles with higher $p_T$ tend to be emitted from the edge of
the source, with $\phi_p \approx \phi_s$.  Together with the finite extent of the overall source, this
implies that, spatially, the emission region is often wider in the direction perpendicular to the particle
motion (indicated by the arrows) than along the motion; this can have strong implications, e.g., for the
difference between $R^2_o$ and $R^2_s$.

\begin{figure}[h]
\epsfig{file=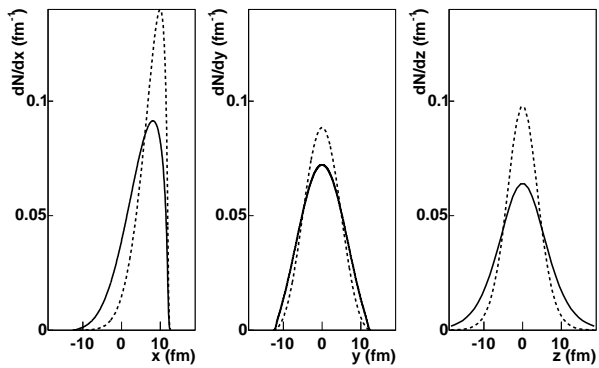,width=8cm}
\caption{
Distribution of the pion spatial distribution in the direction x (left), y (middle) and z (right).  Plain line:
pion momentum $p_x= $ 0.25 GeV/c and $p_y = $ 0. Dash line: pion momentum $p_x= $ 0.5 GeV/c and $p_y = $ 0. Round source parameters of Table~\ref{tab:defaultParams} were used, with the exception of $a_s$ that was set to 0.01.
\label{fig:nogaus}}
\end{figure}

While the blast-wave parameterization provides direct access to the homogeneity region, the radii extracted from two-pion interferometry
are compared to second order moments calculated as shown in
equation~\ref{eq:xtildes-newnotation}. Such comparison is strictly correct only if the homogeneity 
regions are Gaussian distributions. Figure~\ref{fig:nogaus} shows an example of spatial 
distributions along the
3 Cartesian directions for pions with two different momenta, $\overrightarrow{p} $= (0.25 GeV/c, 0, 0) 
and $\overrightarrow{p} $= (0.5 GeV/c, 0, 0); in this case, the $x-$, $y-$ and $z-$ directions correspond
to ``out,'' ``side'' and ``long,'' respectively
 The $z$ distribution is nearly Gaussian; the difference between the $\sigma$ extracted from 
a Gaussian fit and the second order moment is on the order of a few percent. On the other hand, the $x$ and $y$ distributions are clearly 
not Gaussian. To estimate the level of distortion introduced by the non-Gaussian shape, we compare the second order moments with the 
Gaussian $\sigma$ extracted by fitting the peak of the spatial distributons. Indeed, it has been shown in ~\cite{HardtkeVoloshin} that 
calculating correlation function from models and fitting them as experimental data yields radii that are close to the Gaussian $\sigma$ 
extracted by fitting the peak of the spatial distributons. In both the $x-$ and $y-$ directions, 
we find that the Gaussian $\sigma$ are systematically larger than the second 
order moments by up to 20\% (depending on the fit range)
for pions at  $p_T = $ 0.25 GeV/c. This discrepancy diminishes when increasing the transverse momentum or 
the particle mass. E.g. it is on the order of 10\% for pions at $p_T = $ 0.5 GeV/c or for kaons at $p_T = $ 0.25 GeV/c. Thus, comparing 
second order moments with the radii extracted from two-pion correlation functions may 
involve significant systematic errors at low transverse mass.
 This issue may be overcome by calculating correlation functions from the Blast Wave space-time distributions or by relying on imaging 
method to extract space-time distributions from the data~\cite{HBTImaging}. Applying these methods is beyond the scope of this paper and 
we will thus carry on keeping in mind that systematic errors are associated with comparing HBT radii with second order moment at 
low transverse momentum.

On the other hand, resonance decay may introduce a core-halo pattern in the distribution of particle  space-time emission points~\cite{CoreHalo}. Indeed, some resonances decay sufficiently far away 
from the bulk of the system that their decay products emerge beyond the main homogeneity region. Such particles form a halo around the core of the source.  However, to affect the extracted radii, the resonance 
lifetime needs to be short enough for the correlation to take place at a relative momentum
accesible to the experiment. This lifetime is typically between 10 and 100 fm. 
The effect of resonance feed-down on radii measured by two-pion interferometry  has been studied
within a early version 
of the blast wave parameterization~\cite{HeinzResonance}. It was found that the $\omega$ 
meson leads to a halo effect, its $c\tau$ being 23.4 fm. The other resonances have lifetimes that
are either too short  (e.g. $\rho$, $\Delta$) or too long (e.g. $\eta$, $K^0_s$).
 At RHIC energy, thermal models~\cite{Magestro} show that about 10\% of the pions come from $\omega$, which means that this effect should be rather small.
The effect of the very long-lived ($c\tau > 100$~fm) resonances is usually assumed to reduce
the so-called $\lambda$ parameter~\cite{CoreHalo2}, which we do not discuss here.


\begin{figure}[h]
\epsfig{file=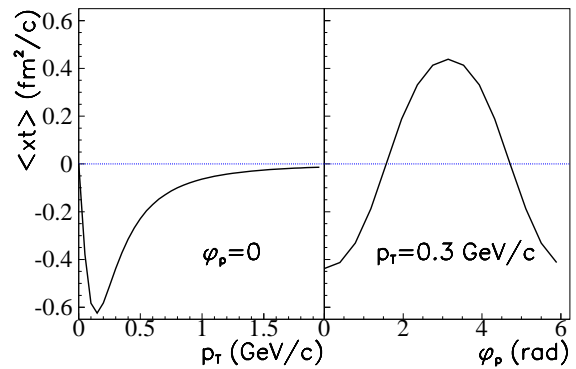,width=8cm}
\caption{(Color online)
Space-time correlation $\langle \tilde{x}\tilde{t} \rangle$ as a function of $p_T$ for $\phi_p=0$ (left
panel), and as a function of $\phi_p$ for $p_T=0.3$~GeV/c.  ``Non-round'' source parameters of Table~\ref{tab:defaultParams}
were used.
\label{fig:xtcorr}}
\end{figure}

Also important are temporal effects, including space-time correlations.  The average $x-t$ correlation,
quantified by $\langle \tilde{x}\tilde{t} \rangle$, for the same anisotropic source, is plotted in
Figure~\ref{fig:xtcorr}, as a function of $p_T$ and $\phi_p$.  In the left panel,
$\phi_p=0^\circ$, and thus $x$ is the ``out'' direction.  We first note that the correlation
is negative, tending to {\it increase} $R^2_o$ for finite $\beta_\perp$ (see Equation~\ref{eq:radii-vs-xmunu});
large and negative ``out''-$t$ correlations at freezeout in some hydrodynamical models~\cite{RG96} have been
a significant component of predicted large $R^2_o/R^2_s$ ratios.
As required by symmetry~\cite{HHLW02}, $\langle \tilde{x}\tilde{t} \rangle$ displays an odd-order
cosine dependence on $\phi_p$; clearly the first-order component dominates here, as shown in the
right panel of Figure~\ref{fig:xtcorr}.  This $\cos(\phi_p)$-dependence
is driven mostly by the $\cos(\phi_s)$ dependence of the $x$ coordinate, coupled with the radial flow which
tends to make $\phi_s \approx \phi_p$ (c.f. Figure~\ref{fig:xyplots}).
However, we also note from the figure
that the {\it scale} of these correlations is $\lesssim 1$~fm$^2$/c, which, as we shall
see, is much smaller than the typical scale of $R^2_o$; hence, space-time correlations do not dominate in the
blast-wave.

More important are the correlation coefficients $\langle \tilde{x}\tilde{y} \rangle$, 
$\langle \tilde{x}^2 \rangle$ and $\langle \tilde{y}^2 \rangle$, and their $p$-dependence.  
$\langle \tilde{x}\tilde{y} \rangle$ quantifies the transverse ``tilt'' of the emission zone 
with respect to the reaction plane.  As required by symmetry~\cite{HHLW02}, this tilt
vanishes at $\phi_p=0$, while, in the present case with $R_y>R_x$, it is positive for
$\phi_p=135^\circ$ (see Figure~\ref{fig:xyplots}).  That 
$\langle \tilde{x}^2 \rangle$ and $\langle \tilde{y}^2 \rangle$ also depend on $\phi_p$ is 
likewise clear from Figure~\ref{fig:xyplots}.  As we shall see, the $\phi_p$-dependence of 
$\langle \tilde{x}\tilde{y} \rangle$,
$\langle \tilde{x}^2 \rangle$, and $\langle \tilde{y}^2 \rangle$, combined with the explicit
$\phi_p$-dependences in Equation~\ref{eq:radii-vs-xmunu}, drive the oscillations in HBT radii which
we will study.

At this point, it is worthwhile to mention that the emission zone (``homogeneity region'')
and the correlation coefficients 
($\langle \tilde{x}^2 \rangle$, $\langle \tilde{y}^2 \rangle$, 
$\langle \tilde{x}\tilde{y} \rangle$, $\langle \tilde{x}\tilde{t} \rangle$, $\langle \tilde{y}\tilde{t} \rangle$)
will vary with $\phi_p$ also for an azimuthally {\it isotropic} source ($R_x=R_y$, $\rho_2=0$).  However, of course,
the measured HBT radii will be $\phi_p$-independent.  This arises due to cancellation and combination
of terms in Equations~\ref{eq:radii-vs-xmunu}.  For $R^2_s$, the first term becomes $\phi_p$-independent, and
the second and third terms {\it combined} are $\phi_p$-independent 
($\left( \langle \tilde{x}^2 \rangle - \langle \tilde{y}^2 \rangle \right) \propto \cos(2\phi_p)$
and $\langle \tilde{x}\tilde{y} \rangle \propto \sin(2\phi_p)$).  For $R^2_o$, the story is the same
for the first three terms, while the fourth term becomes $\phi_p$-independent as 
$\langle \tilde{x}\tilde{t} \rangle \propto \cos(\phi_p)$ and $\langle \tilde{y}\tilde{t} \rangle \propto \sin(\phi_p)$.
Meanwhile, for $R^2_{os}$, the first and second terms cancel each other, as do the two components
of the third term; hence $R^2_{os}=0$ for azimuthally-symmetric sources.
We mention these points here because several of the {\it exact} cancellations and combinations which
hold for an {\it isotropic} source, continue to hold {\it approximately} even when the geometry or
flow field is anisotropic.

We turn now to a detailed study of the observable HBT radius parameters, and the
effects of varying blast-wave parameters.
For a boost-invariant system, the 
symmetry-allowed $\phi_p$ oscillations of the (squared) HBT radii are~\cite{HHLW02}
\begin{eqnarray}
   R_s^2(p_T,\phi_p) &=& R_{s,0}^2(p_T) + {\textstyle2\sum_{n=2,4,6,\dots}} 
   R_{s,n}^2(p_T)\cos(n\phi_p),
 \nonumber\\
   R_o^2(p_T,\phi_p) &=& R_{o,0}^2(p_T) + {\textstyle2\sum_{n=2,4,6,\dots}} 
   R_{o,n}^2(p_T)\cos(n\phi_p),
 \nonumber\\
   R_{os}^2(p_T,\phi_p) &=& 
   {\textstyle2\sum_{n=2,4,6,\dots}} R_{os,n}^2(p_T)\sin(n\phi_p), \\
   R_l^2(p_T,\phi_p) &=& R_{l,0}^2(p_T) + {\textstyle2\sum_{n=2,4,6,\dots}}
   R_{l,n}^2(p_T)\cos(n\phi_p),  \nonumber
\end{eqnarray}
where the HBT radii Fourier coefficients $R^2_{\mu,n}$ ($\mu=o,s,os,l$ and $n=0,2,4,\dots$) are $\phi_p$-independent.

\begin{figure}[h]
\epsfig{file=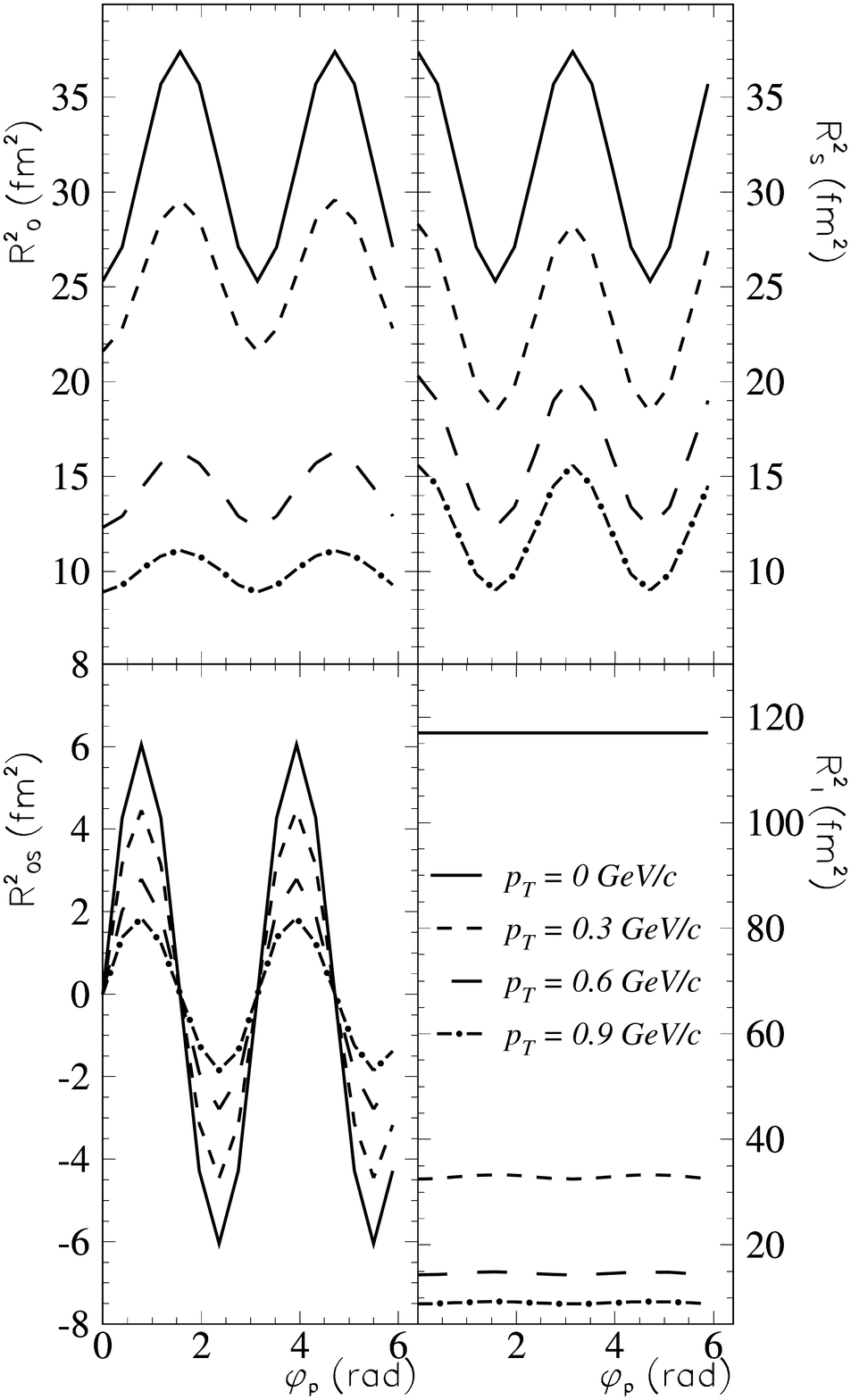,width=8cm}
\caption{
Squared HBT radius parameters calculated with Equations~\ref{eq:xtildes-newnotation} and~\ref{eq:radii-vs-xmunu}
from a blast-wave source with the ``non-round'' default parameters of Table~\ref{tab:defaultParams}.
Squared radii for various cuts in $p_T$ are plotted versus $\phi_p$, the emission angle which respect to the event plane.
\label{fig:HBTradii-oscillate}}
\end{figure}

Figure~\ref{fig:HBTradii-oscillate} shows the $p_T$ and $\phi_p$ dependence of HBT radii calculated
for a blast-wave source with a slightly anisotropic flow field and shape.  In addition to an overall
decrease in the average value of the HBT radii with increasing $p_T$, we observe significant oscillations
in the transverse radii $R^2_o$, $R^2_s$, $R^2_{os}$, and smaller oscillations in $R^2_l$.

The Fourier coefficients may be calculated as
\begin{equation}
\label{eq:HBTradii-FCs}
R^2_{\mu,n}(p_T) = 
\begin{cases}
\langle R^2_\mu(p_T,\phi_p) \cos(n\phi_p) \rangle & (\mu = o, s, l) \\
\langle R^2_\mu(p_T,\phi_p) \sin(n\phi_p) \rangle & (\mu = os) 
\end{cases}.
\end{equation}
In the present model, we find that oscillation amplitudes above 2$^{\rm nd}$ order are very small in all cases considered
($|R^2_{\mu,4}/R^2_{\mu,2}|\lesssim0.01$).
Therefore, the $\phi_p$-dependence of the HBT radii at a given $p_T$ is essentially encapsulated in 
seven numbers: the 0$^{\rm th}$- and 2$^{\rm nd}$-order Fourier coefficients of $R^2_o$, $R^2_s$, and $R^2_l$,
and the 2$^{\rm nd}$-order Fourier coefficient of $R^2_{os}$.  We henceforth explore the evolution of the
$p_T$-dependence of these seven numbers, as model parameters are varied.

\begin{figure}[h]
\epsfig{file=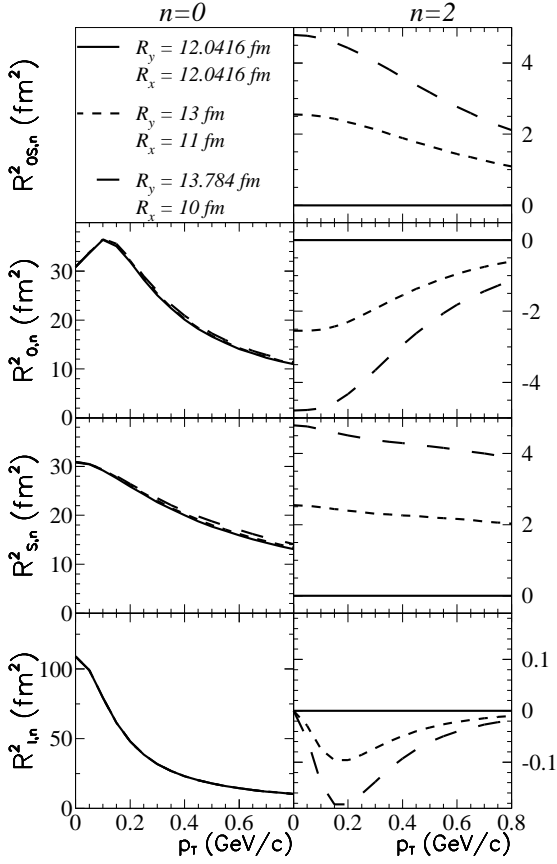,width=8cm}
\caption{
Fourier coefficients of the $\phi_p$-dependence of the squared HBT radii, as calculated
by Equation~\ref{eq:HBTradii-FCs}.  $0^{\rm th}$- and $2^{\rm nd}$-order Fourier coefficients are
plotted in the left and right panels, respectively.
The transverse shape of the source (i.e. $R_y/R_x$) was varied, while $R^2_y+R^2_x$ was held fixed.
``Round'' source values from Table~\ref{tab:defaultParams} are used for the other parameters.
\label{fig:HBT_vary_RyRxrat_fix_Rx2Ry2_NoRatio}}
\end{figure}

Figure~\ref{fig:HBT_vary_RyRxrat_fix_Rx2Ry2_NoRatio} shows the Fourier coefficients for $n=0,2$, corresponding
to a blast-wave source with $T=0.1$~GeV, an isotropic flow field ($\rho_0=0.9$, $\rho_2=0$), a
box profile ($a_s=0$), and time parameters $\tau_0=9$~fm/c, $\Delta\tau = 2$~fm/c.
Here, the average transverse size of the source ($R^2_x+R^2_y$) was held fixed, while the
shape ($R_y/R_x$) was varied. 

\begin{figure}[t]
\epsfig{file=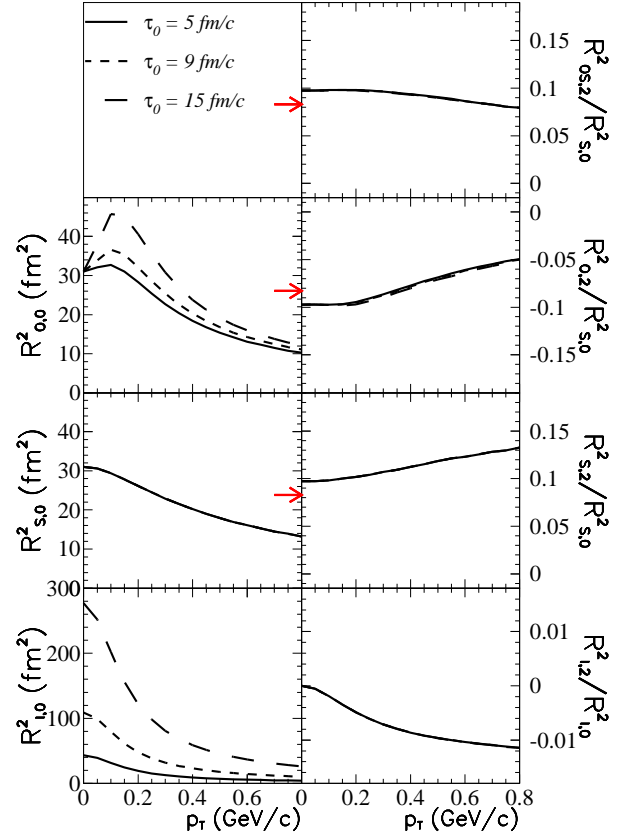,width=8cm}
\caption{(Color online)  
Fourier coefficients of the $\phi_p$-dependence of the squared HBT radii, as calculated
by Equation~\ref{eq:HBTradii-FCs}.
The evolution duration $\tau_0$ is varied.
Values of the other parameters are for a ``non-round'' source, as listed in Table~\ref{tab:defaultParams}.
Arrows indicate values calculated from Equation~\ref{eq:noflow-ratios}; see text for details.
\label{fig:HBT_vary_tau0}}
\end{figure}

\begin{figure}[t]
\epsfig{file=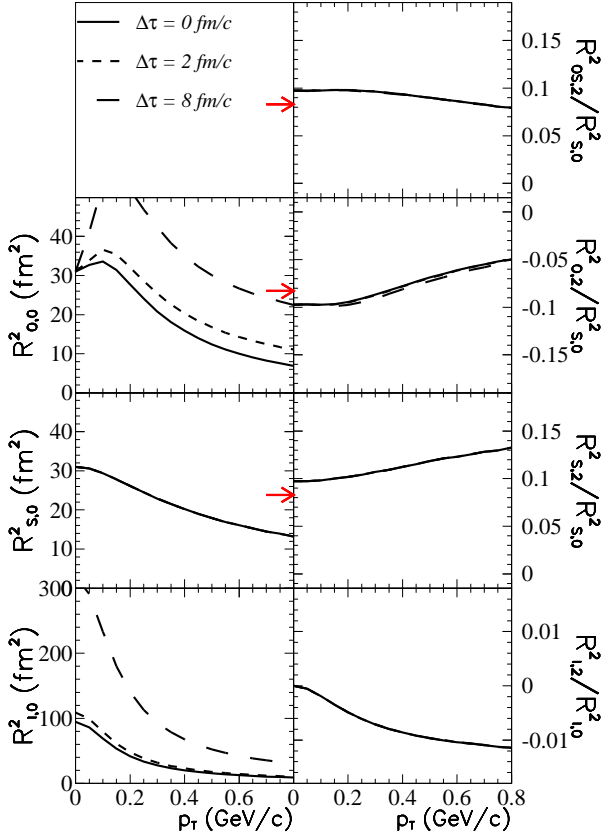,width=8cm}
\caption{(Color online)  
Fourier coefficients of the $\phi_p$-dependence of the squared HBT radii, as calculated
by Equation~\ref{eq:HBTradii-FCs}.
The evolution duration $\Delta\tau$ is varied.
Values of the other parameters are for a ``non-round'' source, as listed in Table~\ref{tab:defaultParams}.
Arrows indicate values calculated from Equation~\ref{eq:noflow-ratios}; see text for details.
\label{fig:HBT_vary_Deltatau}}
\end{figure}

\begin{figure}[t]
\epsfig{file=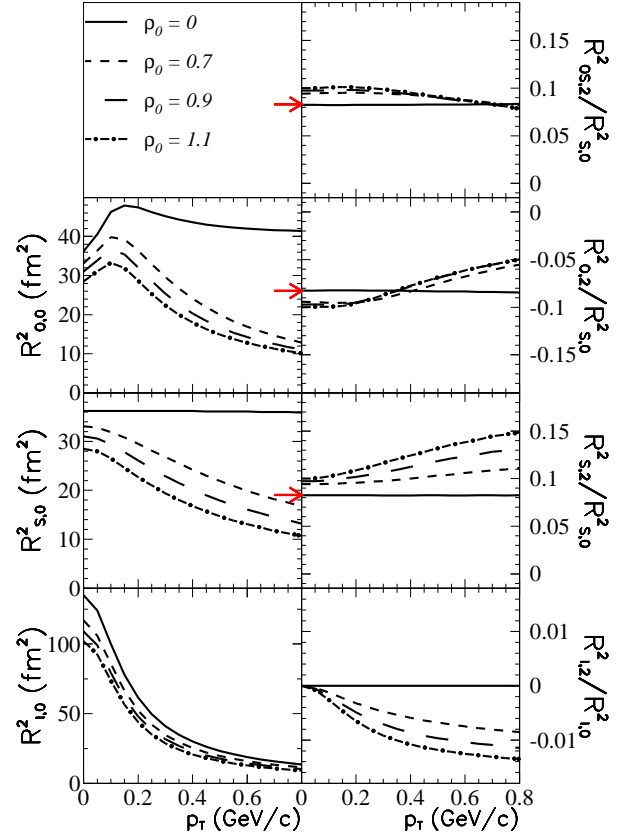,width=8cm}
\caption{(Color online)  
Fourier coefficients of the $\phi_p$-dependence of the squared HBT radii, as calculated
by Equation~\ref{eq:HBTradii-FCs}.
The average radial flow magnitude $\rho_0$ is varied.
Values of the other parameters are for a ``non-round'' source, as listed in Table~\ref{tab:defaultParams}.
Arrows indicate values calculated from Equation~\ref{eq:noflow-ratios}; see text for details.
\label{fig:HBT_vary_rho0}}
\end{figure}

 The $0^{\rm th}$-order Fourier coefficients (corresponding to the
radii $R_o$, $R_s$, and $R_l$ usually measured by experimentalists) are sensitive only to the
average scale, not the shape, of the source.  The average values of the transverse radii $\Ro{0}$ and
$\Rs{0}$ fall with increasing $p_T$ due to radial flow~\cite{WH99,WSH96} (c.f. Figure~\ref{fig:HBT_vary_rho0}).
At intermediate values of $p_T$, $\Ro{0}>\Rs{0}$ due to finite
timescale effects (cf Figures~\ref{fig:HBT_vary_tau0} and~\ref{fig:HBT_vary_Deltatau}), but at high
$p_T$, $\Ro{0}<\Rs{0}$ (i.e. $R_o/R_s < 1$), in qualitative agreement with experimental data~\cite{PhenixHBT,STARHBT}.
The boost-invariant longitudinal flow produces the strong decrease of $\Rl{0}$ with
$p_T$~\cite{WH99,MS88,AS95,HB95,WSH96}.

Richer detail is seen in the {\it oscillations} of the HBT radii,
quantified by Fourier coefficients $\Rmu{2}$ in the right-hand panels of
Figure~\ref{fig:HBT_vary_RyRxrat_fix_Rx2Ry2_NoRatio}.  Here, the elliptical shape of the source is explicitly
clear. 
The signs of the \second-order Fourier coefficients of the transverse radii
directly reflect the out-of-plane-extended source geometry when $R_y>R_x$.  $\Ros{2}$
has a similar geometric interpretation, in terms of the $\phi_p$-evolution of the ``tilt''
of the homogeneity region~\cite{HK02}.
The relatively small oscillations in $R^2_l$ arise not directly from geometry, but instead
from transverse flow gradients, which slightly reduce $R^2_l$~\cite{WSH96}.  In the present
example, the transverse flow increases linearly from 0 (at the center) to $\rho_0=0.9$ (at
the edge of the source), independent of boost angle $\phi_b$.  However, when $R_y>R_x$, 
the flow {\it gradient} is larger for source elements boosted in-plane, leading to slightly
greater reduction of  $R^2_l$ when $\phi_p=0$; hence $\Rl{2}<0$.

Finally, we recall that the flow field is isotropic ($\rho_2=0$) and so all $\phi_p$-dependence arises from
geometry here.  Thus, if the values of $R_y$ and $R_x$ are interchanged 
(corresponding to {\it in-plane}-extended sources), $\Rmu{0}$ would remain
unchanged, and $\Rmu{2}$ would simply change in sign in Figure~\ref{fig:HBT_vary_RyRxrat_fix_Rx2Ry2_NoRatio}.

\begin{figure}[t]
\epsfig{file=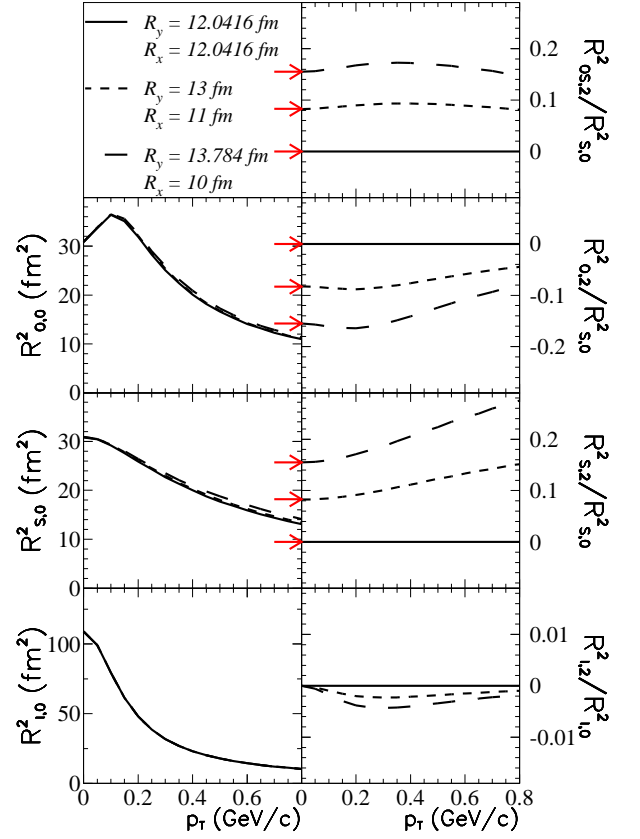,width=8cm}
\caption{(Color online)  
Identical data as in Figure~\ref{fig:HBT_vary_RyRxrat_fix_Rx2Ry2_NoRatio}, except the right
panels show ratios of $2^{\rm nd}$-order and $0^{\rm th}$ order Fourier coefficients
of the $\phi_p$-dependence of the squared HBT radii.
Arrows indicate values calculated from Equation~\ref{eq:noflow-ratios}; see text for details.
\label{fig:HBT_vary_RyRxrat_fix_Rx2Ry2}}
\end{figure}

As discussed above, transverse flow-induced space-momentum correlations tend to decrease
homogeneity lengths as $p_T$ increases.  When combined with other effects (e.g. temporal
effects), non-trivial $p_T$ dependences of the HBT radii result.  The $p_T$-dependences of
the $\phi_p$-averaged values $\Rmu{0}$ have been discussed extensively~\cite{WH99,WSH96}.
Meanwhile, the $p_T$-dependences of the oscillation amplitudes ($\Rmu{2}$) shown in the right panels of
Figure~\ref{fig:HBT_vary_RyRxrat_fix_Rx2Ry2_NoRatio} have not been explored previously and may be non-trivial in principle.

It was suggested~\cite{peter_dan_private} that the $p_T$-dependence of $|\Rmu{2}|$ might be driven
largely by the same effects which generate the $p_T$-dependence of $\Rmu{0}$, and hence the
most efficient and direct way to study the source is to plot $\Rmu{0}$, which encode 
scale information, and then the {\it ratio} of $2^{\rm nd}$- to $0^{\rm th}$-order Fourier
coefficients, which would encode geometric and dynamic anisotropy.  This is an excellent suggestion, though 
consideration must be given to the appropriate scaling.  First, we
consider the transverse radii $R^2_o$, $R^2_s$ and $R^2_{os}$.
The radii $R^2_o$ and $R^2_{os}$ encode both transverse geometry {\it and} temporal information.
As discussed above, space-time (e.g. $x-t$) correlations are small in magnitude, and furthermore
affect the HBT radii in combinations which tend to cancel any $\phi_p$-dependence.  Therefore, we
expect $\Ro{0}$, $\Rs{0}$, $\Rs{2}$, $\Ro{2}$ and $\Ros{2}$ to contain geometric contributions, while temporal
contributions are significant only for $\Ro{0}$.  In this case, the appropriate ratios to study are
$\Ro{2}/\Rs{0}$, $\Rs{2}/\Rs{0}$ and $\Ros{2}/\Rs{0}$.
Indeed, we find numerically that these are the ratios 
least affected by the overall scale of the homogeneity region, which varies both with
$p_T$ and with model parameter.
The oscillation strength $\Rl{2}$ of the longitudinal radius, on the other hand, is entirely
due to implicit $\phi_p$ dependences driven by space-momentum correlations; these same correlations
affect $\Rl{0}$.  Hence, the appropriate ratio to study in this case is $\Rl{2}/\Rl{0}$.

In Figure~\ref{fig:HBT_vary_RyRxrat_fix_Rx2Ry2} we show these ratios for the same sources as were
plotted in Figure~\ref{fig:HBT_vary_RyRxrat_fix_Rx2Ry2_NoRatio}.
The $p_T$-dependence of the ratios is significantly less than that of the oscillation
strengths $\Rmu{2}$, as anticipated, due to the fact that the latter is driven largely by
space-momentum correlations reducing the spatial scale of the homogeneity region.

Going further, we may recall that
in the special case of vanishing space-momentum correlations
($\rho=0$ or $T=\infty$), the transvserse radii oscillate
with identical strengths ($\Ros{2} = \Rs{2} = - \Ro{2}$), and the in-plane and out-of-plane extents of
the source may be directly extracted~\cite{W98,LHW00,HHLW02,E895HBTwrtRP}, as the ``whole source'' is
viewed from every angle.
In that special case, independent of $p_T$
\begin{equation}
\Rs{0} = \frac{1}{2} \left( \langle \tilde{y}^2 \rangle + \langle \tilde{x}^2 \rangle \right) 
       = \frac{1}{8}\cdot (R_y^2+R_x^2) \label{eq:Rs0-noflow}
\end{equation}
\begin{eqnarray}
\Ros{2} = \Rs{2} = -\Ro{2} &=& \frac{1}{4} \left( \langle \tilde{y}^2 \rangle - \langle \tilde{x}^2 \rangle \right)  \nonumber \\
        &=& \frac{1}{16} \left( R_y^2 - R_x^2 \right)  \label{eq:Ros2-Ro2-Rs2-noflow}
\end{eqnarray}
so that
\begin{equation}
\frac{\Ros{2}}{\Rs{0}} = \frac{\Rs{2}}{\Rs{0}} = -\frac{\Ro{2}}{\Rs{0}} = \frac{1}{2}\cdot\frac{R_y^2-R_x^2}{R_y^2+R_x^2} \equiv \frac{\epsilon}{2}
   \label{eq:noflow-ratios}
\end{equation}

\begin{figure}[t]
\epsfig{file=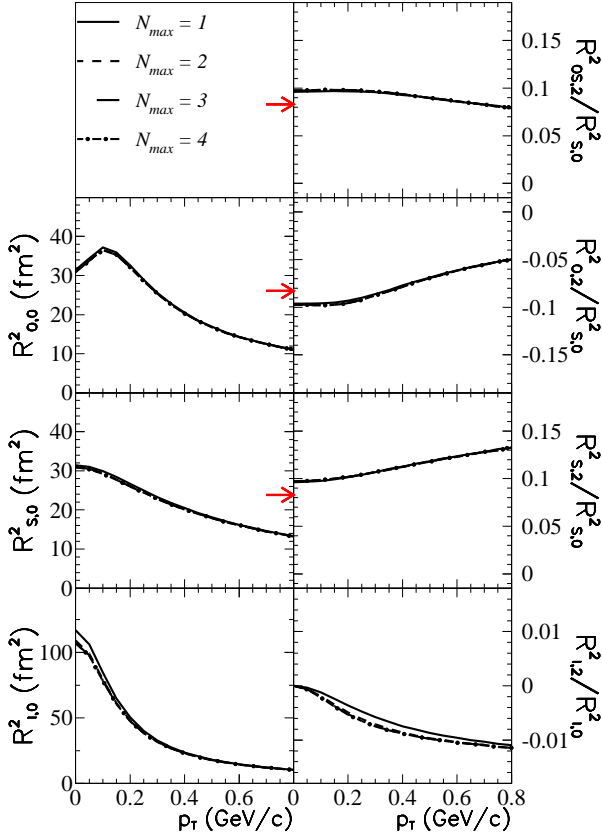,width=8cm}
\caption{(Color online)  
Fourier coefficients of the $\phi_p$-dependence of the squared HBT radii, as calculated
by Equation~\ref{eq:HBTradii-FCs} using several values of $N$, the maximum value of $n$
taken in the summation of Equation~\ref{eq:firstS}.
Values of the blast-wave parameters are for a ``non-round'' source, as listed in Table~\ref{tab:defaultParams}.
Arrows indicate values calculated from Equation~\ref{eq:noflow-ratios}.  See text for details.
\label{fig:HBT_QuantumExpansion}}
\end{figure}

\begin{figure}[t]
\epsfig{file=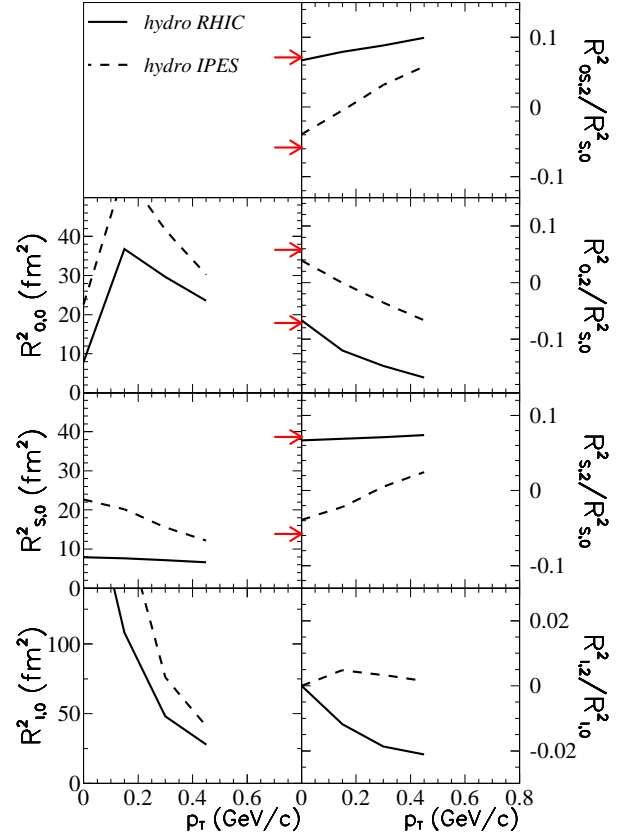,width=8cm}
\caption{(Color online)  
Fourier coefficients of the $\phi_p$-dependence of the squared HBT radii, as calculated
by Equation~\ref{eq:HBTradii-FCs}.
The data are not from blast-wave calculations, but from hydrodynamic calculations of
Heinz and Kolb~\cite{HK02}, for two different equations of state and initial conditions.
Arrows indicate values calculated from Equation~\ref{eq:noflow-ratios} and extracted values
of freezeout edge radii; see text for details.
\label{fig:HBT_KolbHeinz}}
\end{figure}

\begin{figure}[t]
\epsfig{file=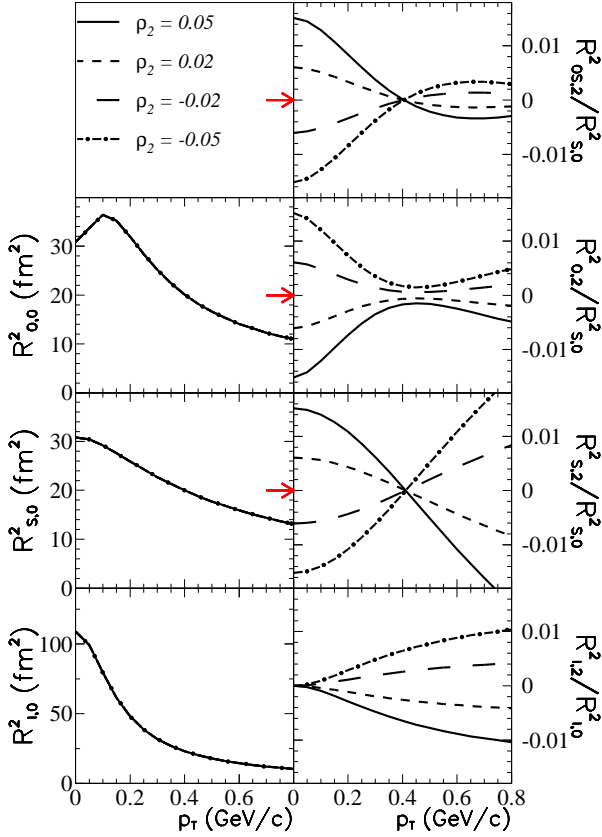,width=8cm}
\caption{(Color online)  
Fourier coefficients of the $\phi_p$-dependence of the squared HBT radii, as calculated
by Equation~\ref{eq:HBTradii-FCs}.
The anisotropy of the flow field ($\rho_2$) is varied.
Values of the other parameters are for a ``round'' source, as listed in Table~\ref{tab:defaultParams}.
Arrows indicate values calculated from Equation~\ref{eq:noflow-ratios}; see text for details.
\label{fig:HBT_vary_rhoa_RyRx12.04}}
\end{figure}

\begin{figure}[t]
\epsfig{file=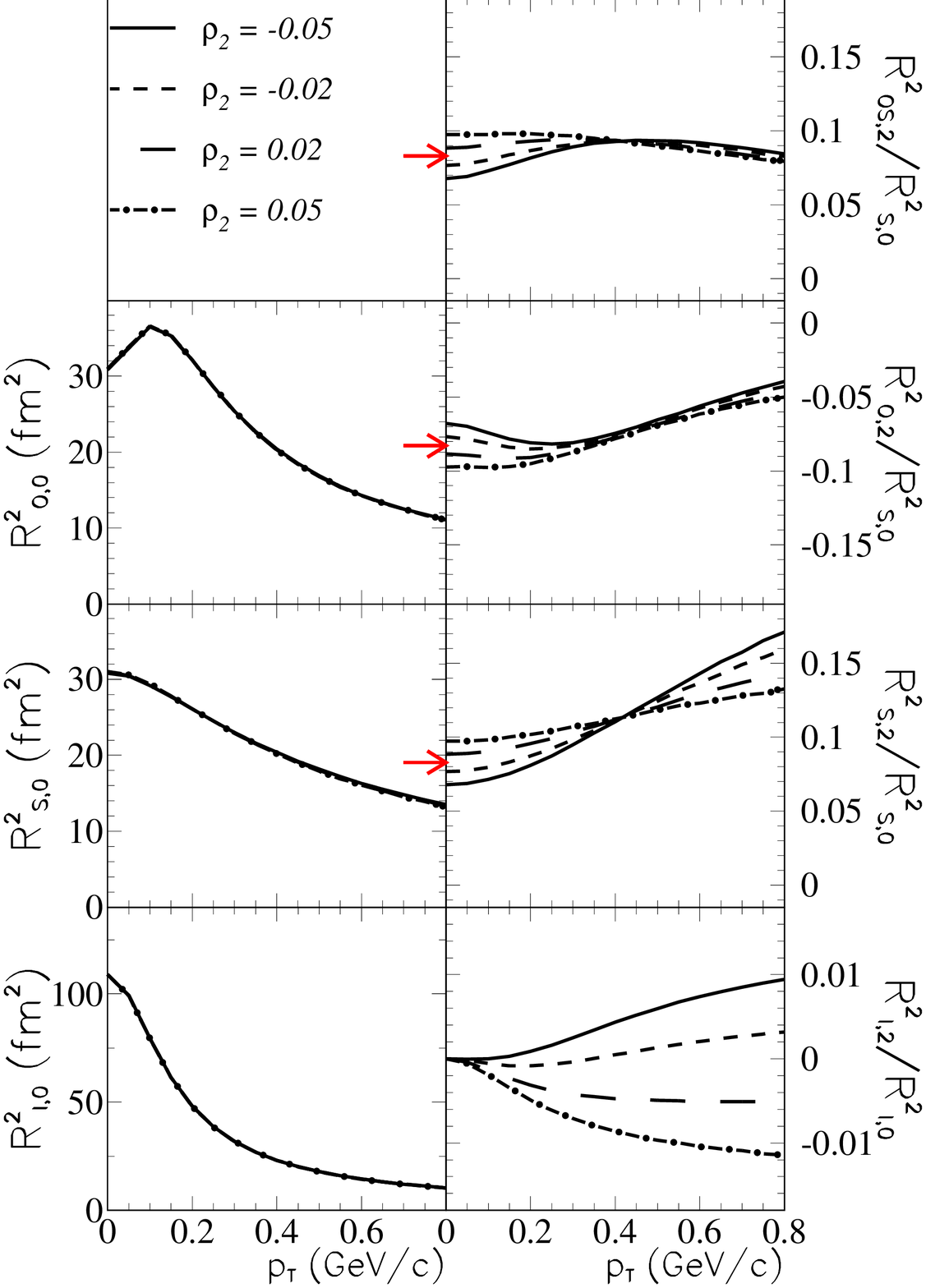,width=8cm}
\caption{(Color online)  
Fourier coefficients of the $\phi_p$-dependence of the squared HBT radii, as calculated
by Equation~\ref{eq:HBTradii-FCs}.
The anisotropy of the flow field ($\rho_2$) is varied.
Values of the other parameters are for a ``non-round'' source, as listed in Table~\ref{tab:defaultParams}.
Arrows indicate values calculated from Equation~\ref{eq:noflow-ratios}; see text for details.
\label{fig:HBT_vary_rhoa_Ry13Rx11}}
\end{figure}

\begin{figure}[h]
\epsfig{file=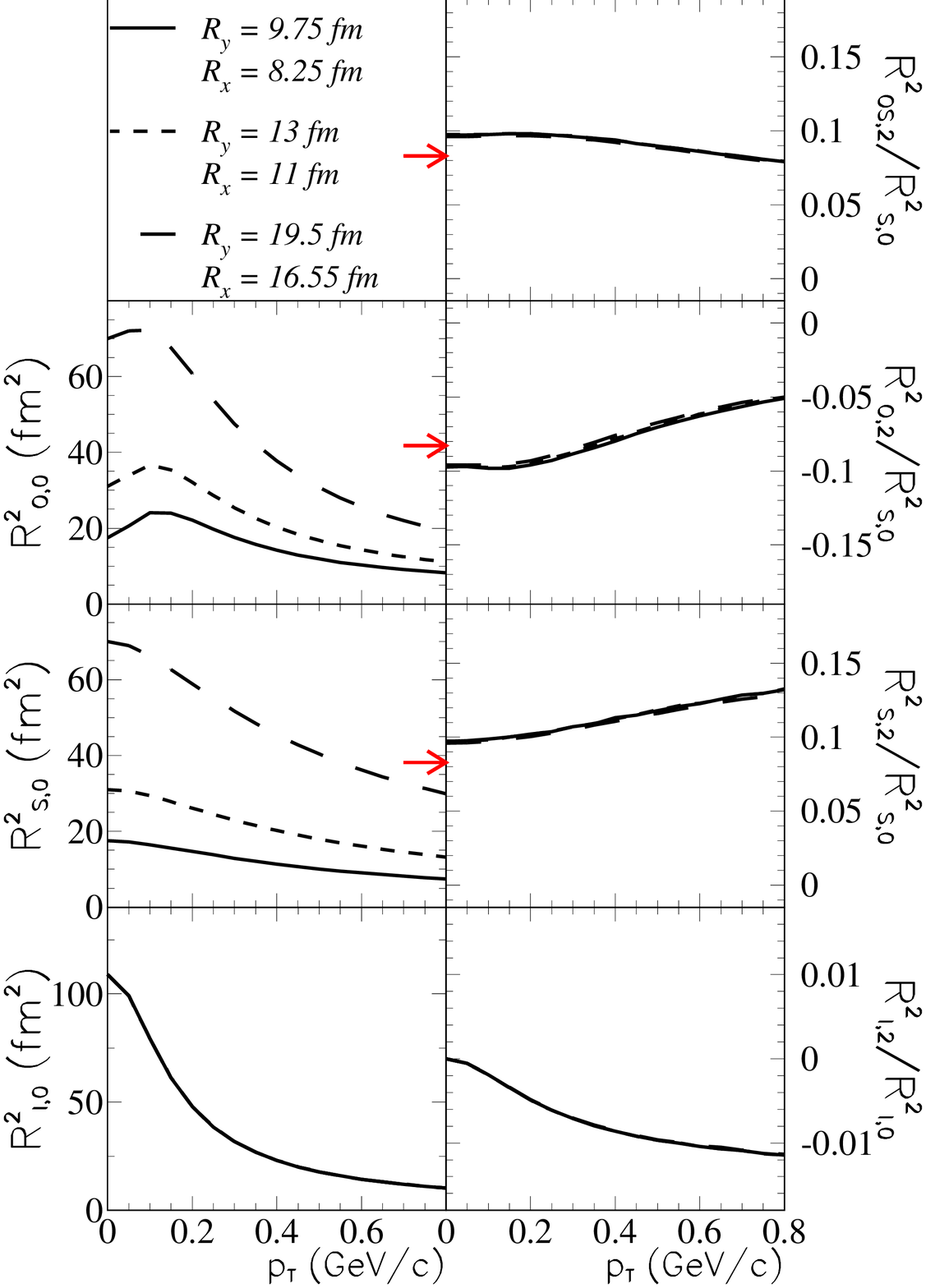,width=8cm}
\caption{(Color online)  
Fourier coefficients of the $\phi_p$-dependence of the squared HBT radii, as calculated
by Equation~\ref{eq:HBTradii-FCs}.
The overall transverse scale ($\sqrt{R_y^2+R_x^2}$) is varied, while keeping $R_y / R_x =13/11$.
Values of the other parameters are for a ``non-round'' source, as listed in Table~\ref{tab:defaultParams}.
Arrows indicate values calculated from Equation~\ref{eq:noflow-ratios}; see text for details.
\label{fig:HBT_vary_Rx2Ry2_fixRyRxrat}}
\end{figure}

\begin{figure}[t]
\epsfig{file=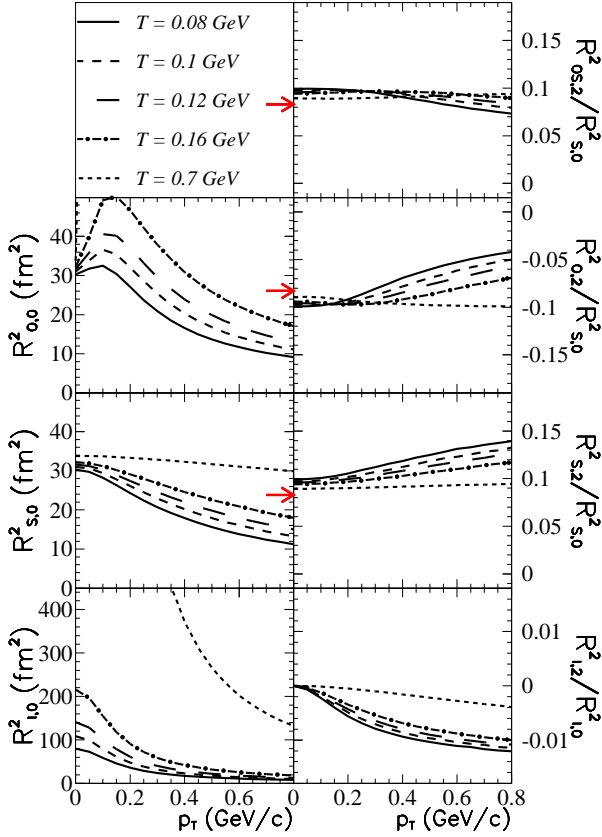,width=8cm}
\caption{(Color online)  
Fourier coefficients of the $\phi_p$-dependence of the squared HBT radii, as calculated
by Equation~\ref{eq:HBTradii-FCs}.
For $T=0.7$~GeV/c, $\Ro{0}$ and $\Rl{0}$ curves partially exceed the displayed scale.
Values of the other parameters are for a ``non-round'' source, as listed in Table~\ref{tab:defaultParams}.
Arrows indicate values calculated from Equation~\ref{eq:noflow-ratios}; see text for details.
\label{fig:HBT_vary_T}}
\end{figure}

\begin{figure}[t]
\epsfig{file=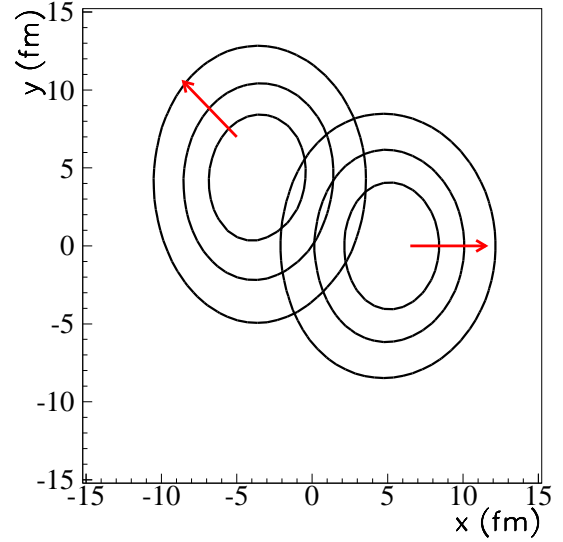,width=8cm}
\caption{(Color online)  
Emission probability contours, plotted on a linear scale,
indicate emission zones for pions with $p_T=0.3$~GeV/c at 
$\phi_p=0^\circ$ and $\phi_p=135^\circ$ (indicated by arrows),
from a blast-wave source with $a_s = 0.3$.
Other parameter values correspond to the ``non-round'' values
listed in Table~\ref{tab:defaultParams}.
\label{fig:xyplots_as0.3}}
\end{figure}

\begin{figure}[h]
\epsfig{file=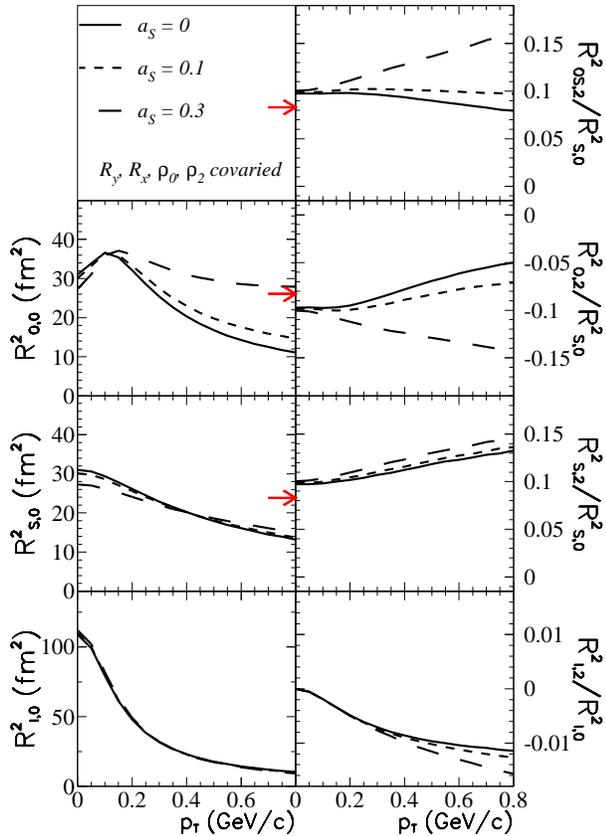,width=8cm}
\caption{(Color online)  
Fourier coefficients of the $\phi_p$-dependence of the squared HBT radii, as calculated
by Equation~\ref{eq:HBTradii-FCs}.
The parameters $T$, $\tau_0$ and $\Delta \tau$ are set to values listed in Table~\ref{tab:defaultParams},
while the parameters $\rho_0$, $\rho_2$, $R_y$ and $R_x$ are co-varied with the surface diffuseness $a_s$;
see text for details.
\label{fig:HBT_vary_as}}
\end{figure}

In the presence of flow, however, HBT radii measured at momentum $\vec{p}$
reflect homogeneity lengths which in principle may vary nontrivially both with $p_T$ and $\phi_p$.
While we find that non-vanishing flow violates the $p_T$-independence of $\Rs{0}$, $\Ro{2}$, $\Rs{2}$ and $\Ros{2}$
(and thus Equations~\ref{eq:Rs0-noflow} and~\ref{eq:Ros2-Ro2-Rs2-noflow}), Equation~\ref{eq:noflow-ratios} remains
remarkably robust.  As seen in the next several Figures, the ratios $\Ro{2}/\Rs{0}$, $\Rs{2}/\Rs{0}$ and $\Ros{2}/\Rs{0}$,
largely independent of $p_T$, provide an estimate 
of the source ellipticity $\epsilon$.  Arrows to the left of the panels for $\Ro{2}/\Rs{0}$, $\Rs{2}/\Rs{0}$ and $\Ros{2}/\Rs{0}$
in Figures~\ref{fig:HBT_vary_RyRxrat_fix_Rx2Ry2}-\ref{fig:HBT_vary_as} indicate $\epsilon/2$ for the sources used.

Now that we have established the quantities to be examined in this Section, we briefly check the importance of using
quantum, rather than classical, statistics in the source function of Equation~\ref{eq:firstS}.  Setting the parameter values to
correspond to the ``non-round'' source of Table~\ref{tab:defaultParams}, we plot in Figure~\ref{fig:HBT_QuantumExpansion} the Fourier coefficients corresponding
to different values of $N$, where the summation in Equation~\ref{eq:firstS} (and Equations~\ref{eq:secondS} and~\ref{eq:bracketBetaPrime}) is
over $n = 1 \ldots N$.  Once again, we find a small difference between the curves for $N=1$ and $N>1$, while inclusion of higher terms
in the summation have essentially no effect.  Blast-wave calculations in this Section correspond to $N=2$.

Recently, Heinz and Kolb, in a hydrodynamic model, calculated HBT radii as a function of $\phi_p$ for non-central collisions~\cite{HK02}.
They used two different Equations of State and initial conditions: one (``RHIC'') is appropriate for soft physics at RHIC energies,
and has successfully reproduced momentum-space observables~\cite{HK01}; the other (``IPES'') assumes extremely high initial energy
densities, perhaps appropriate for collisions at LHC energies.

It is worthwhile to point out that even the
``RHIC'' hydrodynamic calculations fail to reproduce azimuthally-integrated HBT data~\cite{HK01}; here, however, we simply
investigate the connection between the freeze-out geometry and oscillations of the HBT radii.
Both calculations result in a freeze-out configuration,
integrated over $p_T$, which is rather sharp-edged in transverse coordinate-space; thus, we may extract surface radii
$R_x$ and $R_y$ to calculate $\epsilon$.  

Figure~\ref{fig:HBT_KolbHeinz} shows the same quantities as plotted in Figure~\ref{fig:HBT_vary_RyRxrat_fix_Rx2Ry2}, but
extracted from these hydrodynamic calculations.  The ``RHIC'' source, which is geometrically extended out-of-plane
($R_y>R_x$, resulting in a positive $\epsilon$) generates oscillations in the transverse radii with the same phase
as out-of-plane sources in blast-wave calculations.  For this source, the connection between $\epsilon$ and
the radius oscillations (Equation~\ref{eq:noflow-ratios}) is most robust for $\Rs{2}$ and least well-satisfied for
$\Ro{2}$, an effect not observed in the blast-wave.  However, our blast-wave parameterization does not include
explicit $\phi_p$ dependence of the temporal scale, which would affect $\Ro{2}$, and, to a lesser degree, $\Ros{2}$.
Instead of attempting a more sophisticated parameterization, we simply note this fact, and would recommend that an
experimental estimation of the source deformation $\epsilon$ is probably best extracted from $\Rs{2}/\Rs{0}$, which
should be unaffected by azimuthal structure of the temporal scale.  From the study (below) of parameter variations
in the blast-wave, the approximation $\epsilon \approx 2\Rs{2}/\Rs{0}$ is good to $\sim$30\%, for RHIC-type sources.

Figure~\ref{fig:HBT_KolbHeinz} also shows results from the ``IPES'' hydrodynamic calculation.  Here, the freeze-out
shape is extended in-plane~\cite{HK02}, but dynamical effects are so strong in this extreme case, that even the
sign of the transverse radius oscillations changes with $p_T$.  The relationships in Equation~\ref{eq:noflow-ratios}
work only at low $p_T$, and even there only very approximately.  
According to this model, then, geometrical considerations dominate the
   HBT radius oscillations, while dynamical effects begin to dominate at
   much higher energies.

In Figure~\ref{fig:HBT_vary_rhoa_RyRx12.04}, the source geometry is azimuthally isotropic
($R_x=R_y=12.042$~fm), while the flow field is varied from having a stronger boost in-plane ($\rho_2>0$)
to a stronger boost out-of-plane ($\rho_2<0$).  We notice again that the average HBT radius values
$\Rmu{0}$ are unaffected by the anisotropy.  The oscillations ($\Rmu{2}$) are driven by flow gradients.
Naively, one would expect that all HBT radii $R_o$, $R_s$, $R_l$ would be smaller when the emission
angle $\phi_p$ is in the direction of the strongest boost.  For $\rho_2=0.05$, for example, the radii
would be smaller at $\phi_p=0$, the direction of stronger transverse boost; this would correspond to
$\Rmu{2}<0$ ($\mu = o,s,l$).  For $\mu = o$ and $\mu = l$, this is indeed observed, at all $p_T$.
$\Rs{2}$, however, changes sign from positive at low $p_T$, to negative at high $p_T$.  This 
behaviour at low $p_T$ is due to an effect similar to that which led to negative proton $v_2$ at low $p_T$,
even when $\rho_2>0$.  (See discussion surrounding Figure~\ref{fig:v2_vary_rho0_RyRxrat_1.0}.)
In the present case, the particles with $p_T \approx 0$ are more likely to be emitted by source elements
positioned along the $y$-axis, due to the strong in-plane boost for source elements with large spatial
coordinate $x$.  The homogeneity region for $p_T=0$ particles is independent of $\phi_p$,
and has larger extent out-of-plane than in-plane.
Thus we find the counter-intuitive result that $R^2_s(\phi_p=0)>R^2_s(\phi_p=\frac{\pi}{2}) \rightarrow \Rs{2}>0$
at $p_T \approx 0$.  We note that a similar argument holds for $R^2_o$, except that it leads to the
conclusion that $\Ro{2}<0$ at $p_T \approx 0$, and so goes in the same direction as flow-gradient effects.
It is only for $R^2_s$ that the two effects compete.

Finally, comparing the scales on the right-hand panels of Figures~\ref{fig:HBT_vary_RyRxrat_fix_Rx2Ry2}
and~\ref{fig:HBT_vary_rhoa_RyRx12.04}, it is clear that, while the \second-order coefficients
are driven by both anisotropic geometry and flow field, a variation in geometry ($R_y/R_x$) has a stronger
effect on $\Rmu{2}$ than a variation in $\rho_2$, when these parameters are varied by amounts which
would generate a similar effect on elliptic flow $v_2$ (cf Figures~\ref{fig:v2_vary_rhoa_RyRxrat_1.0}
and~\ref{fig:v2_vary_RxRyrat_rhoa0.0}).  Thus, measurement of both $v_2$ and HBT radius oscillations
would allow independent determination of both anisotropic flow strength $\rho_2$ and shape $R_y/R_x$.

In Figure~\ref{fig:HBT_vary_rhoa_Ry13Rx11}, we consider anisotropy simultaneously in both the source
geometry and the flow field.
The source is extended out-of-plane ($R_y=13$~fm and $R_x=11$~fm), and the flow anisotropy ($\rho_2$)
varied.
The \zeroth-order coefficients remain unaffected by the anisotropies, while the $\Rmu{2}$
reflect essentially the cumulative effects of the geometric anisotropy (shown in
Figure~\ref{fig:HBT_vary_RyRxrat_fix_Rx2Ry2}) and the flow field anisotropy (shown in
Figure~\ref{fig:HBT_vary_rhoa_RyRx12.04}), with no strong non-linear coupling between them.
Thus, while {\it in principle} anisotropic flow effects may mask or dominate geometric anisotropy~\cite{HK01},
flow field anisotropies represent small perturbations on the dominant geometric effects in the blast-wave,
using ``realistic'' (cf Section 4) parameters.

In Figure~\ref{fig:HBT_vary_Rx2Ry2_fixRyRxrat} we show the effect of increasing the
transverse size of the source ($\sqrt{R_y^2+R_x^2}$) while keeping the shape ($R_y / R_x$)
and other source parameters fixed.  As expected, the purely-spatial
transverse radius $R_s^2$ (average and oscillation amplitude) increases proportionally with $R_y^2+R_x^2$.
The squared ``outward'' radius parameter,
$R_o^2$, contains both spatial components (which increase with $R_y^2+R_x^2$)
and temporal components (which do not).  Thus, its average value, $\Ro{0}$, increases
almost proportional to $R_y^2+R_x^2$ at low $p_T$ ($\beta_\perp \sim 0$) and less so at higher $p_T$.
Due to the near-cancellation of $\phi_p$-dependence of temporal terms,
the increase in oscillation amplitudes $\Ros{2}$ and $\Ros{2}$ is driven mainly by the spatial terms,
so that $\Ro{2}/\Rs{0}$ and $\Ros{2}/\Rs{0}$ display almost no sensitivity
to $R_y^2+R_x^2$ at any $p_T$.
The longitudinal radius $R_l^2$ is unaffected by variation in the transverse scale.

In Figures~\ref{fig:HBT_vary_tau0} and~\ref{fig:HBT_vary_Deltatau}, we vary the timescale parameters
$\tau_0$ and $\Delta \tau$, respectively.  All dependence of $\Rmu{n}$ on these parameters come 
directly through the dependence of $\langle \tilde{x}_\mu \tilde{x}_\nu \rangle$, which are listed
explicitly in Equations~\ref{eq:xtildes-newnotation}.  After an inspection of those equations, it is unsurprising that 
the effects of varying these parameters are similar.  The space-time correlation coefficients which depend
on the timescale parameters are 
$\langle \tilde{x} \tilde{t} \rangle$,
$\langle \tilde{y} \tilde{t} \rangle$,
$\langle \tilde{t}^2 \rangle$,
and $\langle \tilde{z}^2 \rangle$.  According to Equation~\ref{eq:radii-vs-xmunu}, then,
 $\Rs{0}$ and $\Rs{2}$ are unaffected by variations in
$\tau_0$ and $\Delta \tau$.  $R^2_l = \langle \tilde{t}^2 \rangle$ is directly proportional to 
$\left(3\Delta\tau^2 + \tau_0^2\right)$, so $\Rl{0}$ and $\Rl{2}$ both scale with that quantity.
Turning to the HBT radii with both spatial and temporal contributions, we again find that the
$\phi_p$-dependence of the temporal terms is negligible, so that $\Ro{2}/\Rs{0}$ and $\Ros{2}/\Rs{0}$
are independent of the timescales, while $\Ro{0}$ displays the well-known~\cite{WH99} sensitivity
to timescale.

Thus, we find that, in the blast-wave parameterization, essentially all sensitivity to timescales
comes through the $\phi_p$-independent quantities $\Ro{0}$ and $\Rl{0}$.
However, it is important to point out that an experimental estimate of the freeze-out 
geometric anisotropy, defined in Equation~\ref{eq:noflow-ratios}, from measurements of
the $\Rmu{n}$ {\it would} place an additional constraint on the evolution timescale $\tau_0$.
In particular, the large initial-state anisotropy in coordinate space ($R_y>R_x$) in a non-central collision
will be reduced due to stronger flow in-plane than out-of-plane ($\rho_2>0$ in the present parameterization).
If the source lives for a long time (large $\tau_0$), the system may become round ($R_y=R_x$) or even 
in-plane extended ($R_y<R_x$)~\cite{TLS01}.  A quantitative constraint on $\tau_0$ from the relationship of
the initial to freeze-out anisotropies, however, must be made in the context of a realistic dynamical model
and is beyond the scope of the blast-wave parameterization.


In Figure~\ref{fig:HBT_vary_rho0}, the effect of variations in the $\phi$-averaged (``radial'') flow 
on the HBT radius parameters is shown.
As is well-known~\cite{WH99}, stronger flow reduces the homogeneity lengths, and, indeed,
almost all of the $|\Rmu{n}|$ fall with increasing $\rho_0$.  The one interesting exception is $\Rl{2}$;
while the average scale of the longitudinal radius ($\Rl{0}$) decreases as the flow is increased, 
its $\phi_p$-dependence ($|\Rl{2}|$) {\it increases} (and so, then, does $|\Rl{2}/\Rl{0}|$).
This is because there is no explicit $\phi_p$
dependence in $R^2_l =\langle \tilde{z}^2 \rangle$; any $\phi_p$ dependence is implicit, and thus
is generated by space-momentum correlations~\cite{W98}, which, in this model, arise solely from flow.
In the no-flow limit for a boost-invariant source,
$\langle \tilde{t}^2 \rangle$,
$\langle \tilde{x}^2 \rangle$,
$\langle \tilde{y}^2 \rangle$, and
$\langle \tilde{z}^2 \rangle$ 
are all $\phi_p$-independent constants~\cite{W98,LHW00,HHLW02}.

Indeed, it is for a similar reason that $\Rl{2}$ vanishes for $p_T=0$, independent of model parameter
in Figures~\ref{fig:HBT_vary_RyRxrat_fix_Rx2Ry2}-\ref{fig:HBT_vary_T}.  At $p_T=0$, symmetry demands that
none of the spatial correlation coefficients $\langle \tilde{x}_\mu \tilde{x}_\nu \rangle$ may depend
on $\phi_p$.  Hence, only HBT radii with {\it explicit} $\phi_p$-dependence may exhibit an oscillation
in that limit.
%
%

Finally, we note in Figure~\ref{fig:HBT_vary_rho0} that the oscillation strengths $|\Ro{2}|$ and $|\Rs{2}|$
are somewhat less diminished (at low $p_T$) by increasing radial flow, than is $\Rs{0}$, which measures the
overall spatial
scale of the homogeneity region.  Here, we offer no simple insights on the interplay between the increasing
deformation of the homogeneity region, and its decreasing scale, but simply note that the dependence of the
scaled oscillation strengths on $\rho_0$ is rather small, especially at low $p_T$, even for the unrealistic
case of no average transverse flow ($\rho_0 = 0$).

Figure~\ref{fig:HBT_vary_T} explores the effect on HBT radii of varying the temperature parameter $T$.
Increasing $T$ increases ``thermal smearing'', reducing the flow-induced space-momentum
correlations.  It is well known that this leads to increased homogeneity lengths and HBT radii~\cite{SSH93},
as reflected in the left panels of the Figure.  In the present model, we find a small residual dependence
of $\Rmu{2}$ on $T$ beyond the scaling of $\Rmu{0}$ (right panels).

In the limit of very high temperature, all space-momentum correlations are eliminated, and all
oscillations of HBT radii are again due to the explicit $\phi_p$-dependence in Equations~\ref{eq:radii-vs-xmunu}.
Then, we find that Equations~\ref{eq:Rs0-noflow}-\ref{eq:noflow-ratios} hold, independent of $p_T$.
For the source of Figure~\ref{fig:HBT_vary_T}, according to Equation~\ref{eq:noflow-ratios},
$|\Rmu{2}/\Rs{0}|=0.083$.
%
%
Finally, $\Rl{2}$ decreases with increasing $T$, and must vanish at $T \rightarrow \infty$, when all
space-momentum correlations are destroyed.

Finally, we consider the effect of a finite ``skin thickness'' $a_s$.
As in the discussions surrounding Figures~\ref{fig:Spectra_vary_as} and~\ref{fig:v2_vary_as}, it is
appropriate to scale the flow parameters according to $F(a_s)$ given in Equation~\ref{eq:fraction_vs_as}.
Moreover, it is clearly appropriate to scale the geometric size parameters $R_y$ and $R_x$, as the overall
scale of the source will increase with $a_s$, if these parameters remain fixed.  Less clear is the exact
scaling which would keep, e.g. $\Rs{0}$, independent of $a_s$, especially in the presence of finite flow.
For illustrative purposes, we scale $R_y$ and $R_x$ also by $F(a_s)$, so that $R_y$ = 13.0, 12.21, and 8.84~fm,
for $a_s$ = 0, 0.1, and 0.3, respectively; $R_y/R_x=13/11$ in all cases.  This scaling keeps
$\langle |r| \rangle$, as well as the flow gradient, the same in each case.

Figure~\ref{fig:xyplots_as0.3}
shows homogeneity regions projected onto the $x-y$ plane, for a blast-wave source with $a_s=0.3$, corresponding
to a pseudo-Gaussian geometrical distribution in $\tilde{r}$ (see Figure~\ref{fig:surface}).
Comparing Figures~\ref{fig:xyplots} and~\ref{fig:xyplots_as0.3}, we observe that the lack of a ``hard cut-off''
in coordinate space in the latter case leads to less reduction in the ``out'' direction (i.e. along the
direction of motion).  Thus, the important ratio $\Ro{0}/\Rs{0}$ will be larger at finite $p_T$, when $a_s=0.3$.

We also note that the shape and size of the homogeneity region itself depends much less on $\phi_p$, when $a_s=0.3$.
The homogeneity region for pions emitted to $\phi_p=135^\circ$ is essentially a spatially-translated (and unrotated)
replica of that for pions emitted to $\phi_p=0^\circ$.  Since HBT correlations are insensitive to a spatial
translation of the source, the situation for $a_s=0.3$ is rather similar to the situation in which no flow
is present.  In the no-flow case, the same homogeneity region is measured at all angles, and
all oscillations of the HBT radii arise from the explicit $\phi_p$ dependence in 
Equations~\ref{eq:radii-vs-xmunu}~\cite{W98,LHW00}; for the source in Figure~\ref{fig:xyplots_as0.3},
not the same region, but a (virtually) identical one, is being measured at all angles $\phi_p$.

This has implications for the oscillations of the transverse radii.  Focusing on the homogeneity regions
for $\phi_p=135^\circ$, we observe that while $\langle \tilde{x} \tilde{y} \rangle$ (which quantifies
the ``tilt'' of the homogeneity region relative to the $x$ and $y$ axes) is greater when $a_s=0$,
$R_{os}^2$ (which quantifies the tilt of the homogeneity region relative to the ``out'' and ``side'' directions~\cite{HK02})
is larger for $a_s=0.3$  Therefore, we expect larger values of $\Ros{2}$ for larger $a_s$.

Figure~\ref{fig:HBT_vary_as} shows quantitatively the effects on the HBT radii, when $a_s$ is varied.
As intended by the scaling of flow and size parameters with $F(a_s)$, $\Rs{0}$ remains approximately invariant
when the parameters are covaried.  As discussed above, $\Ro{0}/\Rs{0}$ grows with $a_s$ at finite $p_T$, as
do the magnitudes of the oscillation strengths $|\Ro{2}/\Rs{0}|$, $|\Ros{2}/\Rs{0}|$ and $|\Rs{2}/\Rs{0}|$.
The average value of $R^2_l$ remains roughly constant, while its oscillation amplitude increases slightly
with $a_s$, due to $x-z$ and $y-z$ correlations.

\subsection{Non-identical particle correlations}
\label{sec:non-id}
Final-state interactions between pairs of non-identical particles (e.g. $K-\pi$)
are sensitive to the space-time structure (size, shape, emission duration) of the 
emitting source, analogously to HBT correlations between identical 
particles. Most importantly, the authors of Ref.~\cite{PLBNonId, LedNonId, PRLNonId} show that 
studying correlations between non-identical particles reveals
new information about the average relative space-time 
separation between the emission of the two particles, in the
rest frame of the pair. This unique information may be extremely valuable to
determine the interplay between partonic and hadronic effects. For example, 
the lower hadronic cross-section of kaons compared to pions may cause them to be emitted
earlier and closer to the center of the source than pions. However, if most
of the system evolution takes place at the parton level, the space-time
emission pattern of pions and kaons would be similar. This example is far
from unique as the same kind of argument can be made for protons recalling
that pion-proton hadronic cross section are very large, or conversely to
$\Xi$ and $\Omega$ whose hadronic cross section is expected to be small.
Furthermore, in terms of temporal effects, strangeness 
distillation~\cite{StrangeDistil,earlyStrangeDistil} or other
unique physics may cause some particles to be emitted later than others. 
Preliminary analyses of the  $\pi^{\pm}-K^{\pm}$, 
$\pi^{\pm}-p^{\pm}$, $K^{\pm}-p^{\pm}$ correlation functions have been reported by the STAR 
collaboration in Au-Au collisions at $\sqrt{s_{NN}} = $130 GeV and 200 GeV~\cite{QM03NonId, PiKPRL}. 
These analyses show that
pions, kaons and protons are not emitted at the same average space-time point.

The blast-wave parameterization implicitly assumes that the particle freeze-out conditions (temperature, 
flow profile,  freeze-out time and position)  are the same for all particle species. However, as 
shown earlier, the transverse momentum spectra and elliptic flow or different particle species do 
not look alike. Indeed, the relative contribution of the random emission (quantified by temperature) and
collective expansion depends directly on particle masses. The same phenomena is likely to affect
the particle freeze-out space-time emission distribution. In this section, we will show
that collective flow effects implicit in the blast-wave parameterization induce
a shift between the average freeze-out space-time 
point of different particle species. We will then study how changing the blast-wave parameters affects
these average freeze-out separations.

\subsubsection{Non-identical particle and blast-wave formalism}

Two particles interact when they are close to each other in phase space
in the local pair rest frame. Thus, particle pairs may be correlated
when their relative momentum in the pair rest frame is small. 
For particles with different masses, a small relative momentum in the pair
rest frame means that both particles have similar velocities 
in the laboratory frame and not similar momentum. This point is 
particularly important to realize when studying correlation involving pions. 
Due to the low pion mass, the pion velocity is the same as heavier particles (e.g. kaons or
protons)  when its momentum is much lower than the particle momentum it is associated with.
For example, a proton with a momentum of 1 GeV/c has the same velocity as a 
pion with a momentum of $\approx$0.15 GeV/c.
 
As described in Ref.~\cite{NonIdOutSideLong}, the spatial separation between
particles in the pair rest frame can be projected along three axes,
$\langle r^*_{out} \rangle$,  along the pair transverse momentum,
$\langle r^*_{side} \rangle$, perpendicular to the pair transverse momentum
and $\langle r^*_{long} \rangle$, along the beam axis. 
To study the 
blast-wave prediction, we focus on the limiting case where the relative
momentum between both particles in the pair rest frame is zero, which
means that both particles have the same velocity. Thus
we calculate the separation between particle 1 and 2 in the pair rest
frame $\langle \Delta r^*_{out} \rangle$,   $\langle \Delta r^*_{side} \rangle$ 
and $\langle \Delta r^*_{long} \rangle$, 
at a given pair transverse velocity $\beta_T$, azimuthal angle, $\phi_p$, 
and longitudinal velocity, $\beta_L$:

\begin{eqnarray}
\langle \Delta r^*_{out} \rangle &=& \langle r^*_{out}(1) \rangle -  \langle r^*_{out}(2) \rangle  \\
\langle \Delta r^*_{side} \rangle &=& \langle r^*_{side}(1) \rangle - \langle  r^*_{side}(2) \rangle  \\
\langle \Delta r^*_{long} \rangle &=& \langle r^*_{long}(1) \rangle -  \langle r^*_{long}(2) \rangle  
\end{eqnarray}

With the particle emission points defined as ($x, y, z, t$):

\begin{eqnarray}
\langle r^*_{out} \rangle  &=&  \gamma_T (\langle x \rangle \cos(\phi_p) + \langle y \rangle \sin(\phi_p) - \beta_T \langle t \rangle)  \\        
\langle r^*_{side} \rangle  &=& \langle y \rangle \cos(\phi_p)  - \langle x \rangle \sin(\phi_p)  \\
\langle r^*_{long} \rangle  &=& \gamma_L \langle (z) - \beta_L (t) \rangle  
\end{eqnarray}

Recalling that $\langle \beta_L \rangle = \langle  z/t \rangle $, $\langle r^*_{long}(i) \rangle = 0$. 
Following the notation of Equation~\ref{eq:BPrime}, the 
variables $\langle x \rangle$, $\langle y \rangle$ and $\langle t \rangle$
depend on $\beta_T$ and on the particle mass (m) as:  

\begin{eqnarray}
\langle x \rangle &=& \overline{\left\{x\right\}_{0,0}}(m,\beta_T)  \\
\langle y \rangle &=& \overline{\left\{y\right\}_{0,0}}(m,\beta_T)  \\
\langle t \rangle &=& \frac{\Delta\tau^2+\tau_0^2}{\tau_0} \overline{\left\{1\right\}_{0,1}}(m,\beta_T)  
\end{eqnarray}

\begin{figure}[t]
\epsfig{file=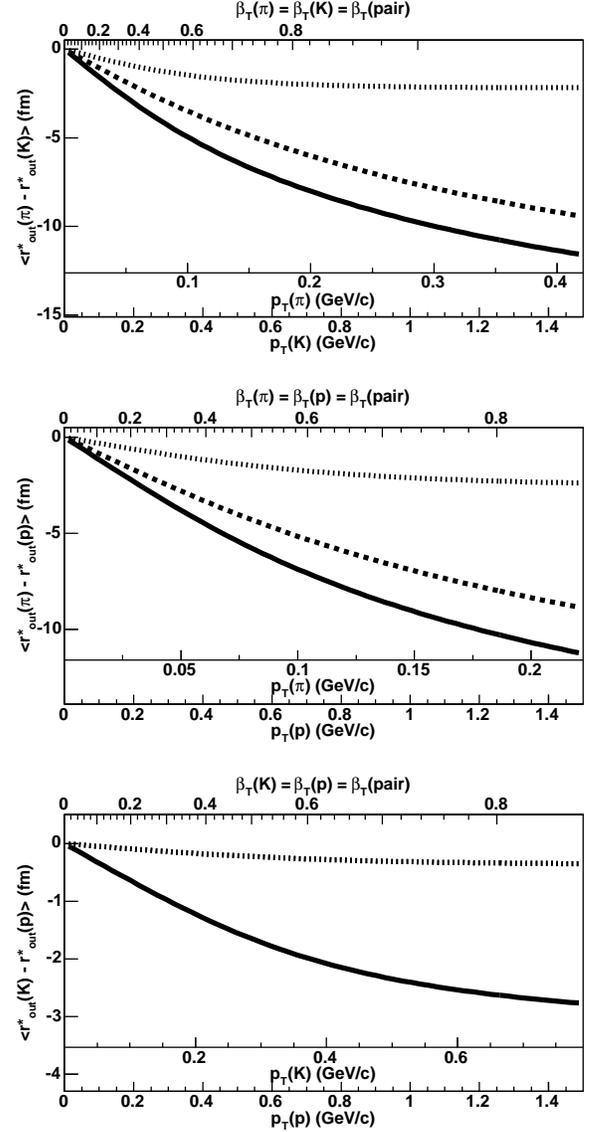,width=8cm}
\caption{
Separation between the average emission point of pions, kaons and protons in the pair
rest frame along the pair transverse velocity. 
The values of the blast-wave parameters are for a round source, as listed in Table~\ref{tab:defaultParams}. Dot line: time shift in the laboratory boosted
to the pair rest frame, dash line: spatial shift in the laboratory boosted to the pair rest frame, plain line: total 
spatial shift  in the pair rest frame.
\label{fig:NonIdPiKP}}
\end{figure}

The average values of $x$ and $y$ will vary with particle mass and particle transverse velocity, which yields
separations between pions and kaons, pions and protons, and protons and kaons in the 
pair rest frame shown as in Figure~\ref{fig:NonIdPiKP}.   
The dashed and dotted lines show the contribution of the spatial ($\Delta r_{out}$) and time 
($\Delta t$) separation boosted to the pair rest frame  along the pair transverse momentum, 
respectively. i.e. the dash line shows
$\langle \gamma_T \Delta r_{out} \rangle$ and the dot line represents 
$ - \langle \gamma_T \beta_T \Delta t \rangle$. The plain line shows
$\langle \Delta r^*_{out} \rangle = \langle \gamma_T (\Delta r_{out} - \beta_T \Delta t) \rangle$, the spatial 
separation 
in the pair rest frame.
When boosting to the pair rest frame, time and spatial shifts in the laboratory frame add up due to
their opposite signs.
The largest (smallest) shift is obtained when 
the mass ratio between both species is the largest (smallest), i.e. for pion-proton (kaon-proton) pairs. 

\begin{figure}[t]
\epsfig{file=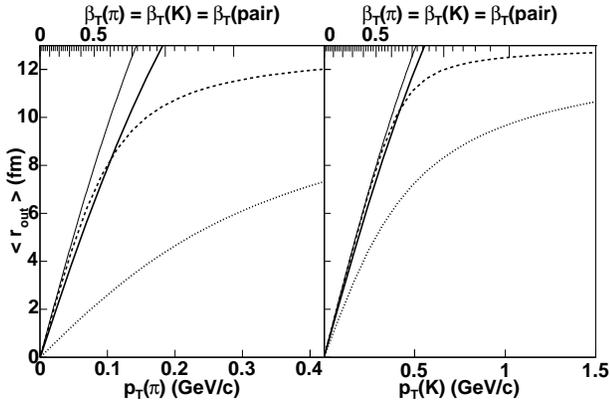,width=8cm}
\caption{
Average emission points of pions (left) and kaons (right) as a function of the pion and kaon velocity 
and momentum. All curve are calculated with $\rho_0$ = 0.9 and $\rho_a$ =0. The parameters $\tau$ and $\Delta t$ are 
irrelevant. Thin plain line: $R_x = R_y$ = 13 fm, T=0 GeV. Thick plain line: infinite system (i.e. $\Omega(r, \phi_s)$ = 1 ), 
T = 0.1 GeV. Dash line: $R_x = R_y$ =13 fm,  T=0.01 GeV. Dot line: $R_x = R_y$ =13 fm, T = 0.1 GeV.
\label{fig:PiKEmPoint}}
\end{figure}

In order to understand the behavior of $\Delta r^*_{out}$, it is instructive to investigate how the average emission points in the 
laboratory frame ($\langle r_{out} \rangle$)  of  pions and kaons behave in various conditions as shown in Figure~\ref{fig:PiKEmPoint}. 
This figure shows the average emission points of pions and kaons in four different configurations: (1) The thin plain line shows the 
flow profile, (Equation~\ref{eq:rho}), which sets the emission point of particles when T = 0 GeV. (2) the dot line is calculated 
assuming an infinite system, i.e. $\Omega(r,\phi_s) = $ 1 and with $T=0.1$~GeV.  (3) the dash line corresponds to a finite system 
as in Equation~\ref{eq:Omega}, with $a_s = $ 0, and (for illustration) an extremely low temperature, $T=0.01$~GeV. (4) the thick plain line is calculated using 
the standard parameters used in Figure~\ref{fig:PiKEmPoint}. Only in case (1) are pions and kaons emitted exactly at the same point.
Since $T=0$~GeV, all particles are always emitted at the same point as set by the flow profile. In case (2), particle emission points
spread around the average emission defined by the flow profile. At small $\eta$ there are as many particles emitted at large $r$,
i.e. large $\rho$ as at small $r$. But when $|\eta|$ is large, the term $e^{- m_T/T cosh \rho(r,\phi_s) cosh(\eta)}$ in
Equation~\ref{eq:secondS} favors small $\rho$, i.e. small $r$. Thus, the average emission point 
$\langle r_{out} \rangle$ is smaller than the one defined by the flow profile. In case (3), the average emission points of pions and 
kaons follow closely the flow profile when the particle rapidity is significantly smaller than $\rho_0$.  Close or beyond 
$\rho_0$, a certain fraction of the particle emission function is truncated due to the system finite size. The particle emission points 
are not allowed to spread beyond the system boundary, hence breaking the balance between particle emitted at small radii and particle 
emitted at large radii. Thus, the particle average emission radius is smaller than the emission radius given by the flow profile.  
Because the temperature is rather small, the average emission points of both species converge rapidly toward the radius of the system.  
The larger temperature in case (4) makes the average emission radii converge slowly toward the system limit. This phenomenon, with 
the addition of the phenomenon described in case (2) reduces very significantly the average particle emission radius compared to the flow 
profile limit.

\begin{figure}[t]
\epsfig{file=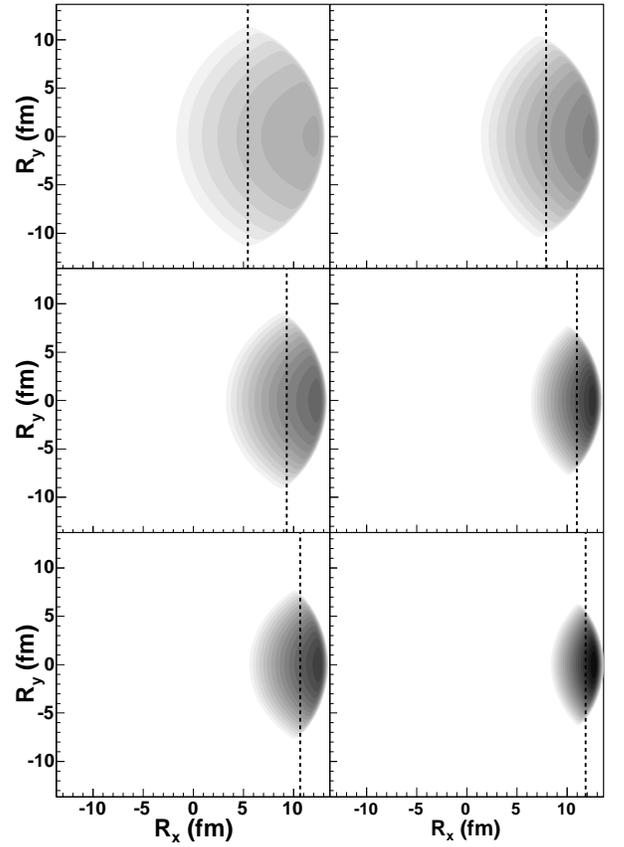,width=8cm}
\caption{
Distribution of the  emission points in the transverse plane, for different particle species emitted at
the same velocity, $\beta_y = 0 $ and $\beta_x = $ 0.907 on the left-hand side and $\beta_x = $ 0.974 on the right
hand side. Top panels are for pions, middle for kaons and bottom for protons.  
The same logarithmic color scale is used for all six panels. The blast-wave parameters are the same as
in Figure~\ref{fig:PiKEmPoint} except that $a_s = 0.01$   The dashed lines show $\langle x \rangle$.
\label{fig:SpatialDensity}}
\end{figure}

The effect of temperature depends on particle mass and momentum. Random smearing is maximal for particles with low mass and momentum 
such as the low $p_T$ pions that are usually associated with kaons or protons in non-identical particle correlation analyses. This 
effect is illustrated in Figure~\ref{fig:SpatialDensity}. It shows the probability of emitting pions at a given point in the transverse 
plane for two different pion momenta $\overrightarrow{p}=(0.15, 0, 0)$ and $\overrightarrow{p} = (0.3, 0, 0)$. Kaon and proton momenta
are calculated so that they have the same velocities as pions.
The region of the system that emits particles of a given momentum shrinks and moves towards the edge of the system as the particle mass, 
and/or momentum increases. The magnitude of the inward radius shift depends on the fraction of the source distribution that is truncated 
due to the system finite size. Thus, the  inward  shift of the average emission radius scales with the source size. This effect yields the
systematic shift between the average emission points of pions, kaons and protons as shown in Figure~\ref{fig:NonIdPiKP} since pion source 
size is the largest and proton the smallest. Light particles are emitted the closer to the center of the source than heavier ones.

In addition to a spatial separation, the blast-wave parameterization induces a time shift between different particle species emitted at the same velocity as shown by the dot line in Figure~\ref{fig:NonIdPiKP}.  Due to random motion, the space-time rapidity ($\eta$) spreads around the momentum rapidity (Y). Because the relationship $t = \tau \cosh(\eta)$ is positive definite, the larger dispersion of $\eta$ for pions  than for kaons or protons leads to a delay of the emission of pions with respect to kaons or protons. Protons are emitted first, then kaons and then pions. The spatial and time shifts have opposite signs in the laboratory frame but they sum up when boosting to the pair rest frame. The plain line in Figure~\ref{fig:NonIdPiKP} shows the added contributions of both shifts. The pion-kaon and pion-proton separation in the pair rest frame ranges from 5 to 15 fm, while the separation between kaons and protons is on the order of 2-4 fm. These shifts are large enough to be measured.

The curves on Figure~\ref{fig:NonIdPiKP} have been obtained by setting the blast-wave parameters to arbitrary values. We will now 
investigate how changing these parameters affects the shift between pions and kaons. The results obtained studying pion-kaon separation 
can be easily extrapolated to pion-proton and kaon-proton separations. Since experimental analyses of non-identical two-particle correlations 
performed to date do not investigate the azimuthal dependences with respect to the reaction plane, we will focus our study on central 
collisions. We will then show that the shift between the average emission points of different particle species 
oscillates with respect to the reaction plane without investigating the effect of varying the parameters in detail.

\subsubsection{Non-identical particle correlations in central collisions}

In central collisions, azimuthal symmetry implies that the particle emission pattern depends only on the relative angle between the position and momentum
$\phi_p - \phi_s$. Thus, setting for convenience $\phi_p=0$, yields:

\begin{eqnarray}
\langle r^*_{out}(i) \rangle  &=& \langle \gamma_T (x_i - \beta_T \Delta t_i) \rangle  \\
\langle r^*_{side}(i) \rangle  &=&  0
\end{eqnarray}

Furthermore, we consider emission from an azimuthally-isotropic
source ($R_x=R_y$ and $\rho_2=0$) only. Hence, the only quantity of interest is $\langle \Delta r^*_{out} \rangle$. 


\begin{figure}[t]
\epsfig{file=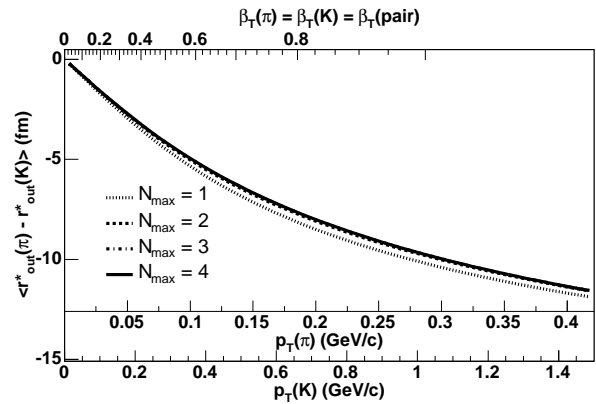,width=8cm}
\caption{
Average separation between $\pi$ and K in the pair rest frame  as a function of the pion and kaon momentum using several values of $N$, the maximum value of $n$
taken in the summation of Equation~\ref{eq:firstS}.
Values of the blast-wave parameters are for a round source, as listed in Table~\ref{tab:defaultParams}.
\label{fig:PiKOutStarVarN}}
\end{figure}

Figure~\ref{fig:PiKOutStarVarN} shows the dependence of the average separation between $\pi$ and K in the pair rest frame  as a function of the pion and kaon momentum using several values of $N$, the maximum value of $n$ taken in the summation of Equation~\ref{eq:firstS}. Pions and kaons are of course treated as bosons. The difference between the average separation  calculated  either by using a Boltzman function (N=1) or by approximating the Bose-Einstein distribution at the $4^{th}$ order
is smaller than 0.5 fm. The maximum relative difference is on the order of 8\% at small transverse momentum. The Bose-Einstein distribution already converges when N=2.


\begin{figure}[t]
\epsfig{file=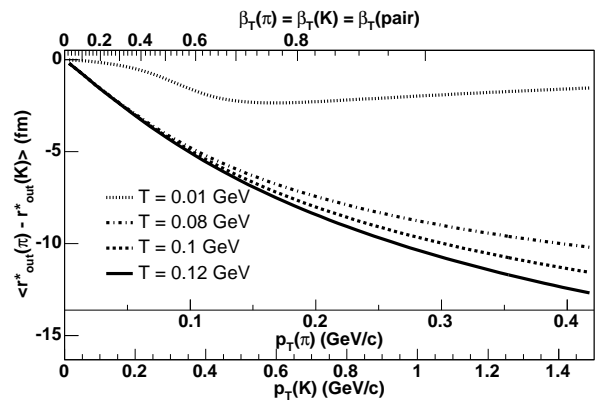,width=8cm}
\caption{
Average separation between $\pi$ and K in the pair rest frame  as a function of the pion and kaon momentum for 
different temperature. 
The values of the other blast-wave parameters are for a round source, as listed in Table~\ref{tab:defaultParams}.
\label{fig:PiKOutStarVarT}}
\end{figure}

\begin{figure}[t]
\epsfig{file=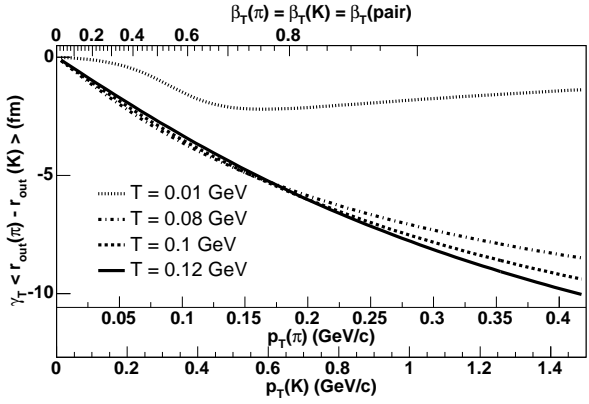,width=8cm}
\caption{
Contribution of the spatial shift to the average
separaration between pions and kaons in the pair rest frame as a function of the 
pion and kaon momentum for 
different temperature. The values of the other blast-wave parameters are for a round source, as listed in Table~\ref{tab:defaultParams}.
\label{fig:PiKOutVarT}}
\end{figure}

\begin{figure}[t]
\epsfig{file=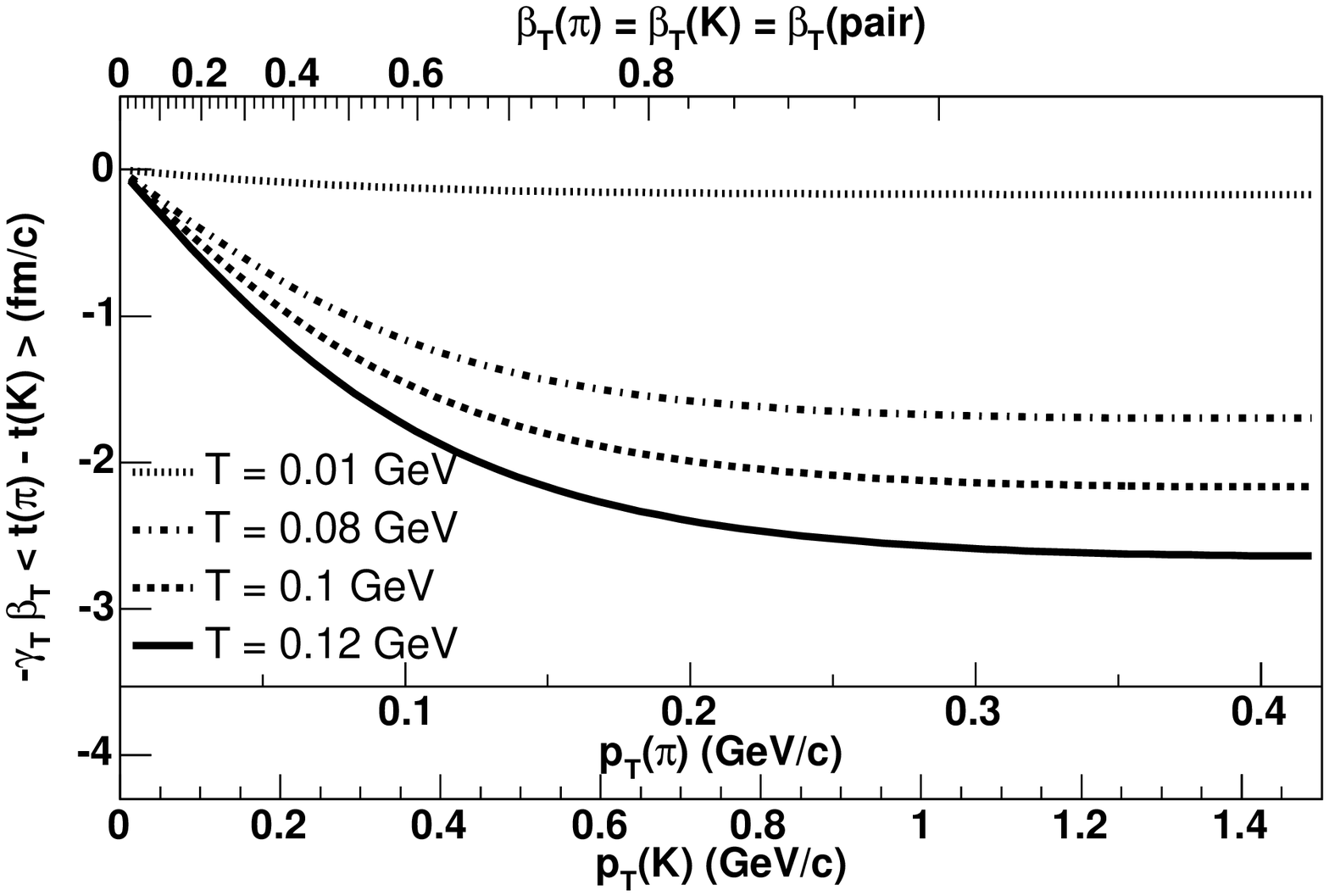,width=8cm}
\caption{
Contribution of the time shift to the average
separaration between pions and kaons in the pair rest frame  as a function of the 
pion and kaon momentum for 
different temperature. The values of the other blast-wave parameters are for a round source, as listed in Table~\ref{tab:defaultParams}.
\label{fig:PiKTimeVarT}}
\end{figure}

Figure~\ref{fig:PiKOutStarVarT} shows the dependence of the spatial shift between pions and kaons in the pair rest frame as a function of 
the pair velocity for different temperature. The shift increases as the temperature increases all the way from 0.01 GeV to 0.12 GeV. When 
the temperature is very low (e.g. 0.01 GeV), pion and kaon emission patterns are dominated by space-momentum correlation independent of 
particle masses. In the limit of zero temperature, there is a unique correspondence between particle velocity and emission point. In that 
case since we consider particles with the same velocity, pions and kaons are emitted from the same point.  On the other hand when the 
temperature is non zero, a fraction of the pions and kaons that would be emitted in a infinite system are truncated, which shifts their 
average emission points inward.  Because the pion source size is significantly larger than the kaon source size due to the pion lower 
mass and momentum, the pion average emission point is more shifted inward than kaon's, which is illustrated in 
Figure~\ref{fig:PiKOutVarT}. This figure shows the contribution of the spatial shift in the 
average separation between the pion and kaon average emission points in the pair rest frame. When the temperature is low (0.01 GeV), both spatial and time separation are small as shown by the dash
lines in figures~\ref{fig:PiKOutVarT} and ~\ref{fig:PiKTimeVarT}. As the temperature increases, the pion emission time increases faster than the kaon emission time; the higher the temperature the larger
the shift between pion and kaon average emission time (after boosting into the pair rest frame). On the other hand the spatial shift varies little  within the temperature range expected
in relativistics heavy-ion collisions (0.08-0.12 GeV). 
At temperature above 10 MeV/c, a fraction of the pion and kaon sources is 
truncated even at low particle velocity. The fraction of the source that is truncated, which leads to 
an inward shift of the average emission radius, varies with transverse momentum and particle mass
but it is relatively insensitive to temperature variation between 0.08 and 0.12 GeV. 
It is interesting to note that the average spatial separation between pion and kaon in the laboratory frame 
actually decreases as the particle velocity rises above 0.8c. However, this decrease is not visible in Figure~\ref{fig:PiKOutStarVarT} and ~\ref{fig:PiKOutVarT} becayse the $\gamma_T$ factor applied when boosting to the pair rest frame rises faster with velocity than the separation between pions and kaons in the laboratory frame decreases.

\begin{figure}[t]
\epsfig{file=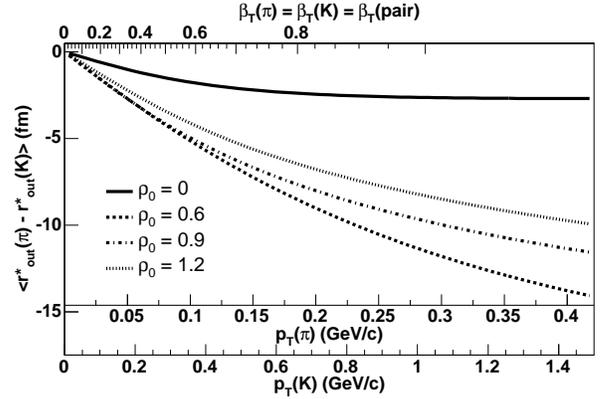,width=8cm}
\caption{
Average shift between the pion and kaon emission point in the pair rest frame as a function of the 
pion and kaon momentum for 
different flow rapidity ($\rho_0$) The values of the other blast-wave parameters are for a round source, as listed in Table~\ref{tab:defaultParams}.
\label{fig:PiKROutStarVarRho}}
\end{figure}

\begin{figure}[t]
\epsfig{file=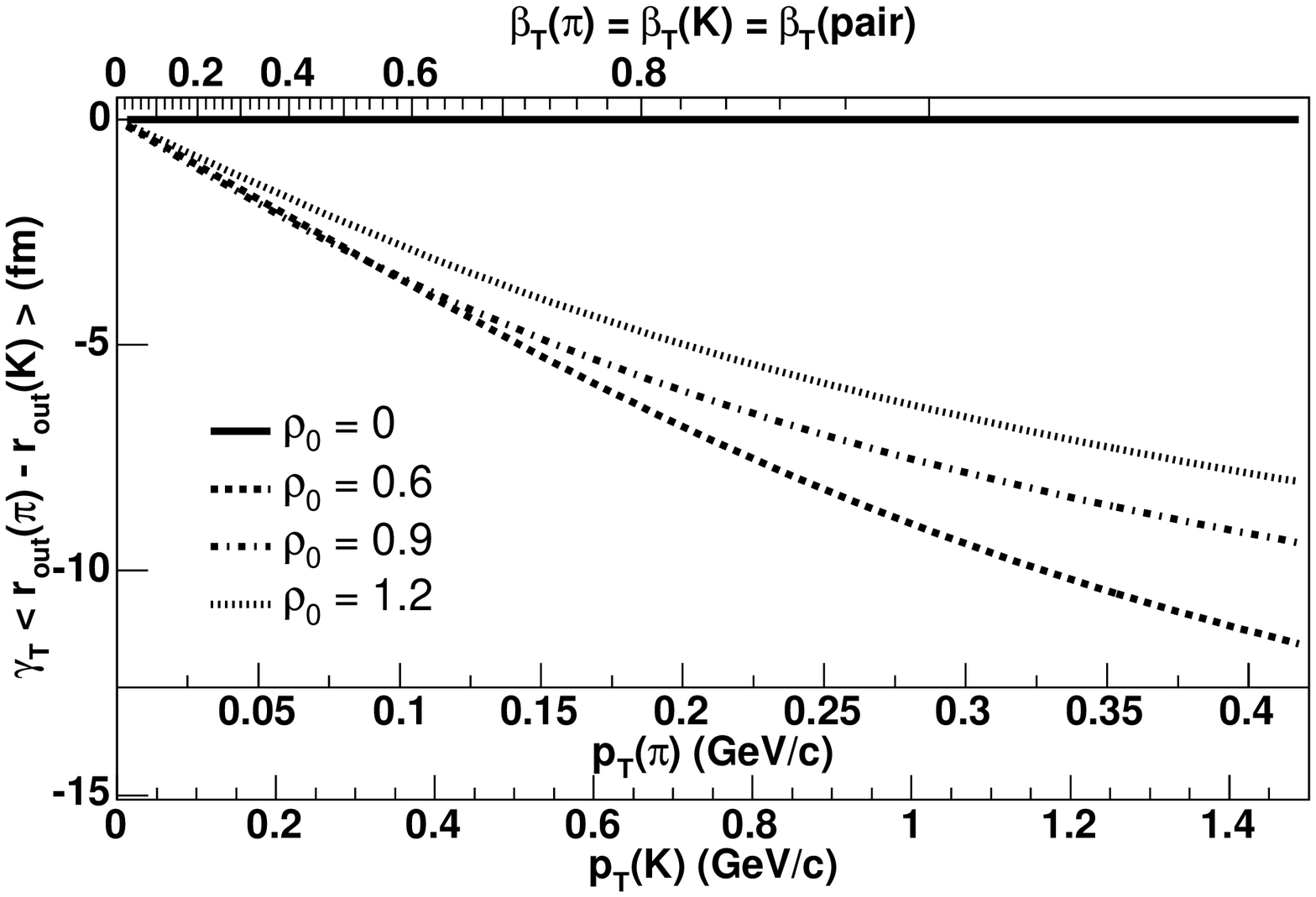,width=8cm}
\caption{
Contribution of the spatial shift to the average
separaration between pions and kaons in the pair rest frame as a function of the 
pion and kaon momentum for 
different flow rapidity ($\rho_0$) The values of the other blast-wave parameters are for a round source, as listed in Table~\ref{tab:defaultParams}.
\label{fig:PiKROutVarRho}}
\end{figure}

\begin{figure}[t]
\epsfig{file=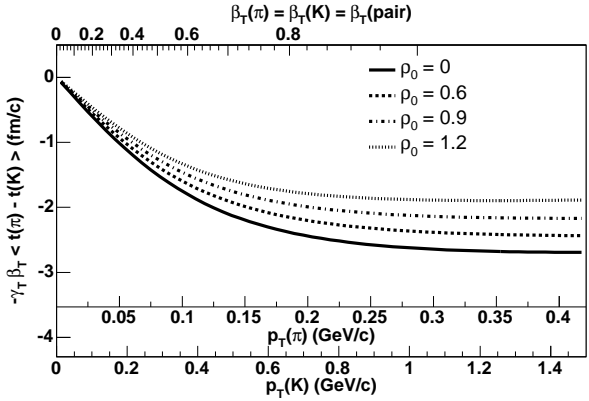,width=8cm}
\caption{
Contribution of the time shift to the average
separaration between pions and kaons in the pair rest frame as a function of the 
pion and kaon momentum for 
different flow rapidity ($\rho_0$) The values of the other blast-wave parameters are for a round source, as listed in Table~\ref{tab:defaultParams}.
\label{fig:PiKTimeVarRho}}
\end{figure}

Changing the maximum flow rapidity ($\rho_0$) also affects the separation between pions and kaons in the pair rest frame as shown in 
Figure~\ref{fig:PiKROutStarVarRho}. When $\rho_0 = 0$, pions and kaons are emitted from the same space point as shown on 
Figure~\ref{fig:PiKROutVarRho} and only the time shift remains (Figure~\ref{fig:PiKTimeVarRho}). Indeed, the time shift depends 
weakly on $\rho_0$. Figure~\ref{fig:PiKTimeVarRho} shows that the contribution of the time shift to
the separaration between pions and kaons in the pair rest frame reaches a plateau when the pion 
momentum reaches 0.25 GeV/c. The magnitude of the time shift starts decreasing in the laboratory 
frame upon reaching this pion momentum but it is compensated by an increase of the boost factor $\gamma_T \beta_T$.
On the other hand, when $\rho_0$ is large enough, a significant spatial separation appears in the pair rest frame, 
which is sensitive to the value of $\rho_0$. The main effect of increasing the flow strength is 
to decrease the pion (and kaon) transverse source size, as shown in Figure~\ref{fig:HBT_vary_rho0}.  In the laboratory frame, the spatial shift between 
pion and kaon average emission point switches from decreasing to increasing as the particle 
velocity increase. The value of the velocity where this switch occurs, depends on the value of $\rho_0$.
The increase of the spatial shift between pions and kaons arises from two effects: the first is the expected
shift of the average emission point of both particles due to the flow profile; the second is that the kaon source
size drops more rapidly than the pion source size.
Then, above a velocity that 
depends on $\rho_0$, the fraction of pion that would be emitted beyond the system limit starts dropping faster than the corresponding 
fraction of kaons, thus the separation between the average emission points of pions and kaons decreases. However, this turn over, which takes place in the laboratory frame is not 
directly visible in Figure~\ref{fig:PiKROutVarRho} because the boost factor $\gamma_T$ compensates it.


\begin{figure}[t]
\epsfig{file=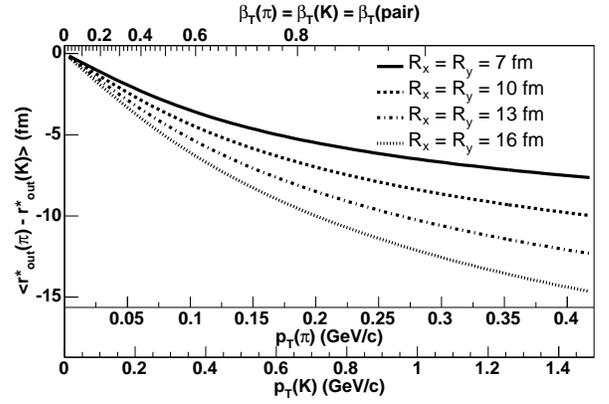,width=8cm}
\caption{
Average shift between the pion and kaon emission point in the pair rest frame as a function of the 
pion and kaon momentum for 
different system radius ($R_x = R_y$). The values of the other blast-wave parameters are for a round source, as listed in Table~\ref{tab:defaultParams}.
\label{fig:PiKROutStarVarR}}
\end{figure}

Figure~\ref{fig:PiKROutStarVarR} shows the sensitivity of the average separation between pion and kaon 
emission  point in the pair rest frame to varying the system radius. This spatial separation scales directly 
with the system radius because it does not modify the fraction of pions or kaons that are truncated
due to the system finite size.   

\begin{figure}[t]
\epsfig{file=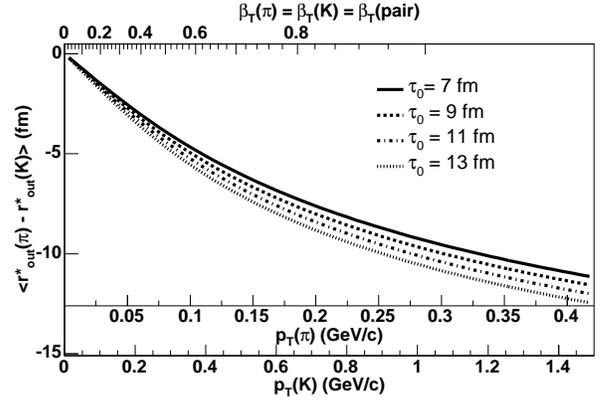,width=8cm}
\caption{
Average shift between the pion and kaon emission point in the pair rest frame as a function of the 
pion and kaon momentum for 
different system proper life time ($\tau_0$). The values of the other blast-wave parameters are for a round source, as listed in Table~\ref{tab:defaultParams}.
\label{fig:PiKROutStarVarTau}}
\end{figure}

Like the system radius, the proper life time $\tau_0$ acts as a scale driving the shift between pion and kaon emission time. 
Figure~\ref{fig:PiKROutStarVarTau} shows the effect of varying  $\tau_0$ on the separation between pion and kaon in the pair 
rest frame. The effect of varying $\tau_0$ is small because the contribution of the time shift to the 
separation in the pair rest frame is significantly smaller than the contribution of the spatial separation.

\begin{figure}[t]
\epsfig{file=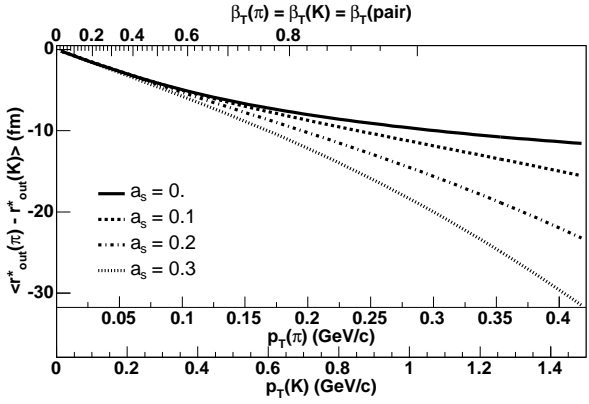,width=8cm}
\caption{
Average shift between the pion and kaon emission point in the pair rest frame as a function of the 
pion and kaon momentum for surface smoothness ($a_s$). 
The values of the other blast-wave parameters are for a round source, as listed in Table~\ref{tab:defaultParams}.
\label{fig:PiKROutStarVarAs}}
\end{figure}

Figure~\ref{fig:PiKROutStarVarAs} shows the effect of varying $a_s$ on the separation between pion and kaon in the pair rest frame. 
Unlike in Figure~\ref{fig:HBT_vary_as}, $\rho_0$ was kept constant. Indeed varying $\rho_0$ 
significantly affects the average emission radii, which hides the effect of changing $a_s$ at low velocity. However, when the velocity 
is larger than $\tanh(\rho_0) = 0.716 c$, the amount of boost and space-momentum correlation than particle acquire depends on $a_s$; 
the larger $a_s$ the larger the boost. Thus, when the pair velocity is lower than $0.716c$, the separation decreases with increasing 
$a_s$ because the fraction of truncated particles decrease. When the pair velocity is larger than $0.716 c$, increasing $a_s$ has the 
same consequence as increasing $\rho_0$.

\subsubsection{Non-identical particle correlations in non-central collisions}

\begin{figure}[t]
\epsfig{file=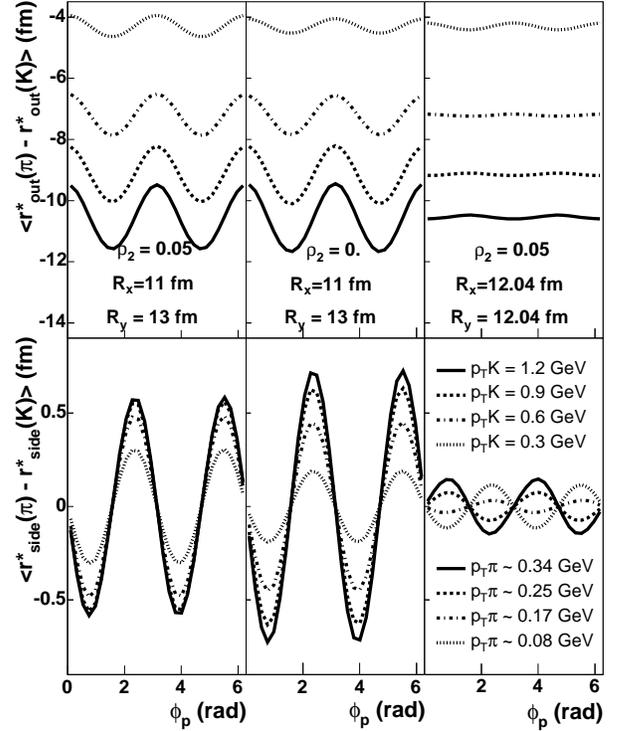,width=8cm}
\caption{
Average separation between pions and kaons in the pair rest frame as a 
function
of their momentum azimuthal angle with respect to the reaction plane. 
The four
different lines are calculated for 4 different kaon momenta and the 
corresponding
4 different pion momenta required so that pions and kaons have the same 
velocity.
The values of the blast-wave parameters used in the left panel are for a 
"non-round" source, as
listed in Table~\ref{tab:defaultParams} .  The same parameter were used 
in the middle
panel except that $\rho_2 = 0$.  The round source parameters are used in 
the right panel
except that $\rho_2 = 0.05$. The separation
along the $out$ and $side$ axes are shown on the upper and lower panel 
respectively.
\label{fig:AsNonIdPiK}}
\end{figure}

As reported in ref~\cite{PRLNonId}, the average space-time separation between different particle species may depend on the particle 
emission angle with respect to the reaction plane. The effect in the blast-wave parameterization is shown in Figure~\ref{fig:AsNonIdPiK}. 
The blast-wave parameters are the same as in Figure~\ref{fig:NonIdPiKP} with the exception of $R_x$ and $\rho_2$ which are varied. Clear 
oscillations of $\langle r^*_{out}(\pi) - r^*_{out}(K) \rangle$ are found when $R_x$ is set to 11 fm, i.e. when the source is out-of-plane 
extended ($R_y = $ 13 fm). Small oscillations of $\langle r^*_{side}(\pi) - r^*_{side}(K) \rangle$ appear as well. The oscillations in both 
directions are on the order of 1-2 fm, which may be measurable. On the other hand, keeping the
source cylindrical but setting the flow modulation parameter $\rho_2$ to 0.05 yields very small oscillations, which will be very 
challenging to probe experimentally. Thus, non-identical two particle correlation analyses with respect to the reaction plane, as 
pion HBT may provide a good handle on the source shape but not on the flow modulation.

\subsection{Summary of effects of parameters}

Above, we have explored the sensitivity of several experimental observables on various
freezeout parameters.
Here, we summarize in an orthogonal manner-- describing briefly the main observable
effects due to an increase in each parameter, if the other parameters are left fixed.

The temperature parameter $T$ quantifies the randomly-oriented kinetic energy component
of the freezeout scenario.  Increasing this energy component leads to decreased slopes of $p_T$
spectra, especially for light-mass particles.  Since random motion destroys space-momentum
correlations, increasing $T$ reduces measured elliptic flow ($v_2$) and increases homogeneity
scales, i.e. $\Rmu{0}$.  On the other hand, for the range of values considered, increasing
temperature increases the average separation in the pair rest frame between 
pions and heavier particles, e.g. $\langle r^*_{out}(K) - r^*_{out}(\pi) \rangle$.

The $\phi$-averaged transverse flow strength is quantified in this model by $\rho_0$.
Increasing this directed energy component decreases slopes of $p_T$ spectra, especially
for heavier particles.  Increasing $\rho_0$ leads to increasing space-momentum correlations,
which reduce $\Rmu{0}$ and $\langle r^*_{out}(K) - r^*_{out}(\pi) \rangle$, 
and at constant $\rho_2$ reduces $v_2$ at high $m_T$.

The parameter $a_s$ quantifies the ``surface diffuseness'' of the spatial density profile.
Taking care to keep the average transverse flow fixed, variation in $a_s$ has little
effect on purely momentum-space observables: $p_T$ spectra and $v_2(m,p_T)$.  On the other
hand, going from a ``box profile'' ($a_s=0$) to a pseudo-Gaussian profile ($a_s=0.3$)
increases the ``out-to-side'' ratio $\Ro{0}/\Rs{0}$ at higher $p_T$, as the homogeneity
region is not constrained by hard geometric ``emission boundaries.''  Furthermore, for
a fixed average transverse flow (and flow gradient), increasing $a_s$ leads to stronger
oscillations in the HBT radius parameters. Increasing $a_s$ also increases the
$\langle r^*_{out}(K) - r^*_{out}(\pi) \rangle$ when the temperature and particle velocity
are such that the pion source size is significantly larger than the kaon's.

Considering variations in the geometric transverse scale of the source
$\sqrt{R_x^2+R_y^2}$, while keeping all else (including $R_y/R_x$) constant,
we conclude that the sensitivity is in the HBT radii and the average
separation between different particle species.  In HBT, the sensitivity to this parameter is 
through $\Ro{0}$ and $\Rs{0}$. The spatial shifts that contribute to the separation 
$\langle r^*_{out}(K) - r^*_{out}(\pi) \rangle$ in the pair rest frame scales
directly with this parameter.
Momentum-space observables such as spectral shapes or $v_2$ are unaffected.

We turn now to variations in the timescale parameters-- the evolution duration
($\tau_0$) and emission duration ($\Delta\tau$).  The only significant variations
are in $\Rl{0}$ and $\Ro{0}$, and these depend differently on $\tau_0$ and $\Delta\tau$,
allowing independent 
probes
of the two parameters. 
The time shifts that contribute to the separation 
$\langle r^*_{out}(K) - r^*_{out}(\pi) \rangle$ in the pair rest frame scales
directly with these parameters ($\tau_0$ and/or $\Delta\tau$).

Non-central collisions may exhibit azimuthally anisotropic geometry and/or dynamics in the
freeze-out configuration.  
These two effects were studied by varying $R_y/R_x$, and $\rho_2$, respectively.
Azimuthally-integrated $p_T$ spectra were negligibly affected, as were the $\phi_p$-averaged
values of HBT radii ($\Rmu{0}$).  Observables designed to probe anisotropy, $v_2$ and $\Rmu{2}$
of course showed strong sensitivity to geometric or dynamical anisotropy, increasing strongly
in magnitude as $R_y/R_x$ deviates from unity and/or $\rho_2$ from zero.

%% file: S4_Fits.tex
\section{Comparison with existing data}
\label{sec:fits}

In the previous section, we reviewed the various experimental
observables that the blast-wave parameterization may be able to
reproduce. The effect of changing blast-wave parameters has
been examined. In this section, we investigate
how well those parameters can be constrained by the experimental 
data. 


Before comparing the blast-wave parameterization with data, we highlight its limitations 
and possibly not valid assumptions:
\begin{itemize}
\item 
{\it Freeze-out from a thermalized system at a constant temperature}. In the blast wave framework, freeze-out is assumed to take place at the same
temperature for every particle. In principle, this condition may be relaxed by fitting separately 
the data for each type of particle. (We do not attempt this here.)
Perhaps a more important assumption, however, is made in discussing ``temperature'' at all; it requires
system thermalization.
If particles are emitted in a non-thermal way our framework may fail to describe the data.
\item
{\it Longitudinal boost invariance}. The blast wave parameterization relies on longitudinal boost 
invariance. Elliptic flow data ($v_2$) published by the PHOBOS collaboration clearly show that 
longitudinal boost invariance is broken for rapidity larger than 1~\cite{PhobosV2}. Thus the blast wave parameterization can only be applied within $|Y|<1$. This assumption will be validated or ruled out by high precision data.
\item
{\it No resonances}. The observables that constrain the blast wave are calculated for each 
particle species using their own masses, ignoring the fact that some particle may be decay products of other 
particles. In fact a large fraction of pions and protons produced at RHIC energy
originate from resonance decays. As mentioned in section 3, pions from $\omega$ meson decay 
distort the pion source space-time distribution but this effect is limited because the fraction
of pions coming from $\omega$ is on the order of 10\%. On the other hand, pion spectra
and to a lesser extent proton spectra may be significantly distorted by resonance 
feed-down. Within the blast wave framework however, it was found in~\cite{Peitzmann,STARSpectra200} that 
accounting for resonances affects the
Blast Wave parameters extracted fit to spectra by no more than 20\%. While 20\% is not negligeable, we feel that the 
gross features of the parameterization are preserved. On the other hand, an investigation of the
effect of resonance feed-down on $v_2$ remains to be performed.
\end{itemize}


We focus on data from Au-Au collisions at $\sqrt{s_{NN}} = $ 130 
GeV  published by the PHENIX and STAR collaborations at RHIC.
The measurements used are listed on top of Table~\ref{tab:FitResult}. 
To avoid uncertainties between different experiments
and to optimize the overlap between centrality bins, we
do not use the full set of data published at RHIC.
Three centrality bins are available. The approximate overlap between the centrality bins 
for the different measures may affect the quality of the fits.  

\begin{table}[t]
\begin{tabular}{cccc}
\hline
\hline
\multicolumn{1}{c}{} &
\multicolumn{1}{c}{Central}  &
\multicolumn{1}{c}{Mid-central} & 
\multicolumn{1}{c}{Peripheral} \\
\hline
data \\
\hline
$\pi$, K, p spectra~\cite{PhenixSpectra130} &
 0-5\%  &
15-30\% &
60-92\%  \\
$\Lambda$ spectra~\cite{StarLaSpectra130}  &
 0-5\% &
20-35\% &
35-75\% \\
pion radii ~\cite{STARHBT} &
0-12\%  &
12-32\% &
32-72\% \\
Elliptic flow ~\cite{STARv2ID} &
0-11\% &
11-45\% &
45-85\% \\
\hline
\hline
$\chi^{2}/$(\#~data~points) \\
\hline
$\pi^+$ \& $\pi^-$ spectra                                	&  7.2/10  		& 26.5/10 		& 13.0/9  \\
$K^+$ \& $K^-$ spectra                                            	&  24.2/22   		& 21.4/22  		& 10.1/10  \\
$p$ \& $\overline{p}$ spectra                               	&  10.6/18   		& 23.2/18  		& 28.0/12 \\
$\Lambda$ \& $\overline{\Lambda}$ spectra  	&  9.5/16    		& 12.8/16  		& 11.0/16 \\
$\pi$ $v_2$                                                                  	&  14.6/12  		& 29.8/12  		&  5.2/12  \\
$p$ $v_2$                                                                    	&  1.6/3    		& 9.2/6   		&  0.8/3  \\
$\pi$  $r_{out}$                                                          	&  1.9/6    		& 0.4/2    		&  0.4/2  \\
$\pi$  $r_{side}$     			            	&  2.7/6   		& 0.07/2    		&  0.06/2 \\
$\pi$  $r_{long}$    			            	&  5.3/6    		& 0.003/2  		& 0.1/2  \\
Total             				            	&  77.6/99 		& 107.7/90 		& 68.7/68 \\
\hline
\hline
parameters \\
\hline
T (MeV)                				&  106 $\pm$ 3           	& 107 $\pm$ 2          	& 100 $\pm$ 5         \\
$ \rho_{0} $        					&  0.89 $\pm$ 0.02     	& 0.85 $\pm$ 0.01   	& 0.79 $\pm$ 0.02 \\
$ \langle \beta_{T} \rangle $  			&  0.52 $\pm$ 0.01     	& 0.50 $\pm$ 0.01   	& 0.47 $\pm$ 0.01 \\
$ \rho_2 $         				               & 0.060 $\pm$ 0.008	& 0.058 $\pm$ 0.005 	& 0.05 $\pm$ 0.01 \\
$ R_{x} (fm) $      				& 13.2 $\pm$ 0.3        	& 10.4 $\pm$ 0.4     	& 8.00 $\pm$ 0.4 \\
$ R_{y} (fm) $      				& 13.0 $\pm$ 0.3        	& 11.8 $\pm$ 0.4     	& 10.1 $\pm$ 0.4   \\
$ \tau (fm/c) $      				& 9.2 $\pm$ 0.4          	&  7.7 $\pm$ 0.9      	& 6.5 $\pm$ 0.6     \\ 
$ \Delta t (fm/c) $ 				& 0.003 $\pm$ 1.3        	&  0.06 $\pm$ 1.3      	&  0.6 $\pm$ 1.8 \\
\hline\hline
\end{tabular}
\caption{
Upper section: data used in the fit.
Middle section: number of $\chi^2/$ data points for each measure. 
Lower section: best fit parameters. 
Note that $\langle \beta_T \rangle$ is not a fit parameter, but it is 
calculated from $\rho_0$. 
\label{tab:FitResult}}
\end{table}

The blast-wave parameters are constrained by simultaneously fitting
transverse momentum spectra, transverse momentum azimuthal
anisotropy ($v_{2}$), and
pion HBT radii. Bose-Einstein and Fermi-Dirac distribution are used when
appropriate using the approximation to the 4$^th$ order as
in Equation~\ref{eq:firstS}, i.e. the value of $n = 4$
in the summation of Equation~\ref{eq:firstS}.
The quality of the fit to each measure is shown by the $\chi^{2}$ per data points
listed in Table ~\ref{tab:FitResult}. 
The number of degrees of freedom is not shown because the number of 
parameters cannot be subtracted
from the number of data points independently for each measure. 
In addition to the 7 blast-wave parameters, there are 8 scaling parameters
 used to normalize $\pi^+$, $\pi^-$, $K^+$, $K^-$, $p$, 
$\overline{p}$, $\Lambda$, and $\overline{\Lambda}$ spectra. These scaling
parameters ($A$) are not minimized by the fitting function but calculated directly.
The $\chi^2$ for a given spectra is given by:
\begin{equation}
\chi^{2} = \sum_{i} (\frac{d_{i} - A c_{i}}{e_{i}})^{2}
\end{equation}
$d_i$ is the measured invariant yield in the bin $i$. $e_i$ is the 
experimental error on $d_i$, and $c_i$ is the calculated invariant yield. The
$\chi^2$ minimum occurs when $d\chi^2/dA$ = 0, hence:
\begin{equation}
A = \frac{\sum_{i} {d_{i}c_{i}/e_{i}^{2}}}{\sum_{i} {c_{i}/e_{i}^{2}}}
\end{equation}

To avoid the region where resonance feed-down may be large
and where the Boltzmann approximation may not be valid, 
fits to transverse momentum spectra are restricted to pions with $p_T > 0.4$ GeV/c.
We further restrict the fit range to $p_T < 1 $ GeV/c for pions, 
$p_T < 1.5 $ GeV/c for kaons, $p_T < 2 $ GeV/c for protons and lambdas,
because hard processes may significantly contribute to the particle
yield above these momenta~\cite{PeitzmannBWHighPt}.

\begin{figure}[h]
\epsfig{file=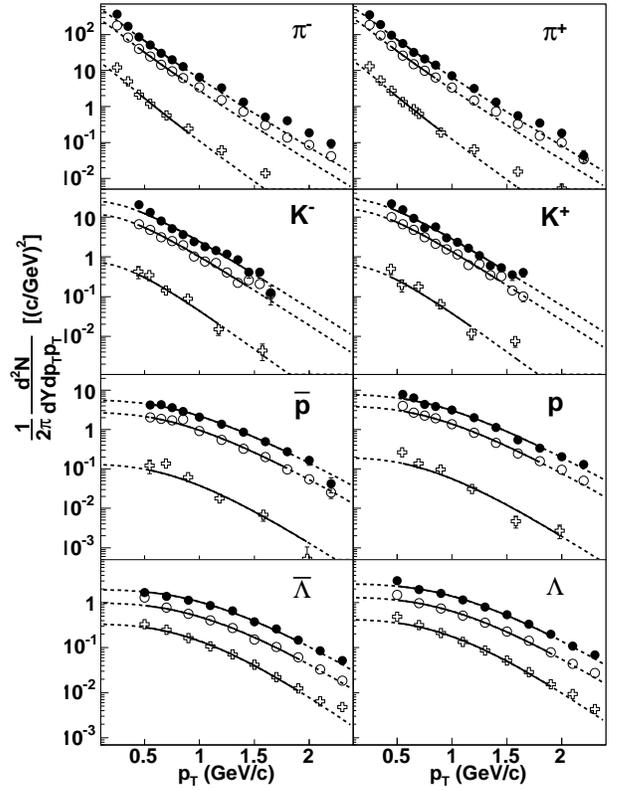,width=8cm}
\caption{Comparison of the data with the blast-wave calculations performed with the
best fit parameters in three centrality bins. The closed circles are central data,
the open circles are mid-central data and the crosses are peripheral data. The plain
lines show the blast wave calculation within the fit range while the dash lines show the 
extrapolation over the whole range.
\label{fig:FitPtSpectra}}
\end{figure}

\begin{figure}[h]
\epsfig{file=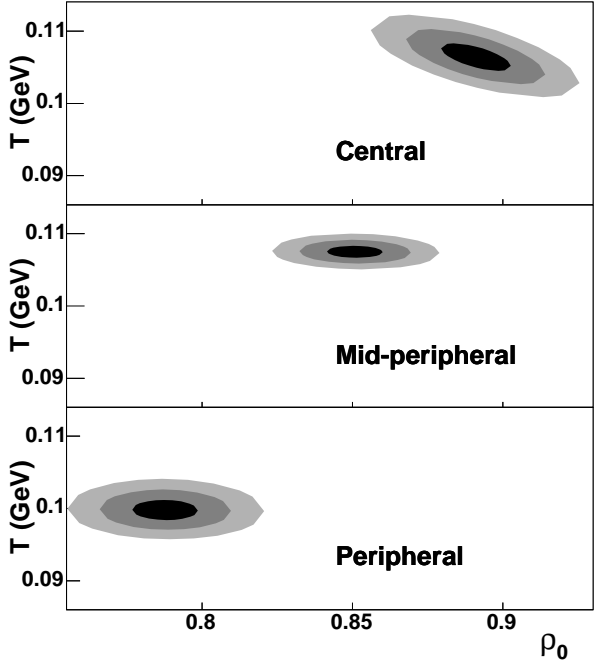,width=8cm}
\caption{Number of $\sigma$ contour of the maximum flow rapidity ($\rho_0$)
with respect to the temperature (T). The shadows show the 1, 2 and 3
 $\sigma$ contours from darkest to lightest.
\label{fig:FitContBetaT}}
\end{figure}

Given the large number of data points from spectra, the temperature (T) and 
the maximum flow rapidity ($\rho_0$) parameters are  constrained by 
transverse mass spectra. A comparison of data and the blast-wave calculation 
with the best fit parameters is shown in 
Figure~\ref{fig:FitPtSpectra}. 
In semi-logarithmic plots, the data and the model fits seem to agree very well.
The $\chi^2$/(\#~data~points) in Table~\ref{tab:FitResult}
show, however, that the fit quality varies between particle
species and centrality bins.  
The correlation between the flow rapidity and temperature is shown
in Figure~\ref{fig:FitContBetaT}.
The extracted temperatures and flow velocities are consistent
with similar studies reported in 
Ref.~\cite{BurwardHoy,Peitzmann,Tomasik,NuKanetaQM01}. The 
differences between these analyses are
due to the use of different data sets, different fit ranges, and different flow and 
spatial profiles.
The study reported in Ref.~\cite{Peitzmann} shows that accounting for resonance feed-down
leads to a temperature 30\% higher and a transverse velocity 15\% lower
than in our study.  

\begin{figure}[h]
\epsfig{file=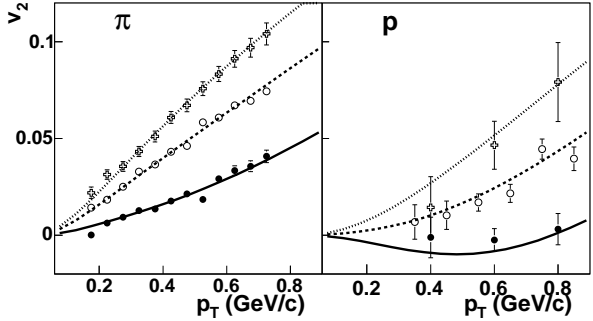,width=8cm}
\caption{Comparison of the $v_2$ data with the blast-wave calculations performed with the best fit parameters in three centrality bins. The closed circles are central data, the open circles are mid-central data and the crosses are peripheral data.
\label{fig:FitV2}}
\end{figure}

\begin{figure}[h]
\epsfig{file=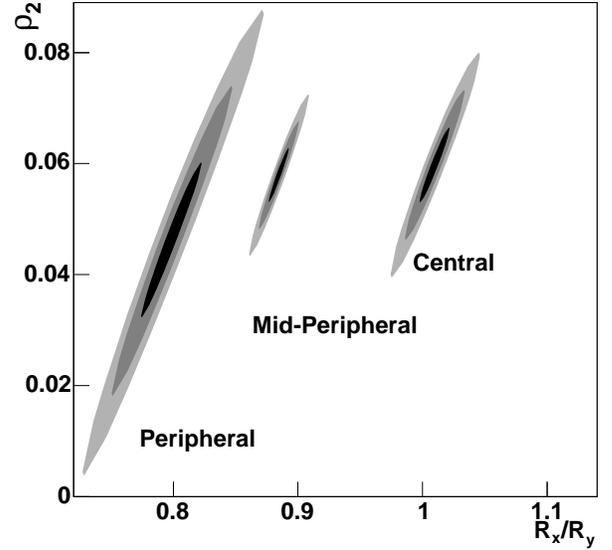,width=8cm}
\caption{Number of $\sigma$ contours of the ratio of the in-plane radius
over the out-of-plane radius ($R_x$/$R_y$) with respect to the flow rapidity oscillation parameter ($\rho_2$). The colors show the 1, 2 and 3 
 $\sigma$ contours from darkest to lightest.
\label{fig:FitContRhoaS2}}
\end{figure}

The fit of $v_2$ mainly constrains the flow rapidity modulation parameter,
$\rho_2$, and the ratio between the radius in-plane and out-of-plane, $R_x/R_y$.
A comparison between the data and the blast-wave calculation with the best fit parameters is shown in Figure ~\ref{fig:FitV2}. The 
correlation between $\rho_2$ and $R_x/R_y$ is shown 
in Figure~\ref{fig:FitContRhoaS2}.
The fit performed in this paper differs from the fit reported
together with the data in Ref.~\cite{STARv2ID}. There are three main differences:
(i) in Ref.~\cite{STARv2ID}, the temperature and flow velocity are free parameters 
while in this paper they are mostly determined by
spectra shapes, (ii) a filled volume and a flow profile are used instead of a shell,
(iii) the fits are performed on three centrality bins instead of the minimum
bias sample. Thus, a direct comparison between the fit parameters extracted
in this paper and those reported in Ref.~\cite{STARv2ID} is difficult.
The freeze-out spatial anisotropy probed by the ratio
$R_x/R_y$, increases when increasing the initial state spatial anisotropy, $i.e.$ 
when going from central to peripheral events. The flow
modulation decreases from central to peripheral events, following the same 
trend as $\rho_0$.

\begin{figure}[h]
\epsfig{file=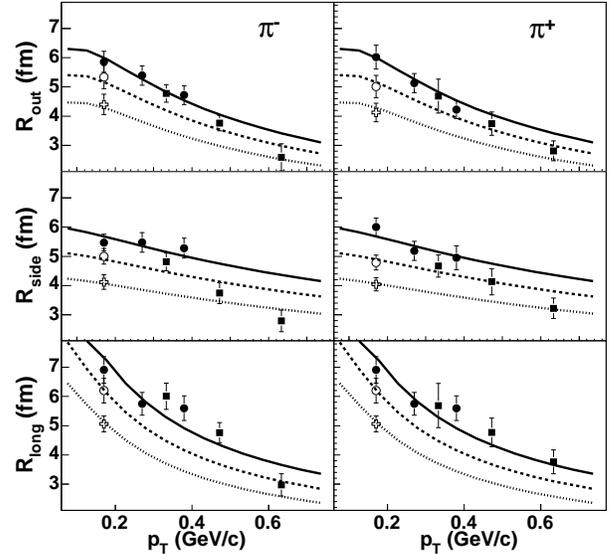,width=8cm}
\caption{Comparison of the pion source data measured by the STAR (circle)
and the PHENIX (box) collaborations with the blast-wave calculations.
Only the STAR data were used to constrain the blast-wave parameters.
The closed circles are the central data,
the open circles are the mid-central data and the crosses are peripheral data.
\label{fig:FitPionHbt}}
\end{figure}

The pion source radii allow us to estimate the parameters
$R_x$ (or $R_y$), $\tau$ and $\Delta t$. As discussed earlier, the ratio
$R_x/R_y$ is constrained by the fit to $v_2$. 
In the mid-peripheral and peripheral bins, the pion source size has been
measured at only one transverse momentum, which weakly constrain the
parameters. The comparison between the data and the blast-wave calculation
performed with the best fit parameters is shown in Figure~\ref{fig:FitPionHbt}.
This figure includes the pion source radii reported by the PHENIX
collaboration~\cite{PhenixHBT} although they weren't used in the fit.
These radii were measured over the 30\% most central events,
which doesn't overlap well with the STAR collaboration centrality bins (0-12\%,
12-32\% and 32-72\%).  Qualitatively the PHENIX data points agree with the
blast-wave fit as they oscillate between the central and mid-peripheral lines.
The  $\chi^{2}$ obtained from the fit is small, because the systematic errors were added in quadrature to the statistical errors. 
The system radius increases from peripheral events to central events
following the growth of the system initial state. 
The final system radius is roughly twice the initial system radius, which
is a clear evidence of the system expansion.
The proper time ($\tau$) is 
surprisingly small. Typically, hydrodynamical calculations reach kinetic freeze-out 
in 15 fm/c, not 9 fm/c~\cite{KolbLatest}. The hydrodynamical 
calculations may be over-predicting the system life-time or the assumption underlying
the extraction of $\tau$ may not be valid.
The assumption of a boost invariant expansion along the beam axis from 
a narrow (less than 1 fm thick)  pancake may be too simplistic.
 Indeed,
it has been shown that accounting for the fluid viscosity breaks the
boost invariant flow~\cite{TeaneyViscosity}. Furthermore, 
three dimensional hydrodynamical
calculations reproduce the measured pseudo-rapidity distribution only
when the initial width of the system along the beam axis extends
over several units of spatial 
pseudo-rapidity~\cite{Hirano3D}. Such initial conditions
lead to a longitudinal expansion that also breaks boost invariant scaling. In these
calculations, even though the system persists
for 15 fm/c, the $R_{long}$ radii is well reproduced.
Thus, the absolute value of  proper life-time, $\tau$, extracted within the blast-wave
parameterization may be questionable.  Along the same lines, the very
short emission duration is surprising. Indeed, a scenario where particles
are emitted in a flash would appear rather unlikely. 
(On the other hand, processes such as supercooling~\cite{CsorgoCsernaiFlash}
have been discussed, and may lead to such a scenario.)
Recent experimental data
~\cite{MercedesQM02} show that the value of the $R_{out}$ radius may 
have been underestimated by the experiments. Thus, since the emission
duration parameters $\Delta t$ strongly depends on $R_{out}$, it is not
appropriate to draw definite conclusions until the experimental issues
are settled.

\begin{figure}[h]
\epsfig{file=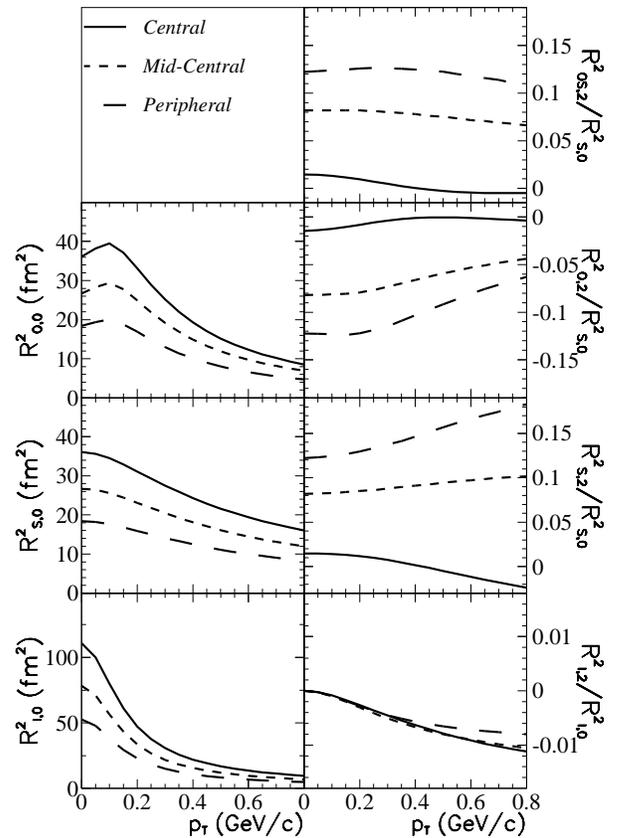,width=8cm}
\caption{Oscillations of the pion source radii obtained with the best fit parameters
in the three centrality bins. 
\label{fig:FitHbtOsc}}
\end{figure}

\begin{figure}[h]
\epsfig{file=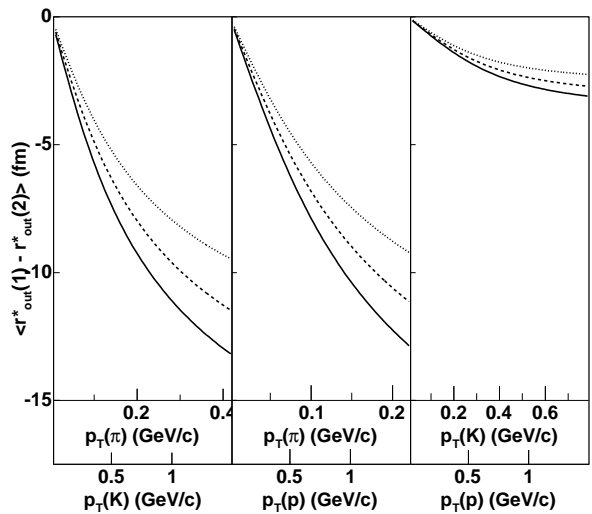,width=8cm}
\caption{Space-time separation between pions and kaons (left), pions and protons (middle) and kaons 
and protons (right) calculated with the best fit parameters in the three centrality bins. 
Plain line: parameters from fit to central data. Dash line: parameters from mid-peripheral data.
Dotted line: parameters from fit to peripheral data.
\label{fig:FitNonId}}
\end{figure}

\begin{figure}[h]
\epsfig{file=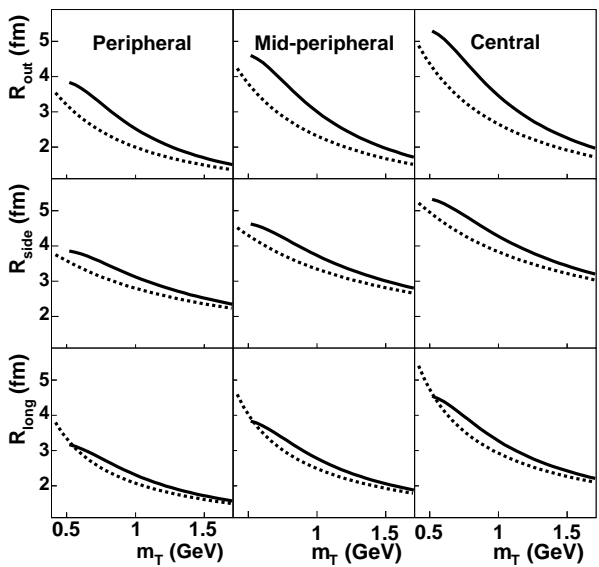,width=8cm}
\caption{Kaon source radii (plain line) compare to the pion radii (dash line). The
blast-wave calculation are performed with the best fit parameters in each centrality
bins.
\label{fig:FitKaonHbt}}
\end{figure}

We have shown that $\pi$, K, p, $\Lambda$ transverse momentum spectra, $\pi$, p elliptic flow, 
and pion source
radii measured in Au-Au collisions at $\sqrt{s_{NN}} = $ 130 GeV are well reproduced by the
blast-wave parameterization. With the exception
of the time-scale parameters  (system proper life time and emission duration), which fall short of any realistic 
model calculations, the fit parameters are within expectations. 
This issue may be resolved when new data on pion source radii become available. 
Based on the published data in Au-Au collisions at $\sqrt{s_{NN}} = $ 130 GeV, we conclude that the 
blast-wave parameterization provides a good description of the system freeze-out stage.

This conclusion will be tested in the future, using the sensitivity of the blast-wave parameterization
to observables that have been presented in the previous section: the oscillation of the pion radii with 
respect to the reaction plane, and the space-time separation between the emission points of 
different particle species. Using the parameters extracted from the fits we 
have calculated the corresponding oscillation of the pion radii 
with respect to the
reaction plane (Figure~\ref{fig:FitHbtOsc}) and 
the separations between the average space-time emission point of pions
kaons and protons (Figure~\ref{fig:FitNonId}).  
We have also calculated the kaon source radii as shown in Figure~\ref{fig:FitKaonHbt} 
since they may become available from the RHIC 
experiments~\cite{MercedesQM02,PHENIXHBTQM02}.
The blast-wave parameterization faces the challenge of simultaneously
reproducing a large variety of observables that will be measured in 
Au-Au collisions at 200 GeV, namely: (1) pion, kaon, proton and $\Lambda$ transverse 
momentum spectra, (2) the elliptic flow of many particle species, 
(3) the pion source radii including the oscillations with respect to the reaction
plane, (4) the kaon source radii, (5) the space-time separation between pion, 
kaon and proton sources. If a satisfactory agreement between these various 
measurements is achieved, it would provide evidence of a
collective expansion that would be very challenging to avoid.
Furthermore, when high quality data becomes available that allows for separate
fits to each particle species, the blast-wave parameterization may 
prove to be an important and handy
tool to assess whether or not the
freeze-out time
 depends on 
the hadronic cross-sections. It will be especially interesting to study 
particles with presumably low hadronic cross sections such as the $\phi$,
$\Xi$, $\Omega$. Since these particles may not be sensitive to hadronic collectivity, they may 
have a higher freeze-out temperature, they may pick up less collective flow,
and freeze-out earlier and from a smaller system than  $\pi$, K, p, $\Lambda$. It may
be possible to distinguish between the collective expansion at the partonic (if any) versus
hadronic stage.

%% file: S5_Conclusion.tex
\section{Summary and Outlook}

We have discussed a scenario of particle kinetic freeze-out
in heavy ion collisions, inspired by the results of full hydrodynamic
calculations of the collision evolution.  The scenario, while simplified,
nevertheless includes several important features which drive experimental
observables of the bulk properties of the system.  This includes random
(``thermal'') motion superimposed upon collective flow fields, an anisotropic
transverse geometry, and a freeze-out distribution in proper time.
The main assumptions of the model are:
longitudinal boost-invariance; 
the same parametric source for all particle types (e.g. $\pi, p$);
and the invariance of source parameters over the kinetic freeze-out
process (or, equivalently, that the model source parameters can
represent the appropriate average of the time-evolving parameter).
We discussed the interplay of the various features, such as space-momentum correlations
depending on the competition between thermal and collective motion and
the source geometry.

The various features (e.g. random thermal motion) were quantified with
several parameters (e.g. $T$) in a mathematical model outlined
in Section~2.  The general class of integrals
which relate the model emission function to the various experimental
observables was identified.

The sensitivity of several observables to the underlying physics
represented by the parameters was studied by systematically varying
each parameter in Section~3.  In addition to momentum-space observables
such as the $p_T$ spectra and elliptic flow, which have been discussed
previously in the context of similar models, we focussed on the sensitivity of
final-state correlations between non-identical particles and 
azimuthally-sensitive HBT, new coordinate-space-sensitive analysis tools
now becoming available with the quality high statistics datasets from RHIC.

The non-trivial interplay between competing physics effects was different
for the different observables.  
It is clear that the model parameter-space can only be constrained by examining
several observables simultaneously; e.g. $p_T$ spectra alone are insensitive
to the source shape ($R_y/R_x$) and only crudely distinguish between high
transverse flow and high temperature.
Furthermore, from Section~3, it is clear that experimental
data can {\it over}-constrain the model parameter space.  If consistency with
$p_T$ spectra and HBT data demand a high temperature and small transverse
boost velocity, this has inescapable consequences for $\pi-K$ correlations
and elliptic flow.  This allows for the ``break-down'' of the parametric
model, and indicates the need to consider new driving physics effects.

In Section~4, the model was used as a functional form in a simultaneous
fit to particle spectra ($\frac{dN}{dp_T}(p_T,m)$), elliptic flow 
($v_2(p_T,m)$), and pion HBT ($R(p_T)$) for three centrality
classes from published RHIC results for Au-Au collisions at
$\sqrt{s_{NN}}=130$~GeV.  While higher-$p_T$ HBT radii from PHENIX are
only qualitatively described, a good fit is obtained to the $p_T$ spectra,
elliptic flow, and low-$p_T$ STAR HBT radii.

Most of the physical parameter values obtained from the fits, as well as their
evolution with collision centrality, fall within reasonable expectations.
The exceptions are the evolution duration $\tau_0$ and emission duration
$\Delta\tau$, which are shorter than most physical models predict.
The $\Delta\tau$ discrepancy may be partially resolved by an improved
experimental treatment of the Coulomb interaction in HBT analyses; this
will be evaluated upon publication of the new results from RHIC.  The
$\tau_0$ value is strongly connected to the model assumption of
boost-invariance; if this assumption is invalid at mid-rapidity at RHIC (e.g. if
there is longitudinal acceleration in the early system evolution),
we expect our fit values to be only lower limits on the evolution duration.

Finally, Section~4 discusses expectations for kaon HBT,
azimuthally-sensitive pion HBT,
and emission space-time separations between non-identical particles
($\pi$, $K$, $p$), based on the fit parameters to the published data.
Expectations for the azimuthally-sensitive pion HBT
are driven by the elliptic flow data, which
indicates $R_y > R_x$ (see Table~\ref{tab:FitResult})-- i.e. the freeze-out
system shape is out-of-plane extended, qualitatively similar to the geometrical
overlap configuration in the entrance channel of the collision.  From
Section~\ref{sec:HBT_radii}, it is clear that this determines the phases
(and amplitudes) of the expected oscillations in the HBT radii-- i.e.
the signs and magnitudes of $R^2_{\mu,2}$.  If azimuthally-sensitive HBT
measurements confirm this out-of-plane freeze-out configuration, it
might provide further evidence of short evolution times $\tau_0$, since
it would imply that the source, which is expanding faster in the reaction
plane than out of it, did not have sufficient time to overcome its initial
geometric anisotropy before freeze-out.  Quantifying the timescale in this
way is outside the scope of the blast-wave model, which does not attempt to
describe the system evolution.

Exploration of simple parameterizations such as the blast-wave is driven
by the desire to connect observations to driving underlying physics.
Furthermore, 
a quantitative interpretation of many measurements (perhaps especially
two-particle interferometry) requires some model assumptions.  Naturally,
one prefers
assumptions which are consistent with {\it other} measurements (e.g. elliptic
flow) in the same system.  Especially since first-principle model calculations
have difficulty consistently explaining the range of observations in the
soft sector of RHIC, it is hoped that insight may be gained from rather simple
theory-{\it inspired} parameterizations such as the blast-wave.  

Inasmuch
as such parameterizations reproduce observations, there is hope that the
driving physics effects have been approximately quantified, and one may
compare the ``underlying'' parameters (e.g. timescales) with first-principles
calculations in an effort to isolate the cause of the discrepancy between
such calculations and observation.  
On the other hand, when simplistic parameterizations cannot, with any set
of parameters, reproduce the main features of the data, it indicates that
something additional is driving observations.  In parameterizations, features
may be implemented and ``turned on and off,'' one at a time, testing for
consistency with a {\it range} of experimental observations.  The insight
gained from such studies can then be fed back into more ``theoretical''
models.  It is hoped that new experimental data will further constrain-- or
break-- the simple parameterization, generating insights into the dynamics
of heavy ion collisions at RHIC.

%% file: BW.bbl
\begin{thebibliography}{99}


\bibitem{QMconferences}
  For reviews and recent developments, see proceedings of the Quark Matter
  conferences:
  QM99 [Nucl. Phys. {\bf A661} (1999)];
  QM01 [Nucl. Phys. {\bf A698} (2002)];
  QM02 [Nucl. Phys. {\bf A715} (2003)].

\bibitem{leptonicObservables}
    E.V. Shuryak, Phys. Lett. {\bf B78} 150 (1978).

\bibitem{gammaProbes}
    E.V. Shuryak, Sov. J. Nucl. Phys. {\bf 28} 408 (1978);
    J. Kapusta, P. Lichard, and D. Seibert, Phys. Rev. {\bf D44} 2774 (1991).

\bibitem{highpTexp}
   BRAHMS Collaboration, Phys. Rev. Lett. {\bf 91} 072305 (2003);
   PHENIX Collaboration, Phys. Rev. Lett. {\bf 87} 052301 (2001);
                   ibid, nucl-ex/0207009;
                   ibid, Phys. Rev. Lett {\bf 91} 072301 (2003);
                   ibid, Phys. Rev. Lett {\bf 91} 072303 (2003);
                   ibid, nucl-ex/0308006;
   PHOBOS Collaboration, Phys. Rev. Lett. {\bf 91} 072302 (2003);
   STAR Collaboration, Phys. Rev. Lett. {\bf 89} 202301 (2002);
                 ibid, Phys. Rev. Lett. {\bf 90} 032301 (2003);
                 ibid, Phys. Rev. Lett. {\bf 90} 082302 (2003);
                 ibid, Phys. Rev. Lett. {\bf 91} 172302 (2003);
                 ibid, Phys. Rev. Lett. {\bf 91} 072304 (2003).

\bibitem{highpTtheory}
    R. Baier, D. Schiff, and B.G. Zakharov, Ann. Rev. Nucl. Part. Sci. {\bf 50} 37 (2000);
    M. Gyulassy, I. Vitev, X.N. Wang, and B. Zhang, nucl-th/0302077.


\bibitem{WH99}
  U.A.~Wiedemann and U.~Heinz,
  Phys.\ Rept.\  {\bf 319}, 145 (1999);
  U.~Heinz and B.V.~Jacak,
  Ann.\ Rev.\ Nucl.\ Part.\ Sci.\  {\bf 49}, 529 (1999).

\bibitem{HanburryBrownTwiss}
  R. Haburry-Brown and R.Q. Twiss, Nature {\bf 177} 27 (1956) and {\bf 178} 1046 (1960).

\bibitem{CascadeFail}
 S. Soff , S. Bass, D. Hardtke, S. Panitkin, Nucl. Phys. {\bf A715} 801 (2003).

\bibitem{HK01}
   U. Heinz and P.F. Kolb, Nucl. Phys. {\bf A702} 269 (2002); hep-ph/0111075.

\bibitem{Hirano_HBT}
   K. Morita, S. Muroya, C. Nonaka, and T. Hirano, Phys. Rev. {\bf C66} 054904 (2002).

\bibitem{Hirano_SpectraV2}
   T. Hirano and K. Tsuda, Phys. Rev. {\bf C66} 054905 (2002).

\bibitem{AMPT}
  B. Zhang, C.M. Ko, B.A. Li, and Z.W. Lin, Phys. Rev. {\bf C61} 067901 (2000);
  Z.W. Lin, S. Pal, C.M. Ko, B.A. Li, and B. Zhang, Phys. Rev. {\bf C64} 011902 (2001).

\bibitem{ZiWei}
  Especially after so-called ``non-flow'' contributions~\cite{STARnonflow} are removed from the experimental data,
  AMPT requires parton cross-sections $\sigma\lesssim 3$~mb to reproduce STAR elliptic flow data~\cite{AMPTv2}
  and $\sigma\approx 10$~mb to reproduce HBT data.  We thank Zi-wei Lin for bringing this to our attention.


\bibitem{STARnonflow}
  STAR Collaboration, C. Adler, {\it et al}, Phys. Rev. {\bf C66} 034904 (2002).

\bibitem{AMPTv2}
  Z.W. Lin and C.M. Ko, Phys. Rev. {\bf C65} 034904 (2002).

\bibitem{AMPT_HBT}
  Z.W. Lin, C.M. Ko, and S. Pal, Phys. Rev. Lett. {\bf 89} 152301 (2002).

\bibitem{BudaLund}
  M. Csan\'{a}d, T. Cs\"{o}rg\H{o}, and B. L\"{o}rstad, nucl-th/0310040;
  M. Csan\'{a}d, T. Cs\"{o}rg\H{o}, B. L\"{o}rstad, and A. Ster, Acta. Phys. Polon. {\bf B35} 191 (2004).


\bibitem{HKHRV01}
  P. Huovinen, P.F. Kolb, U. Heinz, P.V. Ruuskanen, S. Voloshin,
  Phys. Lett. {\bf B503} 58 (2001). 

\bibitem{RQMD}
   H. Sorge, Phys. Rev. Lett. {\bf 78} 2309 (1997).

\bibitem{Humanic}
     T. Humanic, nucl-th/0203004; ibid, Nucl. Phys. {\bf A715} 641c (2003).

\bibitem{MolnarPC}
D. Molnar, M. Gyulassy, Nucl.Phys. {\bf A698} 379  (2002)  and nucl-th/0211017.

\bibitem{BassPC}
S.  Bass, B. Mueller, D. Srivastava, Phys.Lett. {\bf B551}  277 (2003).

\bibitem{HK02}
   U. Heinz and P.F. Kolb, Phys. Lett. {\bf B542} 216 (2002).

\bibitem{TLS01}  
  D. Teaney, J. Lauret, E. Shuryak, nucl-th/0110037 (2001).

\bibitem{TeaneyViscosity}
D. Teaney, nucl-th/0301099.

\bibitem{NA49Fischer}
    H.G. Fischer, Nucl. Phys. {\bf A715} 118c (2003).

\bibitem{ColorGlassCond}
    Y.V. Kovchegov and K.L. Tuchin, Nucl. Phys. {\bf A708} 413 (2002). 
Nucl. Phys. {\bf A717} 249 (2003).

\bibitem{PeitzmannBWHighPt}
T. Peitzmann, nucl-th/0303046.

\bibitem{BurwardHoy} 
J. M. Burward-Hoy for the PHENIX Collaboration, Nucl.Phys. {\bf A715} 498 (2003).

\bibitem{Peitzmann}
T. Peitzmann, Eur.Phys.J. {\bf C26} 539 (2003) .


\bibitem{HW02}
  U. Heinz and S.H.M. Wong, Phys. Rev. {\bf C66} 014907 (2002).


\bibitem{Tomasik}
B. Tomasik, proceedings of 38 Rencontres de Moriond, QCD and hadronic interactions, nucl-th/0304079.

\bibitem{Westfall76}
  G.D. Westfall, J. Gosset, P.J. Johansen, A.M. Poskanzer, W.G. Meyer, H.H. Gutbrod, A. Sandoval, and R. Stock,
  Phys. Rev. Lett. {\bf 37} 1202 (1976).

\bibitem{BGZ78}
   J.P. Bondorf, S.I.A. Garpman, and J. Zimanyi, Nucl. Phys. {\bf A296} 320 (1978).

\bibitem{SR79}
   P.J. Siemens and J.O. Rasmussen, Phys. Rev. Lett. {\bf 42} 880 (1979).

\bibitem{SSH93}
  E. Schnedermann, J. Sollfrank, U. Heinz, Phys. Rev. {\bf C48} 2462 (1993).

\bibitem{BjorkenBI}
  J.D. Bjorken, Phys. Rev. {\bf D27} 140 (1983).

\bibitem{STARv2ID}
  STAR Collaboration, C. Adler, {\it et al}, Phys. Rev. Lett. {\bf 87} 182301 (2001).

\bibitem{VP00}
   S.A. Voloshin and A.M. Poskanzer, Phys. Lett. {\bf B474} 27 (2000).

\bibitem{VoloshinQM03}
   S.A. Voloshin, Nucl. Phys. {\bf A715} 379c (2003); ibid, nucl-th/0202072.

\bibitem{Voloshin97}
   S.A. Voloshin, Phys. Rev. {\bf C55} 1630 (1997).


\bibitem{FabriceBW}
  STAR Collaboration, F. Reti\`{e}re, {\it et al}, nucl-ex/0111013 (2001).


\bibitem{PrattBertsch}
  S. Pratt, T. Cs\"{o}rg\H{o}, and J. Zimanyi, Phys. Rev. {\bf C42} 2646 (1990);
  G. Bertsch, M. Gong, and M. Tohyama, Phys. Rev. {\bf C37} 1896 (1988).

\bibitem{W98}
  U.A. Wiedemann, Phys. Rev. {\bf C57} 266 (1998).

\bibitem{Tomasik9901}
   B. Tom\'{a}\u{s}ik, U.A. Wiedemann, and U. Heinz, Nucl. Phys. {\bf A663} 753 (2000);
   B. Tom\'{a}\u{s}ik and U. Heinz, Phys. Rev. {\bf C65} 031902 (2002).

\bibitem{thanksSergeiVoloshin}
   We thank Sergei Voloshin for correcting an error we had previously made in deriving Eq.~\ref{eq:s2-in-BW}.

\bibitem{LHW00}
  M.A. Lisa, U. Heinz, U.A. Wiedemann, Phys. Lett. {\bf B489} 287 (2000).
\bibitem{HHLW02}
  U. Heinz, A. Hummel, M.A. Lisa, U.A. Wiedemann, Phys. Rev. {\bf C66} 044903 (2002).
\bibitem{E895HBTwrtRP}
  E895 Collaboration, M.A. Lisa {\it et al}, Phys. Lett. {\bf B496} 1 (2000).
\bibitem{BudaLundEllipse}
  T. Cs\"{o}rg\H{o}, S.V. Akkelin, Y. Hama, B. Luk\'{a}cs, and Yu. M. Sinyukov
  Phys. Rev. {\bf C67} 034904 (2003).

\bibitem{PhenixHBT}
  Phenix Collaboration, K. Adcox, {\it et al.}, Phys. Rev. Lett. {\bf 88} 192302 (2002).

\bibitem{MS88}
  A.N. Makhlin and Yu.M. Sinyukov, Z. Phys. {\bf C39} 69 (1988).

\bibitem{AS95}
  S.V. Akkelin and Yu.M. Sinyukov, Phys. Lett. {\bf B356} 525 (1995).

\bibitem{HB95}
  M. Herrmann and G.F. Bertsch, Phys. Rev. {\bf C51} 328 (1995).

\bibitem{WSH96}
  U.A. Wiedemann, P. Scotto, U. Heinz,
  Phys. Rev. {\bf C53} 918 (1996).

\bibitem{CL96}
  T. Cs\"{o}rg\H{o} and B. L\"{o}rstad, Phys. Rev. {\bf C54} 1390 (1996).

\bibitem{WHTW98}
  Y.-F. Wu, U. Heinz, B. Tom\'{a}\u{s}ik, and U. Wiedemann,
  Eur. Phys. J. {\bf C1} 599 (1998).


\bibitem{caviat}
Since particle emission occurs only after the collision begins (i.e. when $\tau > 0$),
the lower limit on the $\tau$ integration in Equations~\ref{eq:FourFoldIntegral} and~\ref{eq:Hs}
ought to be zero instead of $-\infty$.  This distinction vanishes if $\Delta\tau \ll \tau_0$, and
the effect is negligible for the range of parameter values explored here.

\bibitem{S95}   
  Yu. M Sinyukov, In {\it Hot Hadronic Matter: Theory and Experiment}, ed. J. Rafelski.
  New York: Plenum, NATO ASI Series B 346:309 (1995).

\bibitem{HardtkeVoloshin}
D. Hardtke and S. Voloshin, Phys.Rev. {\bf C61} 024905 (2000). 

\bibitem{HBTImaging}
D. Brown and P. Danielewicz, Phys.Rev. {\bf C64} 014902 (2001). P. Danielewicz, D. A. Brown, M. Heffner, S. Pratt and R. Soltz, nucl-th/0407022.

\bibitem{CoreHalo}
S. Nickerson, T. Cs\"{o}rg\H{o} and D. Kiang, Phys.Rev. {\bf C57} 3251 (1998) .

\bibitem{HeinzResonance}
U. Wiedemann and U. Heinz, Phys.Rev. {\bf C56} 3265 (1997).

\bibitem{Magestro}
D. Magestro private communication. Model as described in  P. Braun-Munzinger, K. Redlich, J. Stachel, Invited review for Quark Gluon Plasma 3, eds. R. C. Hwa and Xin-Nian Wang, World Scientific Publishing, nucl-th/0304013.

\bibitem{CoreHalo2}
T. Cs\"{o}rg\H{o}, B. L\"{o}rstad, and J. Zim\'{a}nyi, Z. Phys. {\bf C71} 491 (1996).

\bibitem{RG96}
  D.H. Rischke and M. Gyulassy, Nucl. Phys. {\bf A608} 479 (1996);
  D.H. Rischke, Nucl. Phys. {\bf A610} 88c (1996).

\bibitem{STARHBT}
  STAR Collaboration, C. Adler, {\it et al}, Phys. Rev. Lett. {\bf 87} 082301 (2001).

\bibitem{peter_dan_private}
  P.~Kolb, D.~Magestro, private communications.

\bibitem{PLBNonId}
R. Lednick\'{y}, V.I. Lyuboshitz, B. Erazmus, D. Nouais, Phys. Lett. B  {\bf 373} (1996) 30.
\bibitem{LedNonId}
      R. Lednick\'{y}, nucl-th/0305027; Proc. CIPPQG'01, nucl-th/0112011; 
             Proc. XXXII ISMD, nucl-th/0212089.
\bibitem{PRLNonId}
 S. Voloshin, R. Lednick\'{y}, S. Panitkin, N. Xu, Phys. Rev. Lett.  {\bf 79} (1997) 4766.

\bibitem{StrangeDistil}
D. Ardouin et al., Phys.Lett. B {\bf 446} (1999) 191, nucl-th/0203030.

\bibitem{earlyStrangeDistil}
  C. Greiner, P. Koch, H. St\"{o}cker, Phys. Rev. Lett. {\bf 58} 1825 (1987);
  B. Luk\'{a}cs, J. Zim\'{a}nyi and N.L. Balazs, Phys. Lett. {\bf B183} 27 (1987);
  M. Gyulassy, Phys. Lett. {\bf B286} 211 (1992).

\bibitem{QM03NonId}
F. Reti\`ere (STAR collaboration) , Nucl Phys. A {\bf 715} (2003) 591c, nucl-ex/0212026.

\bibitem{PiKPRL}
J. Adams et al. (STAR collaboration), submitted to Phys. Rev. Lett., nucl-ex/0307025.

\bibitem{NonIdOutSideLong}
R. Lednick\'{y}, Proc. 8th Int. Workshop on Multiparticle Production, Correlations and Fluctuations (1998) 148, 
nucl-th/0304063, and, R. Lednick\'{y}, S. Panitkin, and N. Xu, nucl-th/0304062.

\bibitem{PhobosV2}
B. Back et al., Submitted to Phys. Rev. Lett., nucl-ex/0406021.

\bibitem{STARSpectra200}
J. Adams et al., Phys. Rev. Lett. {\bf 92} (2004) 112301

\bibitem{PhenixSpectra130}
 K. Adcox et al., Phys. Rev. Lett. {\bf 88} 242301 (2002).

\bibitem{StarLaSpectra130}
 C. Adler et al., Phys. Rev. Lett. {\bf 89} 092301 (2002).

\bibitem{NuKanetaQM01}
  N. Xu and M. Kaneta, Nucl. Phys. {\bf A698} 306c (2002).

\bibitem{KolbLatest}
P. Kolb, Proceedings for the 19th Winter Workshop on Nuclear Dynamics, nucl-th/0304036.

\bibitem{Hirano3D}
T.Hirano and K. Tsuda, Phys.Rev. {\bf C66} 041901 (2002).

\bibitem{CsorgoCsernaiFlash}
T. Cs\"{o}rg\H{o} and L. P. Csernai, Phys.Lett. {\bf B333} 494  (1994). hep-ph/9406365.

\bibitem{MercedesQM02}
  STAR Collaboration, M. L\'{o}pez-Noriega, {\it et al}, Nucl. Phys. {\bf A715} 623c (2003).


\bibitem{PHENIXHBTQM02}
PHENIX Collaboration, A. Enokizono, {\it et al}, Nucl. Phys. {\bf A715} 595c (2003).




\bibitem{HKH02}  
   P. Houvinen, P.F. Kolb, and U. Heinz, Nucl. Phys. {\bf A698} 475c (2002).
\bibitem{S02}
  E.V. Shuryak, Phys. Rev. {\bf C66} 027902 (2002).


\end{thebibliography}
